\pdfoutput=1
\documentclass[11pt,letterpaper]{article}

\usepackage[margin=1in]{geometry}

\usepackage{amsmath,amssymb,amsthm,mathtools}
\usepackage{bm}
\usepackage{microtype}
\usepackage{xcolor}
\usepackage{booktabs}
\usepackage{enumitem}
\usepackage{caption}
\usepackage{subcaption}
\usepackage{tikz}
\usetikzlibrary{arrows.meta,calc,positioning,shapes.geometric,shapes.misc,fit,backgrounds,decorations.pathreplacing}
\usepackage{pgfplots}
\pgfplotsset{compat=1.18}
\usepackage[colorlinks=true,linkcolor=blue!55!black,citecolor=green!40!black,urlcolor=blue!55!black]{hyperref}
\usepackage[capitalise,nameinlink]{cleveref}
\usepackage{etoolbox}
\makeatletter
\patchcmd{\l@section}{\addvspace{1.0em \@plus\p@}}{\addvspace{0.45em \@plus\p@}}{}{}
\makeatother

\setlist{itemsep=2pt,topsep=4pt}

\usepackage{aliascnt}
\newtheorem{theorem}{Theorem}[section]

\newaliascnt{lemma}{theorem}
\newtheorem{lemma}[lemma]{Lemma}
\aliascntresetthe{lemma}

\newaliascnt{corollary}{theorem}
\newtheorem{corollary}[corollary]{Corollary}
\aliascntresetthe{corollary}

\newaliascnt{claim}{theorem}
\newtheorem{claim}[claim]{Claim}
\aliascntresetthe{claim}

\newaliascnt{fact}{theorem}
\newtheorem{fact}[fact]{Fact}
\aliascntresetthe{fact}

\newaliascnt{proposition}{theorem}
\newtheorem{proposition}[proposition]{Proposition}
\aliascntresetthe{proposition}

\theoremstyle{definition}
\newaliascnt{definition}{theorem}
\newtheorem{definition}[definition]{Definition}
\aliascntresetthe{definition}

\theoremstyle{remark}
\newaliascnt{remark}{theorem}

\aliascntresetthe{remark}

\crefname{fact}{Fact}{Facts}
\crefname{proposition}{Proposition}{Propositions}
\crefname{claim}{Claim}{Claims}
\crefname{lemma}{Lemma}{Lemmas}
\crefname{corollary}{Corollary}{Corollaries}
\crefname{definition}{Definition}{Definitions}
\crefname{remark}{Remark}{Remarks}

\newcommand{\eps}{\varepsilon}
\newcommand{\R}{\mathbb{R}}

\newcommand{\Fq}{\mathbb{F}_q}
\newcommand{\OPT}{\mathrm{OPT}}
\DeclareMathOperator{\dist}{dist}
\DeclareMathOperator{\cost}{cost}
\DeclareMathOperator{\lev}{lev}
\DeclareMathOperator{\parent}{par}
\newcommand{\up}[2]{#1^{\uparrow}_{#2}}
\newcommand{\dn}[2]{#1^{\downarrow}_{#2}}
\newcommand{\src}[1]{#1^{\mathrm{src}}}
\newcommand{\snk}[1]{#1^{\mathrm{snk}}}
\newcommand{\Hfam}{\mathbb{D}}
\newcommand{\Dec}{\mathcal{D}}
\newcommand{\Pcal}{\mathcal{P}}
\newcommand{\Qcal}{\mathcal{Q}}
\newcommand{\Net}{\mathcal{N}_{\Dec}}
\newcommand{\uOPT}{\underline{\OPT}}
\DeclareMathOperator{\loc}{loc}
\DeclareMathOperator{\bal}{bal}
\DeclareMathOperator{\bag}{bag}

\definecolor{blockcol}{RGB}{52,101,164}
\definecolor{centcol}{RGB}{204,102,0}
\definecolor{rescol}{RGB}{78,154,6}
\definecolor{sepcol}{RGB}{164,0,0}
\definecolor{flowcol}{RGB}{136,138,133}

\title{Constant-Factor Approximation for Capacitated Clustering\\in Low Dimension with $(1+\varepsilon)k$ Centers}
\author{Kangke Cheng$^{1}$\\{\footnotesize\texttt{ke314159@mail.ustc.edu.cn}}
\and Guanlin Mo$^{1}$\\{\footnotesize\texttt{moguanlin@mail.ustc.edu.cn}}
\and Hu Ding$^{1,}$\thanks{Corresponding author.}\\{\footnotesize\texttt{huding@ustc.edu.cn}}}
\date{\normalsize $^{1}$University of Science and Technology of China, Hefei, China}

\begin{document}
\pagenumbering{roman}
\maketitle
\thispagestyle{empty}

\begin{abstract}
Clustering problems such as $k$-median and $k$-means are fundamental in combinatorial optimization. In many of their applications, including load balancing, facility location and redistricting, each center can serve only a bounded number of clients. This leads to \emph{capacitated} clustering, in which a client may be forced to use a distant center because all nearby ones are full. This loss of locality has kept the techniques behind approximation schemes from extending to the capacitated case, and no constant-factor approximation is known in general metrics even for capacitated $k$-median. A long line of work therefore opens $K=k+\lfloor\eps k\rfloor$ centers for a fixed $\eps\in(0,1]$, while respecting the capacities exactly.

We give a deterministic polynomial-time algorithm for uniform soft capacities in low dimension, that is, when the clients or the possible center locations have bounded doubling dimension. With $K$ centers, its ratio is $3-2\eps$ for $k$-median and $9-8\eps$ for $k$-means, or $5-4\eps$ in $\R^d$. Apart from $k$-median in the plane, these are the first polynomial-time ratios bounded by a constant independent of $\eps$; previously, the best ratio was $O(\eps^{-2}\log(1/\eps))$, even for $k$-median. We also obtain ratios that stay bounded as $\eps\to0$ for non-uniform and hard capacities in low dimension, and ratios polylogarithmic in $1/\eps$ in general metrics.

Our approach starts from a simple structural observation: a near-optimal uncapacitated solution can be split into fewer than $2k$ capacity-respecting blocks, so $2k$ centers already suffice, one per block.  Our main technical tool, the \textbf{rescue lemma}, shows that once every block that sends clients to a center of the optimal solution is rescued (given a private center), that center can be closed, so each extra center effectively rescues about two blocks. This lets $\lfloor\eps k\rfloor$ extra centers  rescue nearly an $\eps$-fraction of the blocks, yielding the linear dependence on $\eps$ in our ratios.
\end{abstract}

\newpage
{\small\setcounter{tocdepth}{2}\tableofcontents}
\newpage
\pagenumbering{arabic}

\section{Introduction}\label{sec:intro}

Clustering with capacities arises naturally when resources are limited: a server can run only so many concurrent jobs before response times degrade, a distribution center can supply only a bounded number of retail stores, and a voting district must contain a roughly equal share of the population. In each case, ignoring the capacity constraint leads to solutions that are optimal on paper but infeasible in practice~\cite{CohenAddad20,HWSWZ65,CKY18,BG06,MF14}. We formalize this as follows. In the \emph{capacitated $(k,z)$-clustering} problem, we are given a multiset $P$ of $n$ clients in a metric space, a set $F$ of locations where centers may be opened, a positive integer capacity $U_y$ for every location $y\in F$, and a positive integer $k$. A solution opens at most $k$ centers and assigns every client to an open center so that a center at $y$ serves at most $U_y$ clients. Its cost is the sum, over all clients, of the distance to the assigned center raised to a fixed power $z\ge 1$. The cases $z=1$ and $z=2$ are capacitated $k$-median and capacitated $k$-means. The capacities are \emph{uniform} if $U_y=U$ for all $y$; most of our results concern this case. With \emph{soft} capacities, several centers may be opened at the same location; with \emph{hard} capacities, every location hosts at most one center.

Without capacities, these problems are well understood in low-dimensional spaces: for every fixed $z$, they admit polynomial-time approximation schemes in fixed-dimensional Euclidean spaces and, more generally, in metrics of bounded \emph{doubling dimension}~\cite{ARR98,KR07,CKM19,FRS19,CFS21}. A metric has doubling dimension at most $D$ if every ball can be covered by $2^D$ balls of half the radius, and every subset of $\R^d$ has doubling dimension $O(d)$~\cite{GKL03}. Throughout, \emph{low dimension} means that $P$ or $F$ has doubling dimension at most a fixed constant $D$. Where the distinction matters, we speak of \emph{joint} low dimension if $P\cup F$ has doubling dimension at most $D$, and of \emph{one-sided} low dimension otherwise. The approximation schemes rely on locality. In an optimal uncapacitated solution, every client is served by a nearest open center, so each region of the space interacts with the rest only through points near its boundary. Capacities destroy this property, since a client may have to travel to a far-away center because the nearby centers are full. As a result, much less is known with capacities. With capacities, approximation schemes in doubling metrics are known only in quasi-polynomial time, except for $k$-median in the plane~\cite{CohenAddad20}.

The situation in general metrics is much worse: no constant-factor approximation is known even for capacitated $k$-median. The natural LP relaxation has an unbounded integrality gap even if the capacities or the number of centers may be exceeded by any constant factor smaller than $2$~\cite{Li15,BFRS15}. A long line of work therefore exceeds the capacities by a factor close to $1$ or opens $(1+\eps)k$ centers; we review this work in \Cref{sec:related}. Since the capacitated problem is at least as hard as the uncapacitated one, the known lower bounds for the latter apply here as well: $k$-median cannot be approximated better than $1+2/e$~\cite{JMS02}, and $k$-means is APX-hard in high-dimensional Euclidean space~\cite{ACKS15}. Assuming the Unique Games Conjecture~\cite{Kho02}, the reductions of Bhattacharya, Goyal and Jaiswal~\cite{BGJ21} show that both are NP-hard to approximate within some constant greater than $1$ in high dimensions, even with $(1+c)k$ centers for some $c>0$.\footnote{A proof of the conjecture has been announced~\cite{OAI:The-Unique-Games-Theorem-September-23-2026}; we state the hardness as conditional until it is independently verified.} Both $k$-median and $k$-means are NP-hard in the plane~\cite{MS84,MNV12}.

These barriers suggest a natural relaxation: allow the algorithm a small budget of extra centers beyond $k$, while still respecting every capacity exactly (\emph{exact capacities}). We study this regime for capacitated $(k,z)$-clustering, opening
\begin{equation}\label{eq:K}
K:=k+\lfloor \eps k\rfloor
\end{equation}
centers for a fixed $\eps\in(0,1]$. In low dimension, with $2k$ centers, the problem is easy for uniform soft capacities: splitting each cluster of a near-optimal uncapacitated solution into groups of at most $U$ clients gives fewer than $2k$ capacity-respecting groups, and opening one center for each group gives a nearly optimal cost, as we show in \Cref{lem:2k}.  With fewer than $2k$ centers, this simple solution is no longer available. With $(1+\eps)k$ centers, the best previous polynomial-time ratio for capacitated $k$-median was $O(\eps^{-2}\log(1/\eps))$~\cite{Li16}, growing polynomially in $1/\eps$ even in low dimension.

\subsection{Our Results}\label{sec:results}

In low dimension, we show that $(1+\eps)k$ centers and exact capacities suffice for a constant approximation ratio in deterministic polynomial time for every fixed $z\ge1$. For $k$-median, $k$-means and every other fixed $z\ge1$, these are the first polynomial-time algorithms with $(1+\eps)k$ centers whose ratio is bounded by a constant, except for uniform $k$-median in the plane, where a polynomial-time approximation scheme with $k$ centers is known~\cite{CohenAddad20}. Before, such ratios in polynomial time needed a constant factor more centers~\cite{ABGL15,HXDZ22} or a constant factor larger capacities~\cite{CGTS02,BFRS15}.

Throughout, $\OPT$ denotes the optimal cost with at most $k$ centers under the type of capacities of the respective theorem. An algorithm has \emph{ratio} $\alpha$ if it opens at most  $K=k+\lfloor\eps k\rfloor$ centers, respects the capacities, and produces a solution of cost at most  $\alpha\cdot\OPT$.  All our ratios hold up to an additive $\delta>0$ that can be fixed arbitrarily small; we omit $\delta$ from the stated ratios. All our algorithms are deterministic and run in polynomial time.

\paragraph{Uniform soft capacities.}
Our main result concerns $k$-median and $k$-means. Recall that $P$ is the multiset of clients, $F$ is the set of locations where centers may be opened, and the cost raises every distance to the power $z\ge1$, where $z=1$ for $k$-median and $z=2$ for $k$-means. In Euclidean space $\R^d$, we take $F=\R^d$, as usual.

\begin{theorem}[Uniform soft capacities]\label{thm:soft}
For  every fixed  $\eps\in(0,1]$, there is an algorithm for uniform soft capacities in low dimension with ratio $3-2\eps$ for $k$-median and $9-8\eps$ for $k$-means, or $5-4\eps$ in $\R^d$. More generally, for every fixed $z\ge1$, the ratio is $\beta-(\beta-1)\eps$, where $\beta=3^z$ in general and $\beta=\beta^*_z$ in  $\R^d$; the constant $\beta^*_z\le3^z$ is defined in \Cref{sec:certificates}.
\end{theorem}

 For $k$-median, the ratio in $\R^d$ is the same, since $\beta^*_1=3$.

As $\eps\to0$, the ratio tends to the constant $\beta$. The algorithm, however, requires $\eps>0$, since the degree of its polynomial running time grows as $\eps\to0$, as we discuss at the end of this subsection. 

\paragraph{Non-uniform capacities.}
Our second result allows the capacity $U_y$ to vary across locations, as in the algorithms of Li~\cite{Li16} and Dai and Li~\cite{DaiLi26}. A different algorithm gives ratios that decrease exponentially rather than linearly in $\eps$.

\begin{theorem}[Non-uniform capacities]\label{thm:nonuniform}
For  every fixed  $\eps\in(0,1]$, there is an algorithm for non-uniform capacities in low dimension with ratio $1+2e^{-\eps}$ for $k$-median, with soft or hard capacities, and $1+8e^{-\eps}$ for $k$-means with soft capacities. More generally, for every fixed $z\ge1$, the ratio for non-uniform soft capacities is $1+(3^z-1)e^{-\eps}$.
\end{theorem}

For uniform soft capacities, \Cref{thm:soft} is better, since $e^{-\eps}>1-\eps$.  Along the way, we obtain a bicriteria result of independent interest, which we prove as \Cref{cor:bicriteria}: for every $\delta\in(0,1]$, in every finite metric space, a solution that respects non-uniform capacities and costs at most $(1+\delta)\OPT$ can be computed in polynomial time with $O(k\log(2/\delta))$ centers.

\paragraph{Hard capacities and general metrics.}
Our results also extend from soft to hard capacities and from low dimension to general metrics. With hard capacities, each location hosts at most one center, which makes the problem harder because soft solutions cannot be used directly.

\begin{theorem}[Uniform hard capacities]\label{thm:hard}
For  every fixed  $\eps\in(0,1]$, there is an algorithm for uniform hard capacities in low dimension with ratio $1+2e^{-\eps}$ for $k$-median and $(1+2\sqrt{9-8\eps})^2$ for $k$-means; for $k$-means with $P\cup F\subset\R^d$, the ratio is $11-8\eps+2\sqrt{5-4\eps}$. More generally, for every fixed $z>1$, the ratio is at most $\bigl(1+2(\beta-(\beta-1)\eps)^{1/z}\bigr)^z$, where $\beta=3^z$, and $\beta=\beta^*_z$ if $P\cup F\subset\R^d$ and $P$ has doubling dimension at most $D$.
\end{theorem}

For $k$-means in low dimension, the ratio $(1+2\sqrt{9-8\eps})^2$ is at most $49$, its limit as $\eps\to0$.  For $k$-median, the  ratio $1+2e^{-\eps}$  is that of \Cref{thm:nonuniform}, which also covers hard capacities.

The results above assume low dimension. In general metrics,  we obtain ratios $O(1+\log^z(1/\eps))$  for uniform capacities with $(1+\eps)k$ centers, which we prove as \Cref{thm:general}.  For $k$-median, the ratios are $3+2\ln(1/\eps)$ for soft and $7+2\ln(1/\eps)$ for hard capacities, and  for $k$-means, they are $(3+3\ln(1/\eps))^2$ and $(7+3\ln(1/\eps))^2$.

The ratios above are stated in their simplest form, so that their dependence on $\eps$ is easy to read. For $z>1$, \Cref{tab:results} in \Cref{app:sharper} lists sharper ratios.

\paragraph{Running time.}\label{sec:results-time}
For fixed $z$ and $\delta\le1/2$, the algorithms of \Cref{thm:soft,thm:hard,thm:nonuniform} run in time $(n+|F|)^{\varphi}\cdot(\log\Delta)^{O(1)}$, where $\Delta$ is the aspect ratio of $P\cup F$, the ratio between its largest and its smallest positive distance, and the degree $\varphi$ depends only on $D$, $z$, $\eps$ and $\delta$. If $P\cup F$ has doubling dimension at most $D$, then $\varphi=(D/(\eps\delta))^{O(D/\delta)}$. If only $P$ or only $F$ has doubling dimension at most $D$, as for clients of bounded doubling dimension in a Euclidean space of arbitrary dimension $d$, then $\varphi=(1/(\eps\delta))^{(1/\delta)^{O(D)}}$, independently of $d$. Both bounds are derived in the running-time analysis at the end of \Cref{sec:proof-combine}.

\begin{table}[t]
\centering
\footnotesize
\renewcommand{\arraystretch}{1.12}
\setlength{\tabcolsep}{5pt}
\begin{tabular}{@{}lllll@{}}
\toprule
Setting & Centers & Time & $k$-median & $k$-means\\
\midrule
\multicolumn{5}{@{}l}{\emph{Previous work (general metrics and uniform capacities unless noted)}}\\
\cite{Li16}, soft\textsuperscript{$\dagger$} & $(1+\eps)k$ & poly & $O(\eps^{-2}\log\frac1\eps)$ & ---\\
\cite{DaiLi26}, hard\textsuperscript{$\dagger$} & $(1+\eps)k$ & poly & $O(\eps^{-5})$ & ---\\
\cite{ABGL15}, hard & $2k-1$ & poly & $7+\delta$ & ---\\
\cite{KPR00}, hard & $(5+5/\delta)k$ & poly & $1+\delta$ & ---\\
\cite{HXDZ22}, $\R^d$ & $O(k)$ & poly & --- & $O(1)$\\
\cite{CohenAddad20}, doubling & $k$ & qpoly & $1+\delta$ & $1+\delta$\\
\cite{CohenAddad20}, $\R^2$ & $k$ & poly & $1+\delta$ & ---\\
\midrule
\multicolumn{5}{@{}l}{\emph{This paper, low dimension}}\\
uniform, soft & $(1+\eps)k$ & poly & $3-2\eps$ & $9-8\eps$\\
uniform, soft, $\R^d$ & $(1+\eps)k$ & poly & $3-2\eps$ & $5-4\eps$\\
uniform, hard & $(1+\eps)k$ & poly & $1+2e^{-\eps}$ & $(1{+}2\sqrt{9{-}8\eps})^2$\\
non-uniform, soft & $(1+\eps)k$ & poly & $1+2e^{-\eps}$ & $1+8e^{-\eps}$\\
non-uniform, hard & $(1+\eps)k$ & poly & $1+2e^{-\eps}$ & ---\\
\midrule
\multicolumn{5}{@{}l}{\emph{This paper, general metrics (uniform)}}\\
soft & $(1+\eps)k$ & poly & $3+2\ln\frac1\eps$ & $(3+3\ln\frac1\eps)^2$\\
hard & $(1+\eps)k$ & poly & $7+2\ln\frac1\eps$ & $(7+3\ln\frac1\eps)^2$\\
\bottomrule
\end{tabular}
\caption{ Ratios with exact capacities. Time: poly is polynomial and qpoly is quasi-polynomial time.  \textsuperscript{$\dagger$}Non-uniform. Li~\cite{Li15} gave the first ratio with $(1+\eps)k$ centers, $\exp(O(1/\eps^2))$, for uniform capacities. For uniform capacities, soft and hard are equivalent up to a constant~\cite{Li15}. \Cref{tab:results} in \Cref{app:sharper} lists all ratios.}
\label{tab:comparison}
\end{table}

\begin{figure}[tb]
\centering
\includegraphics[width=0.96\linewidth]{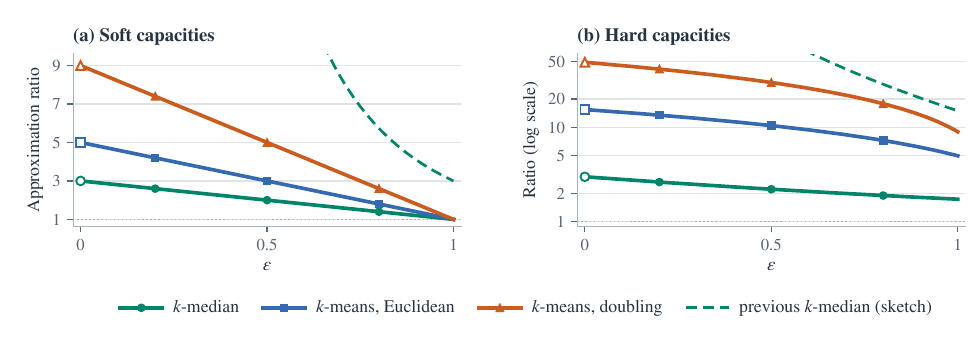}
\captionsetup{font=footnotesize}
\caption{Low-dimensional ratios from \Cref{tab:comparison} with uniform
capacities. Panel~(b) uses a log scale; hollow markers denote
$\eps\to0$. 
The dashed green curve shows $c\,\eps^{-2}(1+\ln(1/\eps))$ for
illustrative constants, sketching the polynomial growth of the
previous $k$-median ratio~\cite{Li16}; Dai and Li's ratio~\cite{DaiLi26}
grows as $O(\eps^{-5})$.}
\label{fig:ratios}
\end{figure}

\paragraph{Comparison.}
\Cref{tab:comparison} compares our ratios for $k$-median and $k$-means with the previous ones that respect the capacities exactly, and \Cref{fig:ratios} plots our low-dimensional ratios for uniform capacities as functions of $\eps$. In low dimension, our ratios are bounded by a constant even as $\eps\to0$. The algorithms of Li~\cite{Li16} and Dai and Li~\cite{DaiLi26} work more broadly, in general metrics with non-uniform capacities, but their ratios $O(\eps^{-2}\log(1/\eps))$ and $O(\eps^{-5})$ grow polynomially in $1/\eps$. In low dimension, \Cref{thm:nonuniform} improves both to $1+2e^{-\eps}$ in their own setting of non-uniform capacities. In general metrics, \Cref{thm:general} gives $3+2\ln(1/\eps)$ for uniform soft $k$-median and $7+2\ln(1/\eps)$ for hard; for $z>1$ it is the first algorithm with $(1+\eps)k$ centers. Cohen-Addad~\cite{CohenAddad20}  gives an approximation scheme with only $k$ centers, but in quasi-polynomial time in doubling metrics; in polynomial time this is known only for $k$-median in the plane. Approximation schemes for capacitated clustering are also known in Euclidean space and in metrics of bounded scatter dimension, which include doubling metrics, but their running time $f(k)\cdot n^{O(1)}$ is not polynomial if $k$ is part of the input~\cite{CL19,BFS24,BGI25}.

\subsection{Organization}\label{sec:organization}
\Cref{sec:prelim} sets up notation. \Cref{sec:approach} gives an overview of  our approach, including the rescue lemma, and states the framework theorem,  which \Cref{sec:proof} proves. \Cref{sec:soft} completes the proof of \Cref{thm:soft} for joint and one-sided low dimension and  in Euclidean space. \Cref{sec:nonuniform} treats non-uniform capacities, and \Cref{sec:hard} treats hard capacities and general metrics. \Cref{sec:related} surveys related work, and  \Cref{sec:open} concludes with open questions.

\section{Preliminaries}\label{sec:prelim}

\begin{definition}[Capacitated $(k,z)$-clustering]\label{def:problem}
An instance consists of a metric space $(V,\dist)$, a set $F\subseteq V$ of \emph{locations} (finite except in the Euclidean setting below), a multiset $P$ of $n$ \emph{clients} at points of $V$, and positive integers $k$ and $U$. The exponent $z\ge1$ is fixed. A \emph{solution} consists of centers, each placed at a location in $F$, and an assignment $\sigma$ that maps every client to a center so that every center serves at most $U$ clients. Its cost is $\cost(\sigma)=\sum_{p\in P}\dist(p,\sigma(p))^z$, where $\dist(p,\sigma(p))$ is the distance from $p$ to the location of $\sigma(p)$. With \emph{soft} capacities, several centers may be placed at the same location; with \emph{hard} capacities, the centers are at distinct locations.
\end{definition}

Capacities are uniform except in \Cref{sec:nonuniform}. $\OPT$ denotes the minimum cost of a solution with at most $k$ centers, with the type of capacities clear from the context. In the \emph{Euclidean setting}, the metric space is $(\R^d,\|\cdot\|)$, centers may be placed anywhere in $\R^d$, and capacities are soft; $\OPT$ is also the infimum of the cost over solutions with distinct centers, since coinciding centers can be moved apart at an arbitrarily small cost, and our algorithm returns distinct centers if $\OPT>0$. We assume $k<n$ and that a solution with at most $k$ centers exists; for soft capacities and in the Euclidean setting, this means $n\le kU$.

As usual, arithmetic operations and evaluations of $\dist(x,y)^z$ count as single steps; \Cref{app:bits} shows that rational arithmetic with polynomially many bits suffices. The rest of this section collects standard facts; proofs are included for completeness.

\paragraph{Doubling dimension.}
For a metric space $(X,\dist)$, $x\in X$ and $r\ge0$, let $B(x,r)=\{y\in X:\dist(x,y)\le r\}$. The doubling dimension of $X$ is at most $D$ if every ball $B(x,r)$ is contained in the union of at most $2^D$ balls of radius $r/2$ centered at points of $X$. The doubling dimension of a multiset of points is that of the set of its points. Subsets of $\R^d$ have doubling dimension $O(d)$~\cite{GKL03}, and \Cref{fact:euclid-doubling} gives the bound $3d$.

\begin{fact}[{\cite[Proposition~1.1]{GKL03}}]\label{fact:packing}
Let $X$ have doubling dimension at most $D$, let $x\in X$, and let $0<s\le r$. Every subset of $B(x,r)$ whose points are pairwise more than $s$ apart has at most $(4r/s)^D$ points.
\end{fact}
\begin{proof}
Applying the definition $\lceil\log_2(2r/s)\rceil$ times covers $B(x,r)$ by at most $2^{D\lceil\log_2(2r/s)\rceil}<(4r/s)^D$ balls of radius at most $s/2$. Each of these balls contains at most one point of the subset.
\end{proof}

\begin{fact}\label{fact:euclid-doubling}
Every finite subset of $\R^d$ has doubling dimension at most $3d$.
\end{fact}
\begin{proof}
Consider a ball $B(x,r)$ and a maximal subset of it whose points are pairwise more than $r/2$ apart. The Euclidean balls of radius $r/4$ around these points are disjoint and contained in a Euclidean ball of radius $5r/4$, so there are at most $5^d\le 2^{3d}$ of them. By maximality, the balls of radius $r/2$ around these points cover $B(x,r)$.
\end{proof}

\paragraph{Minkowski's inequality.}
The triangle inequality and Minkowski's inequality for the $z$-norm give, for every finite list of triples $(x,y,p)$ of points,
\begin{equation}\label{eq:minkowski}
\Bigl(\sum\dist(x,y)^z\Bigr)^{1/z}\le\Bigl(\sum\dist(x,p)^z\Bigr)^{1/z}+\Bigl(\sum\dist(p,y)^z\Bigr)^{1/z},
\end{equation}
where all sums are over the triples of the list.

\paragraph{Transport plans.}
Let $I$ and $J$ be disjoint finite index sets, and let $w_i$ ($i\in I$) and $\ell_j$ ($j\in J$) be nonnegative integers with $\sum_iw_i=\sum_j\ell_j$. An \emph{integral transport plan} is a matrix $\pi$ of nonnegative integers with rows indexed by $I$, columns indexed by $J$, row sums $w_i$ and column sums $\ell_j$. Its \emph{support} $\operatorname{supp}(\pi)$ is the bipartite graph on $I\cup J$ with an edge $ij$ whenever $\pi_{ij}>0$. Given a cost for every pair $ij$, the cost of $\pi$ is the sum of $\pi_{ij}$ times the cost of $ij$ over all pairs.

\begin{fact}[{Cycle-free plans~\cite[Section~11.2]{AMO93}}]\label{fact:cycle}
Every integral transport plan $\pi$ can be transformed in polynomial time into an integral transport plan with the same row and column sums and no larger cost, whose support is a forest contained in $\operatorname{supp}(\pi)$.
\end{fact}
\begin{proof}
A cycle in the support has even length. Adding $1$ to every other entry on the cycle and subtracting $1$ from the others keeps all row and column sums, and the two directions change the cost by opposite amounts. Moving in a direction that does not increase the cost, by the largest amount that keeps the entries nonnegative, removes an edge of the cycle and adds none.
\end{proof}

\paragraph{Uncapacitated clustering.}
We use the following algorithm to compute initial solutions.

\begin{fact}[{Local search for uncapacitated clustering~\cite[Theorem~1]{FRS19}}]\label{thm:frs}
For fixed $D\ge1$, $z\ge 1$ and $\eta>0$, there is a deterministic polynomial-time algorithm that, given $P$ and $F$ such that $P\cup F$ has doubling dimension at most $D$, computes at most $k$ centers at locations in $F$ such that assigning every client to a closest center costs at most $(1+\eta)$ times the minimum cost of such an uncapacitated solution.
\end{fact}

The algorithm is local search with swaps of constant size. Friggstad, Rezapour and Salavatipour state the result for $k$-means~\cite{FRS19}, sketch the extension of the analysis to the sum of $z$-th powers of distances for every fixed $z\ge1$ (Section~5.2), and note that it applies to weighted points, and hence to several clients at the same point (Section~3.3). Independently, Cohen-Addad, Klein and Mathieu~\cite{CKM19} showed that local search is an approximation scheme for $k$-median and $k$-means in Euclidean space of fixed dimension and in minor-free metrics.

\paragraph{The assignment cost of a family.}
We consider finite families $S$ of centers at locations in $F$, several of which may share a location. For such a family, $\cost(S)$ denotes the minimum cost of an assignment of the clients to the centers of $S$ in which every center serves at most $U$ clients. If $|S|U<n$, then $\cost(S)=\infty$. A minimum-cost flow computes $\cost(S)$. For a solution $S^*$, $\cost(S^*)$ is the cost of its assignment.

\section{Overview of Our Approach}\label{sec:approach}

\begin{figure}[tb]
\centering
\tikzset{
  mod/.style={draw, rounded corners=2pt, minimum width=2.3cm, minimum height=0.74cm, align=center, font=\scriptsize, inner sep=1.5pt, line width=0.6pt},
  inp/.style={mod, draw=black!45, fill=black!5},
  cmp/.style={mod, draw=centcol, fill=centcol!12},
  blk/.style={mod, draw=blockcol, fill=blockcol!10},
  frm/.style={mod, draw=sepcol, fill=sepcol!8, line width=0.9pt},
  prj/.style={mod, draw=rescol!75!black, fill=rescol!12},
  grd/.style={mod, draw=violet!70!black, fill=violet!8},
  opt/.style={densely dashed},
  arr/.style={-{Stealth[length=1.6mm]}, line width=0.6pt, black!70},
  lab/.style={font=\footnotesize, anchor=east, align=right}}
\begin{tikzpicture}[x=1cm,y=1cm]
  \node[lab] at (1.45,0) {\Cref{thm:soft}};
  \node[inp] (a1) at (2.75,0) {local search\\ratio $1+\eta$};
  \node[cmp,opt] (a2) at (5.45,0) {compression\\if one side};
  \node[blk] (a3) at (8.15,0) {blocks\\$<2k$ centers};
  \node[frm] (a4) at (10.85,0) {framework\\$\beta=3^z$};
  \draw[arr] (a1)--(a2); \draw[arr] (a2)--(a3); \draw[arr] (a3)--(a4);
  \node[lab] at (1.45,-0.9) {\Cref{thm:soft}, $\R^d$};
  \node[inp] (b1) at (2.75,-0.9) {candidate\\centers};
  \node[cmp] (b2) at (5.45,-0.9) {compression\\of the candidates};
  \node[blk] (b3) at (8.15,-0.9) {blocks\\$<2k$ centers};
  \node[frm] (b4) at (10.85,-0.9) {framework\\$\beta=\beta^*_z$};
  \draw[arr] (b1)--(b2); \draw[arr] (b2)--(b3); \draw[arr] (b3)--(b4);
  \node[lab] at (1.45,-1.8) {\Cref{thm:nonuniform}};
  \node[inp] (d1) at (2.75,-1.8) {local search\\ratio $1+\eta$};
  \node[blk] (d2) at (5.45,-1.8) {sparse start\\$\le2k$ centers};
  \node[grd] (d3) at (8.15,-1.8) {greedy addition\\cost $(1+\eta)\OPT$};
  \node[frm] (d4) at (10.85,-1.8) {framework\\$k+\lfloor\gamma k\rfloor$ centers};
  \node[grd] (d5) at (13.55,-1.8) {greedy addition\\$1+(3^z-1)e^{-\eps}$};
  \draw[arr] (d1)--(d2); \draw[arr] (d2)--(d3); \draw[arr] (d3)--(d4); \draw[arr] (d4)--(d5);
  \node[lab] at (1.45,-2.7) {\Cref{thm:hard}, $z>1$};
  \node[frm] (c1) at (2.75,-2.7) {\Cref{thm:soft}\\ratio $\rho$};
  \node[prj] (c2) at (5.45,-2.7) {projection\\$(1+2\rho^{1/z})^z$};
  \draw[arr] (c1)--(c2);
  \node[lab] at (1.45,-3.6) {\Cref{thm:general}};
  \node[inp] (e1) at (2.75,-3.6) {any algorithm\\ratio $\rho$};
  \node[blk] (e2) at (5.45,-3.6) {blocks\\$<2k$ centers};
  \node[prj,opt] (e3) at (8.15,-3.6) {projection\\$(1+2\rho^{1/z})^z$};
  \node[grd] (e4) at (10.85,-3.6) {greedy deletion\\$+2\ln(1/\eps)$};
  \draw[arr] (e1)--(e2); \draw[arr] (e2)--(e3); \draw[arr] (e3)--(e4);
\end{tikzpicture}
\caption{ How our algorithms are assembled. Each row is the algorithm of one result and runs from left to right; each box is one step, labeled with what it produces. \emph{Local search} computes a near-optimal uncapacitated solution (\Cref{thm:frs}). \emph{Blocks}: its clusters are split into fewer than $2k$ groups of at most $U$ clients, with one center per group (\Cref{sec:approach-blocks}). \emph{Framework}: the algorithm of the framework theorem (\Cref{thm:framework}) reduces the number of centers to $K$. It gives some blocks a private center, using the extra centers, and routes the other blocks to shared centers by a dynamic program; its ratio is $\beta-(\beta-1)\eps$ for the constant $\beta$ in the box. \Cref{thm:near-k} runs it with $\lfloor\gamma k\rfloor$ extra centers. Dashed steps are used only in some cases: \emph{compression} (\Cref{lem:compress}) if only $P$ or only $F$ has bounded doubling dimension, and \emph{projection} (\Cref{lem:soft-to-hard}), which turns a soft solution of ratio $\rho$ into a hard one, for hard capacities. The remaining steps are \emph{candidate centers} in $\R^d$ (\Cref{lem:candidates}), a \emph{sparse start} with at most $2k$ centers (\Cref{lem:sparse}), and \emph{greedy addition} and \emph{deletion} of one center at a time (\Cref{lem:augment,lem:greedy}). For $k$-median, \Cref{thm:hard} uses the algorithm of \Cref{thm:nonuniform}.}
\label{fig:pipeline}
\end{figure}
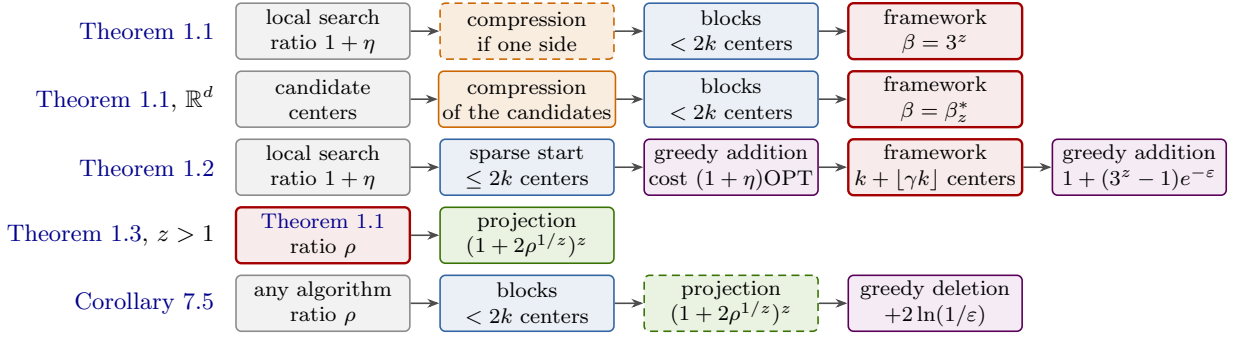

This section  gives an overview of our approach in the setting of \Cref{thm:soft}: soft capacities with a finite set $F$ such that $P\cup F$ has doubling dimension at most $D$. The approach has three steps: splitting the clients into blocks (\Cref{sec:approach-blocks,sec:approach-reference}), rescuing blocks with the extra centers (\Cref{sec:approach-rescue}), and routing the solution through a network (\Cref{sec:approach-network,sec:approach-bad}). The \emph{framework theorem} (\Cref{thm:framework}) at the end of the section combines the three steps, and all our low-dimensional results build on it; \Cref{sec:proof} contains the proofs. \Cref{fig:pipeline} shows how the parts combine into our results. The accuracy parameter $\eta\in(0,1/10]$ is chosen from $\delta$ in each proof, and the other parameters depend on $D$, $z$, $\eps$ and $\eta$; \Cref{tab:params} in \Cref{sec:proof-combine} lists each of them with its role and value.

\subsection{Blocks and Movable Blocks}\label{sec:approach-blocks}

The first step of our framework partitions the clients into capacity-respecting groups, which we call \emph{blocks}. Reassigning every client to a closest center turns a solution with at most $k$ centers into an uncapacitated solution that is no more expensive. Therefore \Cref{thm:frs} gives an uncapacitated solution $S_1$ with at most $k$ centers at locations in $F$ and cost at most $(1+\eta)\OPT$. We split every cluster $Q$ of $S_1$ arbitrarily into $\lceil|Q|/U\rceil$ blocks of at most $U$ clients each. The center of $Q$ is the \emph{representative} of each of these blocks.

\begin{definition}[Block partition]\label{def:blocks}
A \emph{block partition} of $P$ consists of a partition of $P$ into blocks $P_1,\dots,P_m$ with $m\le2k-1$, each of \emph{size} $w_i:=|P_i|\le U$, and a \emph{representative} $a_i\in F$ for every block. The \emph{internal cost} of block $i$ is $A_i:=\sum_{p\in P_i}\dist(p,a_i)^z$, and $A:=\sum_iA_i$.
\end{definition}

\begin{lemma}[{$2k$ centers suffice}]\label{lem:2k}
If the capacities are uniform and $n\le kU$, the blocks of any uncapacitated solution with at most $k$ centers form a block partition with $m<2k$ and total cost $A$ equal to the cost of the solution.
\end{lemma}
\begin{proof}
The representative $a_i$ of a block is the center of the cluster that contains it, so $A$ is the cost of the solution. Since $n\le kU$ and there are at most $k$ clusters,
\begin{equation}\label{eq:block-count}
m=\sum_Q\Bigl\lceil\frac{|Q|}{U}\Bigr\rceil<\sum_Q\Bigl(\frac{|Q|}{U}+1\Bigr)\le 2k .
\end{equation}
\end{proof}

Applied to $S_1$, the lemma gives a block partition with $A\le(1+\eta)\OPT$, and at most $2k-1$ centers, one at each representative, give a feasible solution of this cost. With $K$ centers, at least $m-K$ blocks must share centers with other blocks, so their clients must move to other locations. We measure the cost of this movement by the cost of moving the mass of a block, which consists of $w_i$ units at $a_i$.

\begin{definition}[Movable block]\label{def:movable}
Let $u\ge1$ and $v\ge0$. Block $i$ is \emph{$(u,v)$-movable} if for all $w_i$ locations $y_1,\dots,y_{w_i}\in F$, where $w_i=|P_i|$ is the size of the block, some bijection $\phi\colon P_i\to\{1,\dots,w_i\}$ satisfies
\[
\sum_{p\in P_i}\dist(p,y_{\phi(p)})^z\ \le\ u\cdot A_i+v\cdot\sum_{p\in P_i}\dist(a_i,y_{\phi(p)})^z .
\]
\end{definition}

We think of $y_1,\dots,y_{w_i}$ as \emph{slots}, units of capacity of centers at these locations. The last sum is the same for every bijection: it is the cost of moving the mass of the block to the slots. In words, the clients of a movable block can be matched to any $w_i$ slots at a cost of at most $u$ times the internal cost of the block plus $v$ times the cost of moving its whole mass from the representative to the slots.

\begin{lemma}[Triangle bound]\label{lem:triangle-movable}
Every block of every block partition is $(3^{z-1},(3/2)^{z-1})$-movable.
\end{lemma}
\begin{proof}
Let $\phi$ be any bijection and $p\in P_i$, and write $y=y_{\phi(p)}$. The triangle inequality and the convexity of the $z$-th power give
\[
\dist(p,y)^z\le\Bigl(\frac13\cdot3\dist(p,a_i)+\frac23\cdot\frac32\dist(a_i,y)\Bigr)^z\le3^{z-1}\dist(p,a_i)^z+\Bigl(\frac32\Bigr)^{z-1}\dist(a_i,y)^z .
\]
Summing over $p\in P_i$ proves the lemma.
\end{proof}

\subsection{Following a Reference Solution}\label{sec:approach-reference}

The analysis compares the blocks with a \emph{reference solution} $S^*$, which is any soft solution with at most $k$ centers at locations in $F$. Let $J$ be the set of centers of $S^*$ that serve clients, so $|J|\le k$; center $j\in J$ is at location $c_j$ and serves $\ell_j\le U$ clients. Let $I=\{1,\dots,m\}$ be the set of blocks.

\begin{lemma}[Transport forest]\label{lem:transport}
There is an integral transport plan $\pi$ with row sums $w_i$ ($i\in I$) and column sums $\ell_j$ ($j\in J$) whose support, the bipartite graph of its positive entries, is a forest and whose cost $T:=\sum_{i,j}\pi_{ij}\dist(a_i,c_j)^z$ satisfies
\[
T\le\bigl(A^{1/z}+\cost(S^*)^{1/z}\bigr)^z\le2^{z-1}\bigl(A+\cost(S^*)\bigr).
\]
\end{lemma}
\begin{proof}
Let $\pi^0_{ij}$ be the number of clients of $P_i$ that $S^*$ assigns to center $j$. Applying~\eqref{eq:minkowski} to the triple $(a_i,c_j,p)$ for every client $p\in P_i$ served by $j$ bounds the cost of $\pi^0$ by $(A^{1/z}+\cost(S^*)^{1/z})^z$. The second inequality is $(s+t)^z\le2^{z-1}(s^z+t^z)$ for $s,t\ge0$, which follows from the convexity of $t\mapsto t^z$. \Cref{fact:cycle} turns $\pi^0$ into a plan whose support is a forest, without increasing the cost.
\end{proof}

We fix such a plan $\pi$. It moves $\pi_{ij}$ units of the mass of block $i$ from $a_i$ to $c_j$, and $T_i:=\sum_j\pi_{ij}\dist(a_i,c_j)^z$ is the \emph{moving cost} of block $i$. If every block follows $\pi$, that is, the clients of block $i$ take $\pi_{ij}$ slots of center $j$ for every $j$, then movability gives a solution with the centers of $S^*$ and cost at most $uA+vT$. With
\[
\beta:=u+2^zv,
\]
\Cref{lem:transport} gives, if $A\le(1+\eta)\cost(S^*)$,
\begin{equation}\label{eq:approach-follow}
uA+vT\ \le\ (u+2^{z-1}v)A+2^{z-1}v\cost(S^*)\ \le\ \beta(1+\eta)\cost(S^*).
\end{equation}
For an optimal $S^*$, the factor $\beta$ in~\eqref{eq:approach-follow} is our ratio as $\eps\to0$; for the triangle bound, $\beta=3^{z-1}+2^z(3/2)^{z-1}=3^z$. In Euclidean space, the certificates of \Cref{sec:euclid} replace the triangle inequality and reduce $3^z$ to the smaller constant $\beta^*_z$.

\subsection{Rescuing Blocks}\label{sec:approach-rescue}

The ratio $\beta$ above applies when every block follows the optimal solution. To improve it, we use the extra centers to exempt some blocks from this reassignment.

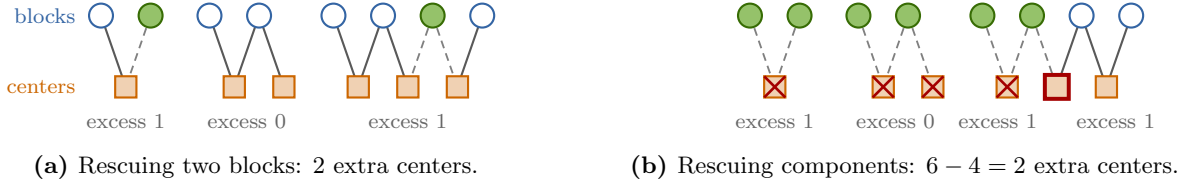
\begin{figure}[t]
\centering
\tikzset{
  blk/.style={circle,draw=blockcol,fill=white,line width=0.8pt,inner sep=0pt,minimum size=9pt},
  rsc/.style={circle,draw=rescol!70!black,fill=rescol!60,line width=0.8pt,inner sep=0pt,minimum size=9pt},
  cen/.style={rectangle,draw=centcol,fill=centcol!30,line width=0.8pt,inner sep=0pt,minimum size=8pt},
  sep/.style={rectangle,draw=sepcol,fill=centcol!30,line width=1.5pt,inner sep=0pt,minimum size=9pt},
  edg/.style={line width=0.8pt,black!65},
  gone/.style={line width=0.7pt,black!50,densely dashed},
  exc/.style={font=\scriptsize,black!60},
}
\newcommand{\closed}[1]{\draw[sepcol,line width=1pt] ($(#1)+(-0.17,-0.17)$) -- ($(#1)+(0.17,0.17)$); \draw[sepcol,line width=1pt] ($(#1)+(-0.17,0.17)$) -- ($(#1)+(0.17,-0.17)$);}
\newcommand{\forestcoords}{%
  \foreach \i/\x in {1/0,2/0.8,3/1.75,4/2.55,5/3.75,6/4.55,7/5.35,8/6.15} \coordinate (b\i) at (\x,1.15);
  \foreach \j/\x in {1/0.4,2/2.15,3/2.95,4/4.15,5/4.95,6/5.75} \coordinate (c\j) at (\x,0);}
\begin{subfigure}[b]{0.48\textwidth}
\centering
\begin{tikzpicture}[scale=0.82]
  \path[use as bounding box] (-1.45,-0.8) rectangle (6.5,1.45);
  \forestcoords
  \foreach \e/\f in {b1/c1,b3/c2,b4/c2,b4/c3,b5/c4,b6/c4,b6/c5,b8/c6} \draw[edg] (\e)--(\f);
  \foreach \e/\f in {b2/c1,b7/c5,b7/c6} \draw[gone] (\e)--(\f);
  \foreach \i in {1,3,4,5,6,8} \node[blk] at (b\i) {};
  \foreach \i in {2,7} \node[rsc] at (b\i) {};
  \foreach \j in {1,...,6} \node[cen] at (c\j) {};
  \node[exc] at (0.4,-0.55) {excess $1$};
  \node[exc] at (2.35,-0.55) {excess $0$};
  \node[exc] at (4.95,-0.55) {excess $1$};
  \node[font=\scriptsize,blockcol,anchor=east] at (-0.25,1.15) {blocks};
  \node[font=\scriptsize,centcol,anchor=east] at (-0.25,0) {centers};
\end{tikzpicture}
\caption{Rescuing two blocks: $2$ extra centers.}
\end{subfigure}\hfill
\begin{subfigure}[b]{0.48\textwidth}
\centering
\begin{tikzpicture}[scale=0.82]
  \path[use as bounding box] (-1.45,-0.8) rectangle (6.5,1.45);
  \forestcoords
  \foreach \e/\f in {b1/c1,b2/c1,b3/c2,b4/c2,b4/c3,b5/c4,b6/c4,b6/c5} \draw[gone] (\e)--(\f);
  \foreach \e/\f in {b7/c5,b7/c6,b8/c6} \draw[edg] (\e)--(\f);
  \foreach \i in {7,8} \node[blk] at (b\i) {};
  \foreach \i in {1,...,6} \node[rsc] at (b\i) {};
  \foreach \j in {1,2,3,4,6} \node[cen] at (c\j) {};
  \node[sep] at (c5) {};
  \foreach \j in {1,...,4} {\closed{c\j}}
  \node[exc] at (0.4,-0.55) {excess $1$};
  \node[exc] at (2.35,-0.55) {excess $0$};
  \node[exc] at (4.0,-0.55) {excess $1$};
  \node[exc] at (5.9,-0.55) {excess $1$};
\end{tikzpicture}
\caption{Rescuing components: $6-4=2$ extra centers.}
\end{subfigure}
\caption{The support of the transport plan $\pi$ is a forest between blocks (circles) and centers of $S^*$ (squares). Rescued blocks are filled, their edges are dashed, and closed centers are crossed out. (a)~Rescuing the two blocks with the largest savings closes no center. (b)~The rescue lemma deletes a few vertices, here one center (bold), which stays open; rescuing all blocks of a small component closes its centers, so the component costs only its excess.}
\label{fig:rescue}
\end{figure}

An extra center can \emph{rescue} a block: a new center at $a_i$ serves exactly the clients of $P_i$. This respects the capacity, since $w_i\le U$, and costs $A_i$ instead of the share $uA_i+vT_i$ of block $i$ in~\eqref{eq:approach-follow}. So rescuing block $i$ reduces the bound~\eqref{eq:approach-follow} by its \emph{saving}
\[
W_i:=(u-1)A_i+vT_i\ \ge0,
\]
and by \Cref{lem:transport}, if $A\le(1+\eta)\cost(S^*)$, the total saving satisfies
\begin{equation}\label{eq:total-saving}
\sum_iW_i=(u-1)A+vT\ \le\ (u-1+2^{z-1}v)A+2^{z-1}v\cost(S^*)\ \le\ (\beta-1)(1+\eta)\cost(S^*).
\end{equation}
Suppose that $\lfloor\eps k\rfloor$ extra centers rescue blocks whose savings add up to at least $(\eps-\eta)\sum_iW_i$. Then the resulting solution achieves the ratio of our theorems: its cost is at most
\begin{equation}\label{eq:approach-goal}
A+(1-\eps+\eta)\sum_iW_i\ \le\ \bigl(\beta-(\beta-1)\eps+O(\beta\eta)\bigr)\cdot\cost(S^*).
\end{equation}

\paragraph{The obstacle.}
The natural strategy rescues the $\lfloor\eps k\rfloor$ blocks with the largest savings, one extra center per block. The centers of $S^*$ that receive mass from these blocks usually also receive mass from other blocks, so this strategy closes no center (\Cref{fig:rescue}(a)). If all savings are equal and $m$ is close to $2k$, the rescued blocks carry only about an $\eps/2$-fraction of $\sum_iW_i$, which gives the weaker ratio $\beta-(\beta-1)\eps/2$.

\paragraph{Closing centers.}
If every block that sends mass to center $j$ of $S^*$ is rescued, then $j$ serves no client and can be closed, which returns one center to the budget; the crossed-out centers in \Cref{fig:rescue}(b) are closed in this way. By \Cref{lem:transport}, the support of $\pi$ is a forest on the $m$ blocks and the $|J|$ centers. If we rescue all blocks of a component of the forest, we may close all centers of the component. So the component costs as many extra centers as its excess, the number of its blocks minus the number of its centers. Consider a budget of $b\le k$ extra centers. In the framework theorem (\Cref{sec:approach-framework}), $b$ is slightly smaller than $\eps k$, because a small part of the extra centers is reserved for the bad pairs of \Cref{sec:approach-bad}. Rescuing all blocks costs $m-|J|\le2k-1-|J|$ centers beyond the $|J|$ centers of $S^*$, while the budget, together with the $k-|J|$ centers that $S^*$ leaves unused, gives $b+k-|J|$ centers beyond them. We have $(b+k-|J|)/(2k-|J|)\ge b/k$, which is equivalent to $(k-|J|)(k-b)\ge0$. So if we could obtain a $b/k$-fraction of the savings in every component, the total saving would be a $b/k$-fraction of $\sum_i W_i$---twice what the naive strategy achieves, because each extra center effectively rescues about two blocks.

This accounting applies only to entire components: rescuing part of a component need not close any center, and one component may contain almost all blocks. We exploit the fact that the support of $\pi$ is a forest, which, unlike a general bipartite graph, has small separators. Fix an integer $L\ge1$, the largest component size we allow; the framework theorem takes $L=\lceil12/\eta\rceil$. We delete fewer than $3k/(L+1)$ of the fewer than $3k$ vertices of the forest so that every remaining component has at most $L$ vertices, rescue the deleted blocks, and keep the deleted centers open (\Cref{fig:rescue}(b)). Then we rescue all components of nonpositive excess, which cost nothing or even free centers, and add the other components greedily in order of saving per unit of excess. The greedy choice wastes at most one component, of excess at most $L$. The following rescue lemma, proved in \Cref{sec:rescue}, shows that compared with the ideal $b/k$-fraction, this procedure loses only an $O(L/k+1/L)$-fraction of $\sum_iW_i$.

\begin{lemma}[Rescue lemma]\label{lem:rescue}
Let $\pi$ be an integral transport plan between disjoint index sets $I$ and $J$ with $|I|\le2k-1$ and $|J|\le k$ whose support is a forest, and let $W_i\ge0$ for $i\in I$. For all integers $L\ge1$ and $0\le b\le k$, there are sets $R\subseteq I$ and $J_0\subseteq J$ such that
\begin{enumerate}[label=(\roman*)]
\item $\pi_{ij}=0$ for all $i\in I\setminus R$ and $j\in J_0$;
\item $|R|+|J|-|J_0|\le k+b$;
\item $\sum_{i\in R}W_i\ge\Bigl(\dfrac{b-L}{k}-\dfrac{3}{L+1}\Bigr)\sum_{i\in I}W_i$.
\end{enumerate}
\end{lemma}

By~(i), rescuing the blocks in $R$ leaves the centers in $J_0$ without clients, so they can be closed, and by~(ii), the result has at most $k+b$ centers. The loss in~(iii) is small if $L$ is large and $k$ is large compared with $L$. For example, if $L=\lceil12/\eta\rceil$, $b\ge(\eps-\eta/4)k-1$ and $k\ge4(L+1)/\eta$, then $(b-L)/k\ge\eps-\eta/4-(L+1)/k\ge\eps-\eta/2$ and $3/(L+1)\le\eta/4$, so~(iii) yields the $(\eps-\eta)$-fraction required in~\eqref{eq:approach-goal}. The framework theorem uses these values, and its algorithm solves instances with smaller $k$ by exhaustive search.

\subsection{Finding a Solution: The Network and the Dynamic Program}\label{sec:approach-network}

The rescued solution is defined in terms of the unknown solution $S^*$. The algorithm instead optimizes over all ways of routing the mass of the blocks through a network built on a \emph{hierarchical decomposition} $\Dec$ of $X=P\cup F$ (\Cref{def:decomposition}). The decomposition consists of partitions of $X$ into \emph{cells} at levels $0,\dots,H$: the cells of level $h$ have diameter at most $2^h$ and refine those of level $h+1$, the cells of level $0$ are single points, and $X$ is the only cell of level $H$. Here $H=O(\log\Delta(X))$ for the ratio $\Delta(X)$ between the largest and the smallest positive distance in $X$.

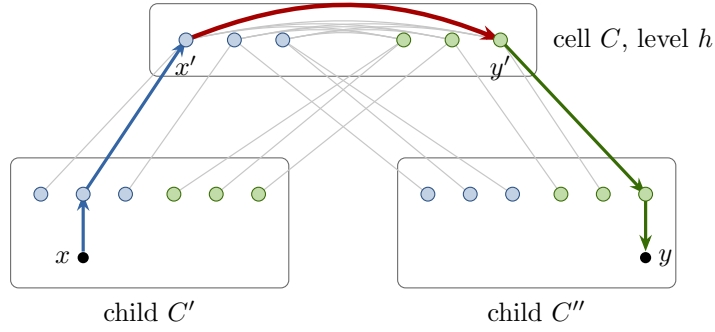
\begin{figure}[t]
\centering
\begin{tikzpicture}[scale=0.8,
  aport/.style={circle,draw=blockcol!80!black,fill=blockcol!30,inner sep=0pt,minimum size=5.2pt},
  dport/.style={circle,draw=rescol!70!black,fill=rescol!35,inner sep=0pt,minimum size=5.2pt},
  pt/.style={circle,fill=black,inner sep=0pt,minimum size=4.2pt},
  faint/.style={black!22,line width=0.45pt},
  asc/.style={-{Stealth[length=2.2mm]},line width=1.2pt,blockcol},
  dsc/.style={-{Stealth[length=2.2mm]},line width=1.2pt,rescol!70!black},
  crs/.style={-{Stealth[length=2.6mm]},line width=1.7pt,sepcol}]
  \draw[black!60,rounded corners=3pt] (2.3,2.55) rectangle (8.7,3.75);
  \draw[black!60,rounded corners=3pt] (0,-0.95) rectangle (4.6,1.2);
  \draw[black!60,rounded corners=3pt] (6.4,-0.95) rectangle (11,1.2);
  \node[font=\small,anchor=west] at (8.8,3.15) {cell $C$, level $h$};
  \node[font=\small,anchor=north] at (2.3,-1.0) {child $C'$};
  \node[font=\small,anchor=north] at (8.7,-1.0) {child $C''$};
  \foreach \k/\x in {1/2.9,2/3.7,3/4.5} {\coordinate (Ca\k) at (\x,3.15);}
  \foreach \k/\x in {1/6.5,2/7.3,3/8.1} {\coordinate (Cd\k) at (\x,3.15);}
  \foreach \k/\x in {1/0.5,2/1.2,3/1.9} {\coordinate (La\k) at (\x,0.6);}
  \foreach \k/\x in {1/2.7,2/3.4,3/4.1} {\coordinate (Ld\k) at (\x,0.6);}
  \foreach \k/\x in {1/6.9,2/7.6,3/8.3} {\coordinate (Ra\k) at (\x,0.6);}
  \foreach \k/\x in {1/9.1,2/9.8,3/10.5} {\coordinate (Rd\k) at (\x,0.6);}
  \foreach \a in {1,2,3} \foreach \d in {1,2,3} \draw[faint] (Ca\a) to[bend left=12] (Cd\d);
  \foreach \a/\b in {1/1,2/1,3/2} {\draw[faint] (La\a) -- (Ca\b); \draw[faint] (Cd\b) -- (Ld\a);}
  \foreach \a/\b in {1/2,2/3,3/3} {\draw[faint] (Ra\a) -- (Ca\b); \draw[faint] (Cd\b) -- (Rd\a);}
  \node[pt] (x) at (1.2,-0.45) {}; \node[font=\small,left=1pt] at (x) {$x$};
  \node[pt] (y) at (10.5,-0.45) {}; \node[font=\small,right=1pt] at (y) {$y$};
  \draw[asc] (x) -- (La2);
  \draw[asc] (La2) -- (Ca1);
  \draw[crs] (Ca1) to[bend left=22] (Cd3);
  \draw[dsc] (Cd3) -- (Rd3);
  \draw[dsc] (Rd3) -- (y);
  \foreach \k in {1,2,3} {\node[aport] at (Ca\k) {}; \node[dport] at (Cd\k) {};
                          \node[aport] at (La\k) {}; \node[dport] at (Ld\k) {};
                          \node[aport] at (Ra\k) {}; \node[dport] at (Rd\k) {};}
  \node[font=\small,below=2pt] at (Ca1) {$x'$};
  \node[font=\small,below=2pt] at (Cd3) {$y'$};
  \begin{scope}[shift={(-0.2,4.55)}]
    \node[aport] at (0,0) {}; \node[font=\footnotesize,anchor=west] at (0.15,0) {ascending copy};
    \node[dport] at (3.0,0) {}; \node[font=\footnotesize,anchor=west] at (3.15,0) {descending copy};
    \draw[asc] (6.35,0) -- (6.85,0); \draw[dsc] (7.0,0) -- (7.5,0); \node[font=\footnotesize,anchor=west] at (7.55,0) {cost $0$};
    \draw[crs] (8.85,0) -- (9.35,0); \node[font=\footnotesize,anchor=west] at (9.4,0) {cost $(\dist(x',y')+4\lambda2^h)^z$};
  \end{scope}
\end{tikzpicture}
\caption{Part of the network: a cell $C$ of level $h$ with two children, each with three portals. Blue and green dots are the ascending and descending copies of the portals of a cell, in the same order. A unit of mass from $x$ climbs to a portal $x'$ of the lowest cell that contains $x$ and $y$, crosses to a portal $y'$ (red), and descends to $y$, at cost $(\dist(x',y')+4\lambda2^h)^z\ge\dist(x,y)^z$; gray lines are other arcs.}
\label{fig:network}
\end{figure}

\paragraph{The network.}
Every cell $C$ of level $h$ receives a set $\Pi_C$ of \emph{portals}, a maximal subset of $C$ whose points are pairwise more than $\lambda2^h$ apart. Here $\lambda\in(0,1)$ is a small constant that controls the spacing of the portals relative to the diameter bound $2^h$; \Cref{sec:approach-bad} fixes its value. By \Cref{fact:packing}, every cell has $|\Pi_C|\le M:=\lceil(4/\lambda)^D\rceil$ portals. Every portal has an \emph{ascending} and a \emph{descending} copy (\Cref{fig:network} and \Cref{def:network}). A unit of mass travels from a point up the hierarchy along ascending copies, each joined to the ascending copy of a portal of the parent cell within distance $\lambda2^{h+1}$. In a cell that contains both end points, it takes one \emph{crossing arc} from the ascending copy of a portal $x'$ to the descending copy of a portal $y'$, and then it travels down along descending copies. Because the objective uses $z$-th powers of distances, the cost of a route cannot be obtained by adding the costs of its arcs. We therefore give all ascending and descending arcs cost $0$ and charge the entire route on its crossing arc, which costs $(\dist(x',y')+4\lambda2^h)^z$. Since the climb and the descent stay within distance $2\lambda2^h$ of the end points, a route from $x$ to $y$ costs at least $\dist(x,y)^z$: the network never underestimates distances. It can, however, overestimate them when every cell that contains both points is much larger than $\dist(x,y)$. For a constant $\Lambda\ge1$, also fixed in \Cref{sec:approach-bad}, we call distinct points $x,y$ \emph{$\Lambda$-bad} if the lowest cell that contains both has a level $h$ with $2^h>\Lambda\cdot\dist(x,y)$. If $(x,y)$ is not $\Lambda$-bad, some route from $x$ to $y$ costs at most $((1+8\lambda\Lambda)\dist(x,y))^z$ (\Cref{lem:routes}), so the stretch is close to $1$ if $\lambda\Lambda$ is small.

\paragraph{Network solutions and the dynamic program.}
A \emph{network solution} declares every block either rescued or \emph{normal} and routes the mass of the normal blocks from their representatives to locations. It opens $\lceil\ell/U\rceil$ centers at a location that receives $\ell$ units and one center for every rescued block, at most $K$ centers in total. Its cost is
\[
\Phi=\sum_{i\text{ normal}}(u-1)A_i+v\cost(f),
\]
where $\cost(f)$ is the cost of the flow $f$. Since routes never underestimate distances, movability turns a network solution into a clustering with at most $K$ centers and cost at most $A+\Phi$ (\Cref{lem:to-clustering}). Apart from the three arcs at the source and the sink of a point, every arc joins portal copies of one cell or of a cell and its parent. So the network has a tree decomposition along the tree of cells with bags of size $O(M)$. A dynamic program over it stores the balance of every vertex of a bag, the inflow minus the outflow, and the number of centers opened so far. It decides which blocks to rescue by a knapsack over the blocks with the same representative. It runs in time $n^{O(M)}$ times a polynomial (\Cref{lem:dp}), which is polynomial for constant $\lambda$.

\subsection{Bad Pairs and the Extra Centers}\label{sec:approach-bad}

The network is accurate only for pairs that are not $\Lambda$-bad, while the dynamic program is polynomial only if $\lambda$, and hence $M$, is a constant. In earlier dynamic programs~\cite{ARR98,Tal04}, every pair pays for the error of the level at which the decomposition cuts it. Since a pair may be cut at any of the $H$ levels, keeping the expected error small requires a number of portals per cell that grows with $H$. Cohen-Addad~\cite{CohenAddad20} treats the badly cut points separately, but still needs $(\log n/\delta)^{O(D)}$ portals per cell, which leads to quasi-polynomial time. We instead rescue the blocks of bad pairs with extra centers, so that a constant number of portals suffices. Call an edge $ij$ of the forest $\operatorname{supp}(\pi)$ \emph{bad} if the pair $(a_i,c_j)$ is $\Lambda$-bad. In a random hierarchical decomposition of a doubling metric, a fixed pair is $\Lambda$-bad with probability $O(D/\Lambda)$. \Cref{sec:decomp} derandomizes this with a hash family: it constructs a family $\Hfam$ of polynomially many decompositions and a distribution on it under which every pair is $\Lambda$-bad with probability $O(D/\Lambda)$. Since the forest has fewer than $3k$ edges, some member of $\Hfam$ has only $O(D k/\Lambda)$ bad edges. We reserve $\lfloor\gamma k\rfloor$ of the extra centers, where $\gamma=\min\{\eps,\eta\}/4$, and choose $\Lambda$ of order $D/\gamma$, so that some member of $\Hfam$ has at most $\lfloor\gamma k\rfloor$ bad edges; the reserved centers rescue the blocks incident to these edges. All other edges of the forest are routed with stretch at most $1+8\lambda\Lambda$, and choosing $\lambda$ of order $\eta/(z\Lambda)$ makes the $z$-th power of this stretch at most $e^\eta$. For fixed $D$, $z$, $\eps$ and $\eta$, both $\lambda$ and $M$ are therefore constants; \Cref{tab:params} lists their exact values.

\subsection{The Framework Theorem}\label{sec:approach-framework}

The other $b=\lfloor\eps k\rfloor-\lfloor\gamma k\rfloor$ extra centers are used in the rescue lemma with $L=\lceil12/\eta\rceil$. Combining the rescue lemma, the network and the decomposition family yields our main structural guarantee. \Cref{sec:proof} collects the parameters in~\eqref{eq:params} and proves the following theorem.

\begin{theorem}[Framework theorem]\label{thm:framework}
Fix $D\ge1$, $z\ge1$, $\eps\in(0,1]$, $\eta\in(0,1/10]$, $u\ge1$ and $v\ge0$, and let $\beta=u+2^zv$. There is a deterministic algorithm with the following guarantee. Its input is an instance with soft capacities such that $P\cup F$ has doubling dimension at most $D$, together with a block partition in which every block is $(u,v)$-movable. It outputs a soft solution with at most $K$ centers whose cost is at most
\[
\bigl(\beta-(\beta-1)\eps+4\beta\eta\bigr)\cdot\cost(S^*)
\]
for every soft solution $S^*$ with at most $k$ centers such that $A\le(1+\eta)\cost(S^*)$. Its running time is polynomial in $n$, $|F|$ and $\log\Delta(P\cup F)$, with a degree that depends on $D,z,\eps$ and $\eta$.
\end{theorem}

The guarantee holds for all such $S^*$ at once, since the algorithm does not depend on $S^*$; \Cref{sec:euclid} uses a reference solution that is not optimal. With the block partition of \Cref{sec:approach-blocks}, \Cref{lem:triangle-movable} and an optimal $S^*$, \Cref{thm:framework} gives \Cref{thm:soft} when $P\cup F$ has doubling dimension at most $D$. \Cref{sec:soft,sec:nonuniform,sec:hard} extend this to the other settings, as outlined  in \Cref{sec:approach-extensions}.

\subsection{Extensions to the Other Settings}\label{sec:approach-extensions}

The other results extend the framework to harder settings. If only one of $P$ and $F$ has bounded doubling dimension, we compress the other side, as described in \Cref{sec:onesided}. In Euclidean space, we replace the triangle inequality by a sharper bound for moving a block: matching its clients to the slots of the optimal solution uniformly at random, the expected cost splits into the internal cost, the moving cost and a cross term, and the cross term vanishes when the representative of the block is an optimal center for it. In \Cref{sec:euclid}, we show that this improves the constant from $3^z$ to $\beta^*_z$. For non-uniform capacities, a rescued block must fit the capacity of its location, so the block construction above no longer works. Instead, \Cref{thm:nonuniform} takes blocks from a near-optimal capacitated solution with $O(k\log(2/\delta))$ centers, which greedy additions provide in every finite metric, and then applies the framework to improve the ratio; \Cref{sec:nonuniform} carries this out. For hard capacities, a projection lemma of Li~\cite{Li15}, which we extend to all $z\ge1$ in \Cref{sec:hard}, selects a hard subfamily from a soft solution and turns a soft ratio $\rho$ into the hard ratio $(1+2\rho^{1/z})^z$; this explains the form of the ratios in \Cref{thm:hard}. In general metrics, where the low-dimensional structure is absent, the framework does not apply. Instead, \Cref{thm:general} starts with one center per block of an uncapacitated solution, for instance of ratio $3$ for $k$-median~\cite{AGKMMP04} or $9$ for $k$-means~\cite{ANSW20}, and greedily deletes centers until $K$ remain. The assignment cost is supermodular: removing a center from a smaller family increases the cost at least as much as removing it from a larger one. Hence, for $k$-median, while $k+q$ centers remain, some deletion increases the cost by at most $2\OPT/q$, and the deletions from about $2k$ to $K$ centers add up to about $2\ln(1/\eps)\OPT$. \Cref{sec:hard} gives the details.

\section{Proof of the Framework Theorem}\label{sec:proof}

This section proves \Cref{thm:framework}. \Cref{sec:rescue} proves the rescue lemma. \Cref{sec:decomp} constructs the hierarchical decompositions. \Cref{sec:network} defines the network and the dynamic program. \Cref{sec:proof-combine} assembles these pieces into the proof of the framework theorem and analyzes the running time.

\subsection{Proof of the Rescue Lemma}\label{sec:rescue}

This subsection proves \Cref{lem:rescue}, the combinatorial core of the framework. As explained in \Cref{sec:approach-rescue}, it shows that $\lfloor\eps k\rfloor$ extra centers can rescue blocks whose savings add up to nearly an $\eps$-fraction of the total.

The proof uses a standard separator for forests.

\begin{lemma}[Forest separator]\label{lem:separator}
Let $G$ be a forest and $L\ge1$ an integer. There is a set $Z$ containing at most a $1/(L+1)$ fraction of the vertices of $G$ such that every component of $G-Z$ has at most $L$ vertices, and $Z$ can be found in linear time.
\end{lemma}
\begin{proof}
Root every tree of $G$ and process the vertices bottom-up. When a vertex $x$ is processed, consider the vertices of its subtree that can be reached from $x$ without passing through a vertex already in $Z$, including $x$ itself. If there are more than $L$ of them, put $x$ into $Z$. Every component of $G-Z$ has a highest vertex $x\notin Z$ and consists of the at most $L$ vertices considered for $x$. When $x$ is put into $Z$, we charge it to the at least $L+1$ vertices considered for $x$. No vertex of $Z$ above $x$ considers these vertices, since its paths to them pass through $x$. The charged sets are therefore disjoint, and $(L+1)|Z|$ is at most the number of vertices of $G$.
\end{proof}

\begin{proof}[Proof of \Cref{lem:rescue}]
\emph{The separator.}
The forest $\operatorname{supp}(\pi)$ has at most $|I|+|J|\le3k-1$ vertices, so \Cref{lem:separator} gives a set $Z$ with $|Z|<3k/(L+1)$ such that every component of $\operatorname{supp}(\pi)-Z$ has at most $L$ vertices. Let $Z_I=Z\cap I$ and $Z_J=Z\cap J$. For a component $Y$ of $\operatorname{supp}(\pi)-Z$, let $I_Y=I\cap Y$ and $J_Y=J\cap Y$, and define its \emph{excess} $d_Y=|I_Y|-|J_Y|$ and its saving $W_Y=\sum_{i\in I_Y}W_i$.

\emph{Rescuing components.}
Every set $\mathcal Y$ of components defines sets $R=Z_I\cup\bigcup_{Y\in\mathcal Y}I_Y$ and $J_0=\bigcup_{Y\in\mathcal Y}J_Y$ that satisfy~(i). If $\pi_{ij}>0$ with $j\in J_Y$, then $i$ lies in $Z_I$ or in $Y$, and thus in $R$. Moreover,
\begin{equation}\label{eq:rescue-count}
|R|+|J|-|J_0|=|J|+|Z_I|+\sum_{Y\in\mathcal Y}d_Y .
\end{equation}

\emph{A small budget.}
If $b<|Z_I|+L$, we take $R=J_0=\emptyset$, which satisfies (i) and (ii). It satisfies~(iii) because its factor is negative: $(b-L)/k<|Z_I|/k<3/(L+1)$.

\emph{A large budget.}
If $b\ge|Z_I|+L$, let $\mathcal Y_0$ be the set of components with $d_Y\le0$. All other components have excess between $1$ and $L$; let $W_+$ and $d_+$ be their total saving and their total excess. By~\eqref{eq:rescue-count}, we may spend the budget
\[
b_+=k+b-|J|-|Z_I|-\sum_{Y\in\mathcal Y_0}d_Y\ \ge\ b-|Z_I|\ \ge\ L
\]
on components of positive excess. We sort these components by $W_Y/d_Y$ in nonincreasing order and let $\mathcal Y_1$ be the longest prefix of total excess at most $b_+$. With $\mathcal Y=\mathcal Y_0\cup\mathcal Y_1$, \eqref{eq:rescue-count} gives~(ii).

\emph{The greedy choice.}
If $\mathcal Y_1$ contains all components of positive excess, then $R=I$, and~(iii) holds because its factor is at most $1$. Otherwise, the first component outside $\mathcal Y_1$ has excess at most $L$, so the total excess of $\mathcal Y_1$ exceeds $b_+-L\ge0$. Since $\mathcal Y_1$ consists of the components with the largest ratios $W_Y/d_Y$,
\[
\sum_{Y\in\mathcal Y_1}W_Y\ \ge\ \frac{b_+-L}{d_+}\,W_+ .
\]

\emph{The ratio.}
Since $\sum_Yd_Y=(|I|-|Z_I|)-(|J|-|Z_J|)$ over all components and $|I|\le2k-1$,
\[
d_+=\sum_Yd_Y-\sum_{Y\in\mathcal Y_0}d_Y\ <\ b_++(k-b)+|Z_J| .
\]
The function $t\mapsto(t-L)/(t+k-b+|Z_J|)$ is nondecreasing for $t>0$, and $b_+\ge b-|Z_I|$, so
\[
\frac{b_+-L}{d_+}\ \ge\ \frac{b_+-L}{b_++k-b+|Z_J|}\ \ge\ \frac{b-|Z_I|-L}{k-|Z_I|+|Z_J|}\ \ge\ \frac{b-|Z|-L}{k}.
\]
By the last inequality, the separator and the rounding lose at most $(|Z|+L)/k$ compared with the fraction $b/k$ of the fractional rescue in \Cref{sec:approach-rescue}. The blocks of $Z_I$ use $|Z_I|$ units of the budget but belong to no component, and the open centers of $Z_J$ add $|Z_J|$ to the denominator. Cross-multiplying turns the last inequality into $(b-|Z_I|-L)|Z_I|+|Z_J|(k-b+L+|Z_J|)\ge0$, which holds since $b\ge|Z_I|+L$ and $b\le k$.

\emph{Conclusion.}
Since $R$ consists of $Z_I$ and the blocks of all components in $\mathcal Y_0\cup\mathcal Y_1$, the two bounds above and $(b-|Z|-L)/k\le1$ give
\[
\sum_{i\in R}W_i\ \ge\ \sum_{i\in Z_I}W_i+\sum_{Y\in\mathcal Y_0}W_Y+\frac{b-|Z|-L}{k}\,W_+\ \ge\ \frac{b-|Z|-L}{k}\sum_{i\in I}W_i .
\]
Finally, $|Z|<3k/(L+1)$ gives~(iii).
\end{proof}

\subsection{Hierarchical Decompositions}\label{sec:decomp}

This subsection constructs the decompositions on which the network of \Cref{sec:network} is built. The network routes mass between two points accurately unless the decomposition cuts the pair at a level much higher than its distance. We therefore need a family of decompositions such that, for every list of pairs, some member of the family makes few pairs of the list $\Lambda$-bad. Randomized decompositions with this property are standard for doubling metrics~\cite{GKL03,Tal04,CFS21}. We give a polynomial-size family in which all levels share one hash function and one radius parameter, so that the family can be enumerated.

Throughout this subsection, $X$ is a finite set with at least two points and doubling dimension at most $D$. Scaling all distances by the same factor changes all costs by the same factor, so we assume that the minimum distance between distinct points of $X$ is $2$. Let $H$ be the smallest integer such that $2^H$ is at least the diameter of $X$; then $H=O(\log\Delta(X))$.

\begin{definition}[Hierarchical decomposition]\label{def:decomposition}
A \emph{hierarchical decomposition} $\Dec$ of $X$ is a sequence of partitions $\Pcal_0,\dots,\Pcal_H$ of $X$ such that $\Pcal_H=\{X\}$, every part of $\Pcal_h$ has diameter at most $2^h$, and $\Pcal_h$ refines $\Pcal_{h+1}$ for every $h<H$. The parts of $\Pcal_h$ are the \emph{cells} of level $h$, and the \emph{parent} of a cell of level $h<H$ is the cell of level $h+1$ that contains it; cells of different levels are different objects even if they contain the same points. For distinct $x,y\in X$, the \emph{cut level} $\lev(x,y)$ is the smallest $h$ such that $x$ and $y$ lie in a common cell of level $h$. For $\Lambda\ge1$, the pair $(x,y)$ is \emph{$\Lambda$-bad} if $2^{\lev(x,y)}>\Lambda\cdot\dist(x,y)$; a pair $(x,x)$ is never bad.
\end{definition}

Since the minimum distance is $2$, the cells of level $0$ are single points.

\begin{lemma}[Decomposition family]\label{lem:family}
There is a family $\Hfam$ of $O(|X|^7)$ hierarchical decompositions of $X$, computable in time polynomial in $|X|$ and $H$, and a probability distribution on $\Hfam$ such that for all $\Lambda\ge1$ and all distinct $x,y\in X$,
\[
\Pr_{\Dec\sim\Hfam}\bigl[(x,y)\text{ is $\Lambda$-bad in }\Dec\bigr]\ \le\ \frac{2^{11}D}{\Lambda}.
\]
\end{lemma}

\begin{corollary}\label{cor:few-bad}
For every $\Lambda\ge1$ and every finite list of pairs of points of $X$, some $\Dec\in\Hfam$ makes at most a $2^{11}D/\Lambda$ fraction of them $\Lambda$-bad (counted with multiplicity).
\end{corollary}
\begin{proof}
By \Cref{lem:family} and linearity of expectation, the expected number of $\Lambda$-bad pairs in the list, counted with multiplicity, is at most $2^{11}D/\Lambda$ times its length. Some member of $\Hfam$ attains at most the expectation.
\end{proof}

\paragraph{The construction.}
The decomposition is built from nets, a hash-based ordering and a random radius parameter. For every level $h<H$, fix a maximal set $X_h\subseteq X$ of points that are pairwise more than $2^{h-3}$ apart; by maximality, every point of $X$ is within distance $2^{h-3}$ of $X_h$. Let $q$ be a prime with $|X|\le q<2|X|$, and identify the points of $X$ with distinct elements of the field $\Fq$, which we write as integers $0,\dots,q-1$. For a polynomial $\psi$ over $\Fq$ of degree at most four, the \emph{$\psi$-order} sorts the points $x$ of $X$ by the pair $(\psi(x),x)$ lexicographically.

For $\tau\in[1/4,1/2]$ and a level $h<H$, the partition $\Qcal_h(\psi,\tau)$ is obtained by assigning every point $x\in X$ to the first point $x'\in X_h$ in $\psi$-order with $\dist(x,x')\le\tau2^h$. Its parts are the nonempty sets of points assigned to the same point of $X_h$. Since $\tau2^h\ge2^{h-2}>2^{h-3}$, every point is assigned, and every part has diameter at most $2\tau2^h\le2^h$. Let $\Pcal_H(\psi,\tau)=\{X\}$. For $h<H$, let $\Pcal_h(\psi,\tau)$ be the common refinement of $\Qcal_h(\psi,\tau),\dots,\Qcal_{H-1}(\psi,\tau)$; its cells are the nonempty intersections of one part of each of these partitions. These partitions form a hierarchical decomposition $\Dec(\psi,\tau)$.

We choose $\psi$ with independent uniformly random coefficients in $\Fq$ and, independently, $\tau$ uniformly at random in $[1/4,1/2]$. For a fixed $\psi$, the decomposition $\Dec(\psi,\tau)$ depends on $\tau$ only through the set of inequalities $\dist(x,x')\le\tau2^h$ that hold. It is therefore constant on every open interval between consecutive values of the set
\[
\{1/4,\,1/2\}\cup\bigl\{\dist(x,x')/2^h\in[1/4,1/2]:\ x\in X,\ x'\in X_h,\ h<H\bigr\}.
\]
For fixed $x$ and $x'$, the condition $\dist(x,x')/2^h\in[1/4,1/2]$ means $2\dist(x,x')\le2^h\le4\dist(x,x')$, which holds for at most two levels $h$. This set therefore has at most $2|X|^2+2$ elements. The family $\Hfam$ consists of $\Dec(\psi,\tau)$ for every polynomial $\psi$ and one value $\tau$ from each of these intervals. It has at most $q^5\cdot(2|X|^2+1)=O(|X|^7)$ members. Since the endpoints of the intervals have probability zero, the random decomposition $\Dec(\psi,\tau)$ is a member of $\Hfam$ with probability one. This defines the distribution on $\Hfam$.

\paragraph{The analysis.}
The proof of \Cref{lem:family} has three steps. \Cref{cl:minwise} shows that a fixed point of a set $N$ is first in $\psi$-order with probability at most $20/|N|$, which is within a constant factor of the probability $1/|N|$ under a uniformly random order. A family of hash functions for which this probability is $(1\pm\epsilon)/|N|$ is called approximately min-wise independent, and Indyk~\cite{Ind01} showed that $O(\log(1/\epsilon))$-wise independence suffices. Since we only need a constant factor, a constant degree suffices: the degree four is chosen so that, after conditioning on $\psi(x)$, any four of the remaining indicators are independent, which bounds the fourth central moment of their sum. \Cref{cl:one-level} uses \Cref{cl:minwise} to bound the probability that one level cuts a small ball, and the proof of \Cref{lem:family} sums these probabilities over the levels.

\begin{claim}\label{cl:minwise}
For every nonempty $N\subseteq X$ and every $x\in N$, the probability that $x$ is the first point of $N$ in $\psi$-order is at most $20/|N|$.
\end{claim}
\begin{proof}
\emph{A fixed value of $\psi(x)$.}
Let $m=|N|$; we may assume $m\ge2$, since otherwise the bound exceeds $1$. By the invertibility of the Vandermonde matrix, the values of $\psi$ at any five distinct points are independent and uniform in $\Fq$. Condition on $\psi(x)$, and let $\chi$ be the number of points $y\in N\setminus\{x\}$ with $\psi(y)<\psi(x)$. If $x$ is first, then $\chi=0$. Given $\psi(x)$, the variable $\chi$ is the sum of $m-1$ indicator variables, any four of which are independent. Each has expectation $\psi(x)/q$, so $\mathbb E\chi=(m-1)\psi(x)/q$. Expanding the fourth power and using four-wise independence, together with the fact that each indicator has variance and fourth central moment at most its mean, gives
\[
\mathbb E[(\chi-\mathbb E\chi)^4]\le\mathbb E\chi+3(\mathbb E\chi)^2.
\]
If $\mathbb E\chi\ge1$, Markov's inequality for the fourth power then gives
\[
\Pr[\chi=0\mid\psi(x)]\le\Pr\bigl[|\chi-\mathbb E\chi|\ge\mathbb E\chi\mid\psi(x)\bigr]\le\frac{\mathbb E\chi+3(\mathbb E\chi)^2}{(\mathbb E\chi)^4}\le\frac{4}{(\mathbb E\chi)^2}.
\]

\emph{Averaging over $\psi(x)$.}
We now average over the $q$ values of $\psi(x)$. Since $q\ge m$, the values with $\mathbb E\chi<1$, that is, $\psi(x)<q/(m-1)$, contribute at most $(q/(m-1)+1)/q\le2/(m-1)$. Using $\sum_{j\ge c}j^{-2}\le2/c$ for $c\ge1$, the values $\psi(x)=j\ge q/(m-1)$ contribute at most
\[
\frac1q\sum_{j\ge q/(m-1)}\frac{4q^2}{(m-1)^2j^2}\ \le\ \frac{4q}{(m-1)^2}\cdot\frac{2(m-1)}{q}=\frac{8}{m-1}.
\]
The total is at most $10/(m-1)\le20/m$.
\end{proof}

\begin{claim}\label{cl:one-level}
For all $x\in X$, $r>0$ and $h<H$, the probability that $\Qcal_h(\psi,\tau)$ separates two points of $B(x,r)$ is at most $2^9D r/2^h$.
\end{claim}
\begin{proof}
\emph{A fixed $\tau$.}
If $r>2^{h-3}$, the bound exceeds $1$. Let $r\le2^{h-3}$ and fix $\tau$. Let $X_-=X_h\cap B(x,\tau2^h-r)$ and $X_+=X_h\cap B(x,\tau2^h+r)$. Since $\tau2^h-r\ge2^{h-3}$ and some point of $X_h$ is within $2^{h-3}$ of $x$, the set $X_-$ is nonempty. Suppose that the first point $x'$ of $X_+$ in $\psi$-order lies in $X_-$. Then $x'$ is within distance $\tau2^h$ of every point of $B(x,r)$. Every point of $X_h$ within distance $\tau2^h$ of some point of $B(x,r)$ lies in $X_+$, so it is $x'$ or comes after $x'$. All of $B(x,r)$ is therefore assigned to $x'$ and lies in one part. By \Cref{cl:minwise}, a union bound over $X_+\setminus X_-$, and $1-1/w\le\ln w$ for $w\ge1$,
\[
\Pr_\psi\bigl[\Qcal_h(\psi,\tau)\text{ separates }B(x,r)\bigr]\le20\,\frac{|X_+\setminus X_-|}{|X_+|}\le20\ln\frac{|X_+|}{|X_-|}.
\]

\emph{Averaging over $\tau$.}
Let $f(s)=\ln|X_h\cap B(x,s)|$, a nondecreasing function that is nonnegative for $s\ge2^{h-3}$. The points of $X_h\cap B(x,5\cdot2^{h-3})$ are pairwise more than $2^{h-3}$ apart. By \Cref{fact:packing}, there are at most $20^D$ of them, so $f(5\cdot2^{h-3})\le D\ln20$. Averaging over $\tau$, that is, over $s=\tau2^h$ uniform in $[2^{h-2},2^{h-1}]$, gives
\[
\mathbb E_\tau\Bigl[\ln\frac{|X_+|}{|X_-|}\Bigr]=\frac{4}{2^h}\int_{2^{h-2}}^{2^{h-1}}\bigl(f(s+r)-f(s-r)\bigr)\,ds
=\frac{4}{2^h}\Bigl(\int_{2^{h-1}-r}^{2^{h-1}+r}f-\int_{2^{h-2}-r}^{2^{h-2}+r}f\Bigr)\le\frac{4}{2^h}\cdot2r\,f(5\cdot2^{h-3}).
\]
The last step uses that $f$ is nondecreasing, that $f\ge0$ on the second interval, and that $2^{h-1}+r\le5\cdot2^{h-3}$. Combining the two bounds, the probability is at most $20\cdot8D\ln20\cdot r/2^h\le2^9D r/2^h$.
\end{proof}

\begin{proof}[Proof of \Cref{lem:family}]
\emph{One level.}
If no partition $\Qcal_l(\psi,\tau)$ with $l\ge h$ separates two points of a ball, neither does their common refinement $\Pcal_h(\psi,\tau)$. By \Cref{cl:one-level} and a union bound over $l$, the partition $\Pcal_h(\psi,\tau)$ therefore separates two points of $B(x,r)$ with probability at most
\[
\sum_{l=h}^{H-1}\frac{2^9D r}{2^l}\ \le\ \frac{2^{10}D r}{2^h}.
\]

\emph{Bad pairs.}
Let $x\ne y$ and $r=\dist(x,y)\ge2$. If $\Lambda r\ge2^H$, then $(x,y)$ is never $\Lambda$-bad, since $\lev(x,y)\le H$. Otherwise, let $h$ be the integer with $2^h\le\Lambda r<2^{h+1}$. Since $2\le\Lambda r<2^H$, we have $1\le h<H$. If $(x,y)$ is $\Lambda$-bad, then $\lev(x,y)>h$, so $\Pcal_h(\psi,\tau)$ separates the points $x,y\in B(x,r)$. By the bound above and $2^{h+1}>\Lambda r$, this happens with probability at most
\[
\frac{2^{10}D r}{2^h}\ <\ \frac{2^{11}D}{\Lambda}.
\]

\emph{Computation.}
The construction enumerates the $O(|X|^7)$ members of $\Hfam$ explicitly, one for every polynomial $\psi$ and each of at most $2|X|^2+1$ values of $\tau$. Since each decomposition is computed in time polynomial in $|X|$ and $H$, so are the family and the distribution.
\end{proof}

\subsection{The Network and the Dynamic Program}\label{sec:network}

This subsection defines the network in which the algorithm routes the blocks, and shows how to optimize over it. \Cref{sec:network-def} defines the network and proves that it never underestimates distances and that it is accurate for pairs that are not $\Lambda$-bad. \Cref{sec:network-sol} defines network solutions, states that an optimal one can be computed by dynamic programming, and converts network solutions into clusterings. \Cref{sec:dp} gives the dynamic program.

Throughout this subsection, $X=P\cup F$ has doubling dimension at most $D$ and minimum distance $2$ between distinct points. We are given a block partition of $P$, parameters $u\ge1$ and $v\ge0$, a hierarchical decomposition $\Dec$ of $X$ with levels $0,\dots,H$, and a parameter $\lambda\in(0,1]$.

\subsubsection{The Network}\label{sec:network-def}

Every cell $C$ of level $h$ receives a set $\Pi_C\subseteq C$ of \emph{portals}: a maximal subset of $C$ whose points are pairwise more than $\lambda2^h$ apart. Since $C$ has diameter at most $2^h$, it lies in $B(x,2^h)$ for every $x\in C$. \Cref{fact:packing} with radius $2^h$ and separation $\lambda2^h$ therefore gives
\begin{equation}\label{eq:portals}
|\Pi_C|\le M:=\bigl\lceil(4/\lambda)^D\bigr\rceil .
\end{equation}
A cell of level $0$ is a single point, which is its only portal. For a cell $C$ of level $h<H$ with parent $C^+$ and a portal $x'\in\Pi_C$, we fix a portal $\parent(x')\in\Pi_{C^+}$ with $\dist(x',\parent(x'))\le\lambda2^{h+1}$. It exists because $x'\in C^+$ and $\Pi_{C^+}$ is maximal.

\begin{definition}[Network]\label{def:network}
The network $\Net$ has two vertices $\up{x'}C$ and $\dn{x'}C$ for every cell $C$ and every portal $x'\in\Pi_C$, and two vertices $\src x$ and $\snk x$ for every point $x\in X$. Its arcs and their costs are:
\begin{itemize}
\item \emph{ascending arcs} $\up{x'}C\to\up{\parent(x')}{C^+}$ and \emph{descending arcs} $\dn{\parent(x')}{C^+}\to\dn{x'}C$, of cost $0$, for every cell $C$ with parent $C^+$ and every $x'\in\Pi_C$;
\item \emph{crossing arcs} $\up{x'}C\to\dn{y'}C$, of cost $(\dist(x',y')+4\lambda2^h)^z$, for every cell $C$ of level $h$ and all $x',y'\in\Pi_C$;
\item arcs $\src x\to\up x{\{x\}}$, $\dn x{\{x\}}\to\snk x$ and $\src x\to\snk x$, of cost $0$, for every $x\in X$, where $\{x\}$ is the cell of level $0$ that contains $x$.
\end{itemize}
\end{definition}

Since ascending arcs increase the level, descending arcs decrease it, and crossing arcs go from ascending to descending copies, $\Net$ is acyclic. A path from $\src x$ to $\snk y$ either is the arc $\src x\to\snk y$ (if $x=y$), or it climbs along ascending arcs, uses exactly one crossing arc, and descends along descending arcs.

\begin{lemma}[Routes]\label{lem:routes}
Let $x,y\in X$.
\begin{enumerate}[label=(\alph*)]
\item Every path from $\src x$ to $\snk y$ has cost at least $\dist(x,y)^z$.
\item Let $\Lambda\ge1$. If $x=y$ or the pair $(x,y)$ is not $\Lambda$-bad, then some path from $\src x$ to $\snk y$ has cost at most $\bigl((1+8\lambda\Lambda)\dist(x,y)\bigr)^z$.
\end{enumerate}
\end{lemma}
\begin{proof}
\emph{Portals above $x$ and $y$.}
For $l=0,\dots,H$, let $C_l$ be the cell of level $l$ that contains $x$, and let $x_0=x$ and $x_l=\parent(x_{l-1})\in\Pi_{C_l}$. Every portal has a single parent, so the ascending arcs from $\up x{\{x\}}$ visit exactly $\up{(x_0)}{C_0},\up{(x_1)}{C_1},\dots$. Since $\dist(x_{h-1},x_h)\le\lambda2^h$ for $h\ge1$,
\[
\dist(x,x_l)\le\sum_{h=1}^l\lambda2^h<\lambda2^{l+1}.
\]
Similarly, let $C'_l$ be the cell of level $l$ that contains $y$, and let $y_0=y$ and $y_l=\parent(y_{l-1})$, so that $\dist(y,y_l)<\lambda2^{l+1}$. A path that descends to $\dn y{\{y\}}$ must visit $\dn{(y_l)}{C'_l},\dots,\dn{(y_0)}{C'_0}$ for some $l$.

(a) The arc $\src x\to\snk y$ exists only if $x=y$, and it costs $0$. Any other path climbs to $\up{(x_l)}{C_l}$, crosses to $\dn{(y_l)}{C'_l}$ with $C'_l=C_l$, and descends. Its cost is $(\dist(x_l,y_l)+4\lambda2^l)^z$, and by the triangle inequality and the bounds above,
\[
\dist(x,y)\le\dist(x,x_l)+\dist(x_l,y_l)+\dist(y_l,y)<\dist(x_l,y_l)+4\lambda2^l.
\]

(b) For $x=y$, use the arc $\src x\to\snk x$. Otherwise, let $l=\lev(x,y)$. Then $C_l=C'_l$, and $2^l\le\Lambda\dist(x,y)$ since $(x,y)$ is not $\Lambda$-bad. The path that climbs to $\up{(x_l)}{C_l}$, crosses to $\dn{(y_l)}{C_l}$, and descends to $\snk y$ costs
\[
(\dist(x_l,y_l)+4\lambda2^l)^z\le(\dist(x,y)+8\lambda2^l)^z\le\bigl((1+8\lambda\Lambda)\dist(x,y)\bigr)^z. \qedhere
\]
\end{proof}

\subsubsection{Network Solutions}\label{sec:network-sol}

In a network solution, every block is either rescued or normal. Normal blocks inject their clients as mass at their representatives, and a location that receives $\ell$ units of mass opens $\lceil\ell/U\rceil$ centers.

\begin{definition}[Network solution]\label{def:network-solution}
A \emph{network solution} $(R,f)$ consists of a set $R\subseteq I$ of \emph{rescued} blocks and an integral flow $f\ge0$ on the arcs of $\Net$ such that every portal vertex has equal inflow and outflow, every vertex $\src x$ has outflow $s_x:=\sum_{i\notin R:\,a_i=x}w_i$, and every vertex $\snk y$ has an inflow $\ell_y$ with $\ell_y=0$ if $y\notin F$. Its number of centers is $|R|+\sum_{y\in F}\lceil\ell_y/U\rceil$, and it is \emph{feasible} if this number is at most $K$. Its cost is
\[
\Phi(R,f):=\sum_{i\notin R}(u-1)A_i+v\cost(f),
\]
where $\cost(f):=\sum_e\cost(e)f(e)$ is the cost of the flow, summed over all arcs $e$ of $\Net$.
\end{definition}

\begin{lemma}[Dynamic program]\label{lem:dp}
A feasible network solution of minimum cost, if one exists, can be computed in time $(2n+1)^{O(M)}\cdot\mathrm{poly}(|X|,H)$.
\end{lemma}

\begin{lemma}[From network solutions to clusterings]\label{lem:to-clustering}
Suppose that every block is $(u,v)$-movable. Given a feasible network solution $(R,f)$, one can compute in polynomial time a soft solution with at most $K$ centers at locations in $F$ whose cost is at most $A+\Phi(R,f)$.
\end{lemma}
\begin{proof}
\emph{Centers and paths.}
Since $\Net$ is acyclic, $f$ decomposes into $\sum_xs_x$ paths of one unit each, every path going from some $\src x$ to some $\snk y$ with $y\in F$. For every $y\in F$ with $\ell_y>0$, open $\lceil\ell_y/U\rceil$ centers at $y$ and distribute the $\ell_y$ paths that end at $\snk y$ among them, at most $U$ per center. For every location $x$, distribute the $s_x$ paths that start at $\src x$ among the normal blocks with representative $x$, so that block $i$ receives $w_i$ paths. Let $y_{i,1},\dots,y_{i,w_i}$ be their end points, and let $\sum_y$ denote the sum over $y=y_{i,1},\dots,y_{i,w_i}$. By \Cref{lem:routes}(a), $\sum_y\dist(a_i,y)^z$ is at most the total cost of the paths of block $i$. Summing over the normal blocks gives
\[
\sum_{i\notin R}\sum_y\dist(a_i,y)^z\le\cost(f).
\]

\emph{The clustering.}
Assign the clients of every normal block $i$ to the centers of its paths by a minimum-cost bipartite matching. By \Cref{def:movable}, its cost is at most $uA_i+v\sum_y\dist(a_i,y)^z$. Serve every rescued block $i$ by a new center at $a_i$, at cost $A_i$. Every center serves at most $U$ clients, and there are at most $K$ centers since $(R,f)$ is feasible. By the bound above, the total cost is at most
\[
\sum_{i\in R}A_i+\sum_{i\notin R}\Bigl(uA_i+v\sum_y\dist(a_i,y)^z\Bigr)\le A+\Phi(R,f). \qedhere
\]
\end{proof}

\subsubsection{The Dynamic Program}\label{sec:dp}

We prove \Cref{lem:dp}. The network follows the tree of cells, which gives a tree decomposition of small width.

\paragraph{Tree decompositions.}
A tree decomposition of a directed graph is a tree together with a set of vertices of the graph, called a bag, for every node of the tree. Both endpoints of every arc must lie in a common bag, and for every vertex, the nodes whose bags contain it must form a nonempty subtree. Its width is the largest bag size minus one. A \emph{nice} tree decomposition with introduce-arc nodes is rooted and has empty bags at the root and at the leaves. Every other node is of one of four types. An introduce node or a forget node has one child, and its bag differs from the bag of the child by one vertex. An introduce-arc node has one child with the same bag and is labeled with an arc between two bag vertices; every arc is the label of exactly one node. A join node has two children with the same bag as itself. A tree decomposition can be turned in polynomial time into a nice one of the same width whose number of nodes is polynomial in the number of nodes, the width, and the number of arcs~\cite[Section~7.3]{CFKLMPPS15}.

\paragraph{A tree decomposition of the network.}
Start with the tree whose nodes are the cells of $\Dec$, in which every cell is adjacent to its parent. The bag of a cell $C$ is $\bag(C)=\{\up{x'}C,\dn{x'}C:x'\in\Pi_C\}$. Subdivide every edge between a cell $C$ and its parent $C^+$ by a node with bag $\bag(C)\cup\bag(C^+)$. To the node of every cell $\{x\}$ of level $0$, attach a new leaf with bag $\bag(\{x\})\cup\{\src x,\snk x\}$. Crossing arcs lie in a bag $\bag(C)$, ascending and descending arcs between $C$ and its parent $C^+$ lie in $\bag(C)\cup\bag(C^+)$, and the three arcs at $x$ lie in the new leaf. A vertex $\up{x'}C$ or $\dn{x'}C$ lies exactly in the bags of $C$, of the subdivision nodes next to $C$, and possibly of the leaf attached to $C$, which form a subtree. The vertices $\src x$ and $\snk x$ lie in one bag. This gives a tree decomposition of $\Net$ with $O(H|X|)$ nodes, and by~\eqref{eq:portals} its bags have size at most $4M$. We turn it into a nice tree decomposition with introduce-arc nodes of the same width.

\paragraph{Local tables.}
For a location $x$, let $I_x=\{i:a_i=x\}$. For integers $0\le s\le n$ and $0\le q\le K$, let $\loc_x(s,q)$ be the minimum of $\sum_{i\in I_x\setminus R_x}(u-1)A_i$ over all $R_x\subseteq I_x$ with $|R_x|=q$ and $\sum_{i\in I_x\setminus R_x}w_i=s$. If no such $R_x$ exists, let $\loc_x(s,q)=\infty$. A knapsack recursion over the blocks of $I_x$ computes all values $\loc_x(s,q)$ in time $O(|I_x|\,nK)$, together with optimal sets $R_x$.

\paragraph{States.}
Consider a node of the nice tree decomposition; the arcs introduced below it and the vertices forgotten below it are fixed. A partial solution at this node assigns a flow value in $\{0,\dots,n\}$ to every arc introduced below the node and a number $q_x$ to every forgotten vertex $\src x$. On these arcs, every forgotten portal vertex must have equal inflow and outflow, every forgotten $\src x$ an outflow $s_x$ with $\loc_x(s_x,q_x)<\infty$, and every forgotten $\snk y$ an inflow $\ell_y\ge0$ with $\ell_y=0$ if $y\notin F$. The state of the partial solution is the pair $(\bal,k')$. Here $\bal$ maps every vertex of the bag to its balance, that is, its inflow minus its outflow on the introduced arcs, and $k'=\sum q_x+\sum\lceil\ell_y/U\rceil$ counts the centers at the forgotten sources and sinks. The cost of the partial solution is $\sum\loc_x(s_x,q_x)+v\sum\cost(e)f(e)$, over the forgotten sources and the introduced arcs. The table of the node stores the minimum cost of a partial solution for every state with balances in $\{-n,\dots,n\}$ and $k'\le K$.

\paragraph{Transitions.}
The tables are computed bottom-up. A leaf has the single state with empty $\bal$ and $k'=0$, of cost $0$. An introduce node sets the balance of the new vertex to $0$. An introduce-arc node for an arc $e$ tries every flow value $f(e)\in\{0,\dots,n\}$. It decreases the balance of the tail of $e$ and increases the balance of its head by $f(e)$, and it adds $v\cdot\cost(e)\cdot f(e)$ to the cost. A forget node for a portal vertex keeps the states in which this vertex has balance $0$. A forget node for $\src x$ keeps the states in which $\src x$ has balance $-s$ for some $s\ge0$, chooses $q$, and adds $q$ to $k'$ and $\loc_x(s,q)$ to the cost. A forget node for $\snk y$ keeps the states in which $\snk y$ has a balance $\ell\ge0$, where $\ell=0$ if $y\notin F$, and adds $\lceil\ell/U\rceil$ to $k'$. A join node combines pairs of states of its children by adding their balances and their counters $k'$. This is correct because every arc is introduced at exactly one node. The minimum cost over the states of the root with $k'\le K$ is the minimum cost of a feasible network solution, and an optimal solution is recovered by storing the choices.

\paragraph{Correctness of the value range.}
In a network solution, $f$ decomposes into at most $n$ unit paths in the acyclic network. Since every path visits a vertex at most once, every arc carries at most $n$ units, and the total inflow and the total outflow of every vertex are at most $n$. The balance on any set of arcs therefore lies in $\{-n,\dots,n\}$, and restricting the tables to this range loses no network solution.

\paragraph{Running time.}
Since a bag has at most $4M$ vertices, a table has at most $(2n+1)^{4M}(K+1)$ entries. A join node takes time quadratic in the table size, and the other nodes take time linear in it times $O(n)$. Since $K\le2n$ and the nice tree decomposition has $\mathrm{poly}(|X|,H)$ nodes, the total time is $(2n+1)^{O(M)}\cdot\mathrm{poly}(|X|,H)$. This proves \Cref{lem:dp}.

\subsection{Putting the Parts Together}\label{sec:proof-combine}

The algorithm of \Cref{thm:framework} takes the block partition and the pair $(u,v)$ as input and uses the parameters
\begin{equation}\label{eq:params}
\gamma=\frac{\min\{\eps,\eta\}}{4},\qquad L=\Bigl\lceil\frac{12}{\eta}\Bigr\rceil,\qquad k_0=\Bigl\lceil\frac{4(L+1)}{\eta}\Bigr\rceil,\qquad \Lambda=\frac{2^{13}D}{\gamma},\qquad \lambda=\frac{\eta}{8z\Lambda}.
\end{equation}
\begin{table}[t]
\centering
\small
\renewcommand{\arraystretch}{1.1}
\begin{tabular}{@{}lll@{}}
\toprule
Parameter & Role & Value\\
\midrule
$\delta$ & additive loss in the ratio of a theorem & given\\
$\eta$ & accuracy of the framework & of order $\delta/3^z$, at most $1/10$\\
$\gamma$ & fraction of $k$ reserved for blocks on bad edges & $\min\{\eps,\eta\}/4$\\
$b$ & extra centers used by the rescue lemma & $\lfloor\eps k\rfloor-\lfloor\gamma k\rfloor$\\
$L$ & size of the components in the rescue lemma & $\lceil12/\eta\rceil$\\
$k_0$ & exhaustive search for $k<k_0$ & $\lceil4(L+1)/\eta\rceil$\\
$\Lambda$ & threshold for $\Lambda$-bad pairs & $2^{13}D/\gamma$\\
$\lambda$ & spacing of the portals, relative to the cell diameter bound & $\eta/(8z\Lambda)$\\
$M$ & upper bound on the number of portals per cell & $\lceil(4/\lambda)^D\rceil$\\
\bottomrule
\end{tabular}
\caption{The parameters of the framework theorem, for fixed $D$, $z$ and $\eps$.}
\label{tab:params}
\end{table}
\Cref{tab:params} lists their roles. In brief, $\eta$ controls all approximation errors and is chosen from $\delta$ in each proof. Of the other parameters, $\gamma$, $b$ and $L$ belong to the rescue lemma, and $\Lambda$, $\lambda$ and $M$ to the network. The two groups are linked through $\gamma$. A larger reserve of $\lfloor\gamma k\rfloor$ centers for bad edges allows a smaller threshold $\Lambda=2^{13}D/\gamma$ for the ratio between the cut level of a pair and its distance, hence a coarser portal spacing $\lambda$, fewer portals $M$ per cell and a faster dynamic program; but it leaves fewer extra centers $b$ for the rescue lemma. The choice $\gamma=\min\{\eps,\eta\}/4$ makes $\Lambda$, $\lambda$ and $M$ constants that depend only on $D$, $z$, $\eps$ and $\eta$, while the rescue lemma still saves an $(\eps-\eta)$-fraction.

With these parameters fixed, the algorithm proceeds in three steps.
\begin{enumerate}[label=\arabic*.,leftmargin=1.6em]
\item If $k<k_0$, try every multiset of at most $k$ locations, each with an optimal assignment computed by a minimum-cost flow, and return the best solution.
\item Otherwise, scale the distances so that the minimum distance in $X=P\cup F$ is $2$. For every decomposition $\Dec$ in the family $\Hfam$ of \Cref{lem:family}, compute a network solution of minimum cost in $\Net$ by \Cref{lem:dp} with portal parameter $\lambda$.
\item Turn the cheapest network solution found into a clustering by \Cref{lem:to-clustering}.
\end{enumerate}

\begin{proof}[Proof of \Cref{thm:framework}]
If $k<k_0$, the algorithm returns an optimal soft solution with at most $k$ centers, whose cost is at most $\cost(S^*)$. This suffices because $\beta-(\beta-1)\eps\ge1$. Now let $k\ge k_0$, and fix a soft solution $S^*$ with at most $k$ centers and $A\le(1+\eta)\cost(S^*)$. Let $\pi$ be the transport plan of \Cref{lem:transport} for $S^*$, with moving costs $T_i$ and savings $W_i=(u-1)A_i+vT_i$. By~\eqref{eq:total-saving}, $\sum_iW_i\le(\beta-1)(1+\eta)\cost(S^*)$. The proof chooses the rescued blocks, finds a decomposition with few bad edges, and routes the remaining mass through the network.

\emph{Rescued blocks.}
We apply \Cref{lem:rescue} to $\pi$, with $L$ as in~\eqref{eq:params} and $b=\lfloor\eps k\rfloor-\lfloor\gamma k\rfloor$; since $\gamma\le\eps\le1$, we have $0\le b\le k$. This gives sets $R_0\subseteq I$ and $J_0\subseteq J$ with properties (i)--(iii). Since $b\ge\eps k-\gamma k-1$, $\gamma\le\eta/4$, $3/(L+1)\le\eta/4$ and $(L+1)/k\le\eta/4$,
\[
\frac{b-L}{k}-\frac{3}{L+1}\ \ge\ \eps-\gamma-\frac{L+1}{k}-\frac{3}{L+1}\ \ge\ \eps-\eta,
\]
and~(iii) gives
\[
\sum_{i\in R_0}W_i\ \ge\ (\eps-\eta)\sum_iW_i .
\]

\emph{Few bad edges.}
We apply \Cref{cor:few-bad} to the list of the pairs $(a_i,c_j)$ over the edges $ij$ of $\operatorname{supp}(\pi)$. This forest has $|I|+|J|<3k$ vertices, so the list has fewer than $3k$ pairs. By the choice of $\Lambda$, some $\Dec\in\Hfam$ makes fewer than $2^{11}D\cdot3k/\Lambda=3\gamma k/4$ of these pairs $\Lambda$-bad. This number is an integer smaller than $\gamma k$, so it is at most $\lfloor\gamma k\rfloor$. Let $R_{\mathrm{bad}}$ be the set of blocks incident to the edges with $\Lambda$-bad pairs, so $|R_{\mathrm{bad}}|\le\lfloor\gamma k\rfloor$. Let $R=R_0\cup R_{\mathrm{bad}}$.

\emph{Routing.}
We define a network solution $(R,f)$ in $\Net$. For every edge $ij$ of $\operatorname{supp}(\pi)$ with $i\notin R$, the pair $(a_i,c_j)$ is not $\Lambda$-bad, and we send $\pi_{ij}$ units from $\src{(a_i)}$ to $\snk{(c_j)}$ along a path given by \Cref{lem:routes}(b). Since the row sums of $\pi$ are the sizes $w_i$, every vertex $\src x$ sends $\sum_{i\notin R:\,a_i=x}w_i$ units, as required.

\emph{Feasibility.}
We show that $(R,f)$ has at most $K$ centers. Let $\ell'_j=\sum_{i\notin R}\pi_{ij}\le\ell_j\le U$ be the remaining load of center $j\in J$. By~(i), $\ell'_j=0$ for $j\in J_0$. The inflow $\ell_y$ of $\snk y$ is the sum of $\ell'_j$ over the centers $j$ with $c_j=y$. Since $\ell'_j\le U$, the number $\lceil\ell_y/U\rceil$ is at most the number of these centers with $\ell'_j>0$. By~(ii), the number of centers is therefore at most
\[
|R|+\sum_{y\in F}\Bigl\lceil\frac{\ell_y}{U}\Bigr\rceil\ \le\ |R_{\mathrm{bad}}|+|R_0|+\bigl(|J|-|J_0|\bigr)\ \le\ \lfloor\gamma k\rfloor+k+b=K.
\]

\emph{The cost of the flow.}
By \Cref{lem:routes}(b) and $8\lambda\Lambda=\eta/z$, every unit sent along the edge $ij$ costs at most $((1+\eta/z)\dist(a_i,c_j))^z\le e^\eta\dist(a_i,c_j)^z$. Summing over all units gives $\cost(f)\le e^\eta\sum_{i\notin R}T_i$. With $R_0\subseteq R$ and the bound on $\sum_{i\in R_0}W_i$, this gives
\[
\Phi(R,f)\le\sum_{i\notin R}\bigl((u-1)A_i+e^\eta vT_i\bigr)\le e^\eta\sum_{i\notin R_0}W_i\le e^\eta(1-\eps+\eta)\sum_iW_i .
\]

\emph{Conclusion.}
Since $(R,f)$ is a feasible network solution for one of the decompositions that the algorithm tries, the algorithm finds a network solution of cost at most $\Phi(R,f)$. By \Cref{lem:to-clustering} and~\eqref{eq:total-saving}, its output costs at most
\[
A+e^\eta(1+\eta)(1-\eps+\eta)(\beta-1)\cost(S^*).
\]
For $\eta\le1/10$, we have $e^\eta(1+\eta)\le1+2.2\eta$. Since $0\le1-\eps+\eta\le1.1$, it follows that
\[
e^\eta(1+\eta)(1-\eps+\eta)\le1-\eps+3.5\eta .
\]
Together with $A\le(1+\eta)\cost(S^*)$, the cost is at most
\[
\bigl(1+\eta+(\beta-1)(1-\eps)+3.5(\beta-1)\eta\bigr)\cost(S^*)\le\bigl(\beta-(\beta-1)\eps+4\beta\eta\bigr)\cost(S^*). \qedhere
\]
\end{proof}

\paragraph{Running time.}\label{sec:running-time}
For $k<k_0$, there are at most $(|F|+1)^{k_0}$ multisets, each handled by one minimum-cost flow computation. Otherwise, the family $\Hfam$ has $O(|X|^7)$ members. By \Cref{lem:dp}, each dynamic program takes time $(2n+1)^{O(M)}\cdot\mathrm{poly}(|X|,H)$, where $H=O(\log\Delta(X))$ and $M=\lceil(4/\lambda)^D\rceil$. By~\eqref{eq:params}, $4/\lambda=2^{18}zD/(\eta\gamma)$, so
\[
M=\bigl(zD/(\eta\min\{\eps,\eta\})\bigr)^{O(D)}.
\]
The proofs of \Cref{thm:soft} also run the local search of \Cref{thm:frs}, whose swaps involve
\[
D^{O(D)}\cdot(2^z/\eta)^{O(2^zD/\eta)}
\]
centers. Since $k_0=O(1/\eta^2)$, the running time is polynomial with degree at most
\[
\Bigl(\frac{2^zD}{\eta\min\{\eps,\eta\}}\Bigr)^{O(2^zD/\eta)}.
\]
For \Cref{thm:soft} with $P\cup F$ of bounded doubling dimension, $\eta$ is of order $\delta/3^z$, so for fixed $z$ and $\delta\le1$ the degree is $(D/(\eps\delta))^{O(D/\delta)}$. If only $P$ or only $F$ has bounded doubling dimension, \Cref{lem:compress} replaces $D$ by a doubling dimension $D'$ that is exponential in $D$ (\Cref{sec:proof-soft-onesided,sec:euclid-proof}). For fixed $z$ and $\delta\le1/2$, $D'=(1/\delta)^{O(D)}$, so the degree becomes $(D'/(\eps\delta))^{O(D'/\delta)}=(1/(\eps\delta))^{(1/\delta)^{O(D)}}$. For centers anywhere in $\R^d$, $\eta$ is only polynomial in $\delta$ for fixed $z$ (\Cref{sec:euclid-proof}), and the candidate centers multiply the exponent of $n$ by a factor polynomial in $1/\delta$. The degree is still $(1/(\eps\delta))^{(1/\delta)^{O(D)}}$ for fixed $z$.

\section{Uniform Soft Capacities}\label{sec:soft}

This section completes the proof of \Cref{thm:soft}. When $P\cup F$ has bounded doubling dimension, the framework theorem applies directly (\Cref{sec:proof-soft}). When only one side has bounded doubling dimension, the compression lemma of \Cref{sec:onesided} reduces the problem to one in which both sides do. \Cref{sec:euclid} treats Euclidean space, where certificates sharpen the movability constants.

\subsection{Proof of \texorpdfstring{\Cref{thm:soft}}{Theorem 1.1} for Joint Low Dimension}\label{sec:proof-soft}

Suppose that $P\cup F$ has doubling dimension at most $D$, and let $\eta=\min\{1/10,\ \delta/(4\cdot3^z)\}$. By \Cref{thm:frs} and \Cref{sec:approach-blocks}, compute a block partition with $A\le(1+\eta)\OPT$. By \Cref{lem:triangle-movable}, every block is $(3^{z-1},(3/2)^{z-1})$-movable, with $\beta=3^z$. Applying \Cref{thm:framework} with an optimal soft solution as $S^*$ gives a solution with at most $K$ centers. Since $4\cdot3^z\eta\le\delta$, its cost is at most $(3^z-(3^z-1)\eps+4\cdot3^z\eta)\OPT\le(3^z-(3^z-1)\eps+\delta)\OPT$. \qed

\subsection{One-Sided Low Dimension}\label{sec:onesided}

The framework theorem needs $P\cup F$ to have bounded doubling dimension. This subsection shows that it suffices for one of $P$ and $F$ to have bounded doubling dimension; the other side can be compressed. We also use the compression lemma in the Euclidean setting (\Cref{sec:euclid}), where the other side is a set of candidate centers.

\begin{lemma}[Compression]\label{lem:compress}
Let $Z$ and $Y$ be finite sets in a metric space, where $Z$ has doubling dimension at most $D\ge1$, and let $\eta_c\in(0,1/4]$. One can compute in polynomial time a map $\mu\colon Y\to Y$, which maps every point $y$ to its \emph{proxy} $\mu(y)$, such that for all $x\in Z$ and $y\in Y$,
\begin{equation}\label{eq:compress}
\dist(x,\mu(y))\le(1+3\eta_c)\dist(x,y)\qquad\text{and}\qquad\dist(x,y)\le(1+3\eta_c)\dist(x,\mu(y)),
\end{equation}
and $Z\cup\mu(Y)$ has doubling dimension at most
\[
D':=\log_2\bigl(2^{4D}+6(768/\eta_c)^D N\bigr)=O\bigl((32/\eta_c)^D\log^2(2/\eta_c)\bigr),
\]
where $N=(1+11/\eta_c)^{(32/\eta_c)^D+\lceil\log_2(2/\eta_c)\rceil(16/\eta_c)^D}$.
\end{lemma}

\begin{figure}[b]
\centering
\begin{tikzpicture}[scale=0.66,
  qpt/.style={circle,fill=black!45,inner sep=0pt,minimum size=2.2pt},
  netpt/.style={cross out,draw=centcol!90!black,line width=0.9pt,inner sep=0pt,minimum size=5pt},
  ypt/.style={circle,fill=blockcol,inner sep=0pt,minimum size=5pt},
  sig/.style={blockcol!55,line width=0.5pt}]
  \begin{scope}
  \clip (-10.2,-1.25) rectangle (10.2,2.15);
  \draw[black!55,dashed] (0,0) circle (2.4);
  \draw[black!40,dashed] (0,0) circle (4.8);
  \draw[black!30,dashed] (0,0) circle (9.0);
  \end{scope}
  \foreach \x in {-9.6,-9.2,...,9.6} \node[qpt] at (\x,{0.3*sin(18*\x)}) {};
  \foreach \x in {-2.0,-1.2,-0.4,0.4,1.2,2.0} \node[netpt] at (\x,{0.3*sin(18*\x)}) {};
  \foreach \x in {-4.0,-2.8,2.8,4.0} \node[netpt] at (\x,{0.3*sin(18*\x)}) {};
  \foreach \x in {-8.4,-6.0,6.0,8.4} \node[netpt] at (\x,{0.3*sin(18*\x)}) {};
  \node[circle,draw=black,fill=white,line width=0.8pt,inner sep=0pt,minimum size=5.5pt] (q) at (0,0) {};
  \node[font=\small,below=3pt] at (q) {$q$};
  \node[ypt] (y) at (0.85,1.25) {};
  \coordinate (xy) at (0.8,{0.3*sin(14.4)});
  \draw[black!70,densely dotted,line width=0.7pt] (y) -- (xy) -- (q);
  \node[circle,fill=black,inner sep=0pt,minimum size=3.4pt] at (xy) {};
  \node[font=\small,below right=0pt and -2pt] at (xy) {$x_y$};
  \node[ypt] (y2) at (-0.4,1.1) {};
  \node[font=\small,above right=-1pt] at (y) {$y$};
  \node[font=\small,above left=-1pt] at (y2) {$y'$};
  \foreach \x in {-2.0,0.4,2.0,-4.0,4.0,-8.4,8.4} \draw[sig] (y) -- (\x,{0.3*sin(18*\x)});
  \node[font=\scriptsize,black!70,anchor=south] at (1.9,1.75) {$8r$};
  \node[font=\scriptsize,black!60,anchor=south] at (4.45,1.75) {$16r$};
  \node[font=\scriptsize,black!50,anchor=south] at (8.8,1.75) {$32r$};
  \node[font=\small,anchor=west] at (9.85,0.1) {$Z$};
\end{tikzpicture}
\caption{The compression of \Cref{lem:compress} (not to scale). The points $y$ and $y'$ have distance about $r$ to the low-dimensional set $Z$ (gray dots) and the same net point $q$. This point lies within distance $\eta_cr$ of a closest point $x_y$ of $Z$ to $y$, so $\dist(y,q)<3r$. The ball of radius $8r$ around $q$ and the rings around it (dashed) contain net points of $Z$ (crosses), whose spacing grows with the distance from $q$. If $y$ and $y'$ have the same rounded distances to all these net points (thin lines), they get the same proxy. Their distances to every point of $Z$ then differ by a factor of at most $1+3\eta_c$.}
\label{fig:compress}
\end{figure}
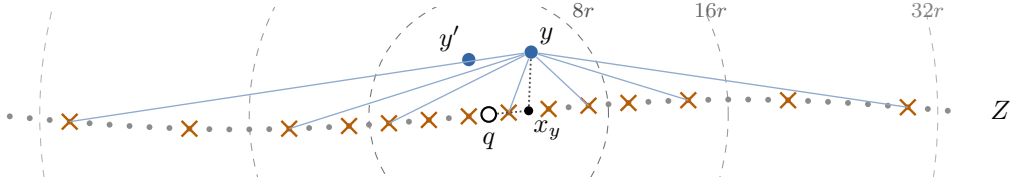

By~\eqref{eq:compress} with $x=y$, every point of $Z\cap Y$ is mapped to itself. \Cref{fig:compress} illustrates the construction.

\begin{proof}
The construction uses that a point $y$ at distance about $r$ from $Z$ is within $3r$ of a net point $q$. For a point $x\in Z$ outside the ball $B(q,8r)$, the distance $\dist(x,y)$ is within a factor of $2$ of $\dist(x,q)$. It therefore suffices to locate $x$ up to an error proportional to $\dist(x,q)$. This allows nets whose spacing doubles from ring to ring, and points beyond radius $16r/\eta_c$ need no net at all.

\emph{Construction.}
For a point $y$, let $\dist(y,Z)=\min_{x\in Z}\dist(y,x)$. Points $y\in Y$ with $\dist(y,Z)=0$ lie in $Z$, and we set $\mu(y)=y$. The other points are partitioned into \emph{layers}. The layers are the smallest positive value of $\dist(y,Z)$ over $y\in Y$ and the numbers obtained from it by repeated doubling, and $y$ belongs to layer $r$ if $r\le\dist(y,Z)<2r$. For every layer $r$, fix a maximal subset $Z_r$ of $Z$ whose points are pairwise more than $\eta_cr$ apart. For $y$ in layer $r$, let $x_y$ be a point of $Z$ closest to $y$, and let $\nu(y)\in Z_r$ be a point with $\dist(x_y,\nu(y))\le\eta_cr$. By the triangle inequality,
\begin{equation}\label{eq:anchor}
\dist(y,\nu(y))\le\dist(y,Z)+\eta_cr<3r .
\end{equation}
For a layer $r$ and a point $q\in Z_r$, consider the ball $Z\cap B(q,8r)$ and, for $j=1,\dots,\lceil\log_2(2/\eta_c)\rceil$, the ring $\{x\in Z:2^{j+2}r\le\dist(x,q)<2^{j+3}r\}$ around $q$. Fix a maximal subset of the ball whose points are pairwise more than $\eta_cr$ apart, and a maximal subset of the $j$-th ring whose points are pairwise more than $2^{j+1}\eta_cr$ apart. The \emph{signature} of a point $y$ in layer $r$ with $\nu(y)=q$ is an integer vector. It lists $\lfloor\dist(y,x)/(\eta_cr)\rfloor$ for every point $x$ of the subset of the ball, and $\lfloor\dist(y,x)/(2^{j+1}\eta_cr)\rfloor$ for every point $x$ of the subset of the $j$-th ring, for all $j$. A \emph{group} consists of the points of $Y$ with the same layer $r$, the same point $\nu(y)$ and the same signature. We fix one point of every group as the proxy of all points of the group. Computing $\mu$ takes polynomial time.

\emph{Distances.}
We show that two points $y,y'$ of the same group, with layer $r$ and $q=\nu(y)=\nu(y')$, satisfy
\[
|\dist(x,y)-\dist(x,y')|\le3\eta_c\min\{\dist(x,y),\dist(x,y')\}\qquad\text{for every }x\in Z,
\]
which implies~\eqref{eq:compress}. By~\eqref{eq:anchor}, $\dist(y,y')<6r$. If $\dist(x,q)\le8r$, some point $x'$ of the subset of the ball has $\dist(x,x')\le\eta_cr$, and $|\dist(x',y)-\dist(x',y')|<\eta_cr$ since the signatures agree. By the triangle inequality, $|\dist(x,y)-\dist(x,y')|<3\eta_cr$, and both distances are at least $r$, the smallest distance from the layer to $Z$. If $x$ lies in the $j$-th ring, the same argument gives $|\dist(x,y)-\dist(x,y')|<3\cdot2^{j+1}\eta_cr$, while both distances are at least $2^{j+2}r-3r\ge2^{j+1}r$. Finally, if $x$ lies beyond the last ring, then $\dist(x,q)\ge16r/\eta_c$. Both distances then exceed $16r/\eta_c-3r\ge2r/\eta_c$, while $|\dist(x,y)-\dist(x,y')|\le\dist(y,y')<6r$.

\emph{Points per net point.}
We show that every point $q\in Z_r$ equals $\nu(y)$ for at most $N$ points $y\in\mu(Y)$ of layer $r$. Such a point $y$ lies outside $Z$ and is determined by $\nu(y)$ and its signature. The subset of the ball has at most $(32/\eta_c)^D$ points by \Cref{fact:packing}, and the subset of each ring at most $(16/\eta_c)^D$. Every coordinate of a signature is smaller than $11/\eta_c$, since $\dist(y,x)<3r+8r$ for $x$ in the ball and $\dist(y,x)<3r+2^{j+3}r\le\frac{19}{4}\cdot2^{j+1}r$ for $x$ in the $j$-th ring. A signature therefore takes at most $N$ values.

\emph{Covering a ball.}
We now bound the doubling dimension: we cover a ball $B(y_0,\tau)$ in $Z\cup\mu(Y)$ by balls of radius $\tau/2$ around points of $Z\cup\mu(Y)$. Let $x_0$ be a point of $Z$ closest to $y_0$.

\emph{Case $\dist(y_0,Z)\le2\tau$.} For every point of the ball within distance $\tau/8$ of $Z$, a closest point of $Z$ lies in $B(x_0,\tau/8+\tau+2\tau)$. Halving the radius four times covers this ball of $Z$ by $2^{4D}$ balls of radius at most $\tau/4$ around points of $Z$. The balls of radius $\tau/2$ around the same points cover all these points of the ball. Every other point $y$ of the ball lies in $\mu(Y)$, and its layer $r$ satisfies $\tau/16<r\le\dist(y,Z)\le3\tau$, so there are at most six such layers. For a fixed layer $r$, the points $\nu(y)$ lie in $Z_r\cap B(x_0,3r+\tau+2\tau)\subseteq Z_r\cap B(x_0,12\tau)$ and are pairwise more than $\eta_cr>\eta_c\tau/16$ apart. By \Cref{fact:packing}, there are at most $(768/\eta_c)^D$ of them, and at most $(768/\eta_c)^D N$ points $y$ of this layer. We cover each such point by the ball of radius $\tau/2$ around it.

\emph{Case $\dist(y_0,Z)>2\tau$.} Since $\dist(\cdot,Z)$ is $1$-Lipschitz, every point of the ball has distance more than $\dist(y_0,Z)/2$ and less than $3\dist(y_0,Z)/2$ from $Z$. The ball therefore lies in $\mu(Y)\setminus Z$, and the layers of its points lie between $\dist(y_0,Z)/4$ and $3\dist(y_0,Z)/2$; there are at most three such layers. For a fixed layer, the points $\nu(y)$ with $y$ in the ball lie in $B(x_0,6\dist(y_0,Z))$ and are pairwise more than $\eta_c\dist(y_0,Z)/4$ apart. Every layer again contributes at most $(96/\eta_c)^D N$ points, and we cover each by a ball of radius $\tau/2$ around it.

In both cases, at most $2^{4D}+6(768/\eta_c)^D N$ balls of radius $\tau/2$ suffice, so $Z\cup\mu(Y)$ has doubling dimension at most~$D'$, where
\[
D'=O\bigl((32/\eta_c)^D\log^2(2/\eta_c)\bigr),
\]
since $\log N=O\bigl((32/\eta_c)^D\log^2(2/\eta_c)\bigr)$.
\end{proof}

\subsubsection{Proof of \texorpdfstring{\Cref{thm:soft}}{Theorem 1.1} for One-Sided Low Dimension}\label{sec:proof-soft-onesided}

The case that $P\cup F$ has doubling dimension at most $D$ is proved in \Cref{sec:proof-soft}. For the other cases, let $\eta=\min\{1/10,\ \delta/(5\cdot3^z)\}$ and $\eta_c=\min\{1/4,\ \eta/(18z)\}$, and let $D'$ be the bound of \Cref{lem:compress} for this $\eta_c$. Since $6z\eta_c\le\eta/3$ and $\eta\le1/10$,
\[
(1+3\eta_c)^{2z}\le e^{6z\eta_c}\le e^{\eta/3}\le1+\eta/2 .
\]
For an instance whose points have doubling dimension at most $D'$, the algorithm of \Cref{sec:proof-soft} with accuracy $\eta$ computes a solution with at most $K$ centers. Its cost is at most $3^z-(3^z-1)\eps+4\cdot3^z\eta$ times the optimal cost of the instance. Since $3^z-(3^z-1)\eps\le3^z$ and $\eta\le1/4$, the product below exceeds $3^z-(3^z-1)\eps$ by at most $4\cdot3^z\eta+(1+4\eta)3^z\eta/2\le5\cdot3^z\eta\le\delta$:
\begin{equation}\label{eq:onesided-ratio}
\bigl(3^z-(3^z-1)\eps+4\cdot3^z\eta\bigr)(1+\eta/2)\le3^z-(3^z-1)\eps+\delta .
\end{equation}

\emph{Clients of low dimension.}
Suppose that the clients have doubling dimension at most $D$. Apply \Cref{lem:compress} with the set of points of the clients as the low-dimensional side, $Y=F$ and accuracy $\eta_c$. Moving every center of an optimal solution from its location $y$ to $\mu(y)$ gives a solution with locations in $\mu(F)$ and cost at most $(1+3\eta_c)^z\OPT$. Since $P\cup\mu(F)$ has doubling dimension at most $D'$, the algorithm of \Cref{sec:proof-soft} with accuracy $\eta$ computes a solution with at most $K$ centers at locations in $\mu(F)\subseteq F$ and cost at most $(3^z-(3^z-1)\eps+4\cdot3^z\eta)(1+3\eta_c)^z\OPT$. By~\eqref{eq:onesided-ratio}, this is at most $(3^z-(3^z-1)\eps+\delta)\OPT$.

\emph{Locations of low dimension.}
Suppose that $F$ has doubling dimension at most $D$. Apply \Cref{lem:compress} with $F$ as the low-dimensional side, $Y$ the set of points of the clients and accuracy $\eta_c$. Move every client $p$ to $\mu(p)$; several clients may now share a point. The points of the resulting instance form the set $F\cup\mu(Y)$, of doubling dimension at most $D'$. Keeping the assignment of an optimal solution shows that its optimal cost is at most $(1+3\eta_c)^z\OPT$. Compute a solution of it with the algorithm of \Cref{sec:proof-soft} with accuracy $\eta$, and move every client back to its original point while keeping its center. By~\eqref{eq:compress}, this multiplies the distance of every client to its center by at most $1+3\eta_c$. The cost is therefore at most $(3^z-(3^z-1)\eps+4\cdot3^z\eta)(1+3\eta_c)^{2z}\OPT$, which is at most $(3^z-(3^z-1)\eps+\delta)\OPT$ by~\eqref{eq:onesided-ratio}. \qed

\subsection{Euclidean Space}\label{sec:euclid}

This subsection proves \Cref{thm:soft} for centers anywhere in $\R^d$. \Cref{sec:certificates} shows that certificates make blocks movable when their representatives are optimal or nearly optimal centers. \Cref{sec:candidates} constructs candidate centers, and \Cref{sec:euclid-proof} assembles the algorithm. For a finite multiset $Q$ of points of $\R^d$, let $\Psi_Q(c)=\sum_{p\in Q}\|p-c\|^z$, where every point is counted with its multiplicity. A minimizer of $\Psi_Q$ is an \emph{optimal center} of $Q$. For $z>1$, the function $\Psi_Q$ is convex and differentiable, with gradient
\[
\nabla\Psi_Q(c)=-z\sum_{p\in Q}\|p-c\|^{z-2}(p-c),
\]
where terms with $p=c$ are zero.

\paragraph{The case of $k$-means.}
Take the representative $a_i$ of block $i$ to be an optimal center of $P_i$, so that $\Psi_{P_i}(a_i)=A_i$; for $k$-means, $a_i$ is the centroid of $P_i$. Match the clients to the slots $y_1,\dots,y_{w_i}$ by a uniformly random bijection and expand
\[
\|p-y\|^2=\|p-a_i\|^2+2\langle p-a_i,\,a_i-y\rangle+\|a_i-y\|^2 .
\]
The expectation of the middle term, summed over $p\in P_i$, is $\frac{2}{w_i}\langle\sum_p(p-a_i),\sum_y(a_i-y)\rangle=0$, where $y$ runs over the slots. Since some bijection costs at most the expectation, every block is $(1,1)$-movable, and $\beta=5$ instead of $3^2=9$.

\subsubsection{Certificates and Movability}\label{sec:certificates}

For $z\ge1$, we call $(u,g,v)\in\R^3$ a \emph{$z$-certificate} if
\begin{equation}\label{eq:certificate}
|1+t|^z\le u+gt+v|t|^z\qquad\text{for all }t\in\R .
\end{equation}
The name reflects its use: by \Cref{lem:cert-vector,thm:stability}, a $z$-certificate $(u,g,v)$ certifies that every block whose representative is an optimal center of the block is $(u,v)$-movable, since the term with coefficient $g$ becomes a cross term that vanishes at an optimal center. Let $\beta^*_z$ be the smallest value of $u+2^zv$ over all $z$-certificates. Then $\beta^*_1=3$, $\beta^*_2=5$ and $\beta^*_z<3^z$ for $z>1$; \Cref{app:constant} characterizes $\beta^*_z$ and compares it with $3^z$ (\Cref{fig:Bz}). For $z=1$, the triangle inequality gives the certificate $(1,0,1)$, and every block is $(1,1)$-movable for any representative; so let $z>1$.

\begin{lemma}[Certificates extend to vectors]\label{lem:cert-vector}
Let $z>1$, and let $(u,g,v)$ be a $z$-certificate. Then $u\ge1$, $v\ge1$, and for all $p,a,y\in\R^d$,
\[
\|p-y\|^z\le u\|p-a\|^z+g\|p-a\|^{z-2}\langle p-a,\,a-y\rangle+v\|a-y\|^z,
\]
where the middle term is zero if $p=a$.
\end{lemma}
\begin{proof}
\emph{Reduction to a cosine.}
We show $u,v\ge1$ and reduce the claim to an inequality for the cosine of one angle. The certificate inequality at $t=0$ gives $u\ge1$, and dividing it by $|t|^z$ and letting $t\to\infty$ gives $v\ge1$. If $p=a$, the claim follows from $v\ge1$. Otherwise, both sides are homogeneous of degree $z$ under scaling of the three points, so we may assume $\|p-a\|=1$. Let $t=\|a-y\|$, and let $\chi\in[-1,1]$ be the cosine of the angle between $p-a$ and $a-y$ (any value if $t=0$). Then $\|p-y\|^2=1+t^2+2t\chi$, and we must show that
\[
f(\chi):=(1+t^2+2t\chi)^{z/2}-gt\chi\ \le\ u+vt^z .
\]
At $\chi=1$ and $\chi=-1$, this is the certificate inequality at $t$ and at $-t$. If $z\ge2$, then $f$ is convex in $\chi$, so its maximum over $[-1,1]$ is attained at $\chi=\pm1$.

\emph{The case $1<z<2$.}
It remains to show $f(\chi)\le u+vt^z$ for $1<z<2$ and $t>0$, since $f(\chi)=1\le u$ for $t=0$. In this case, $f$ is concave, with derivative $f'(\chi)=t(z\,r(\chi)^{z-2}-g)$, where $r(\chi)=(1+t^2+2t\chi)^{1/2}$ increases from $|1-t|$ to $1+t$. If $g\le0$, then $f$ is nondecreasing and maximized at $\chi=1$. If $g>0$, let $r^*=(z/g)^{1/(2-z)}$. Unless $|1-t|<r^*<1+t$, the function $f$ is monotone on $[-1,1]$ and maximized at $\chi=\pm1$. Otherwise, it is maximized where $r(\chi)=r^*$, and using $g=z(r^*)^{z-2}$, the maximum equals
\[
(r^*)^z-\frac g2\bigl((r^*)^2-1-t^2\bigr)=\Bigl(1-\frac z2\Bigr)(r^*)^z+\frac g2+\frac g2\,t^2 .
\]
As a function of $t$, this maximum minus $vt^z$ has derivative $t^{z-1}(gt^{2-z}-zv)$, which changes sign at most once, from negative to positive. Over the interval $|1-r^*|\le t\le1+r^*$ of values of $t$ with $|1-t|\le r^*\le1+t$, it is therefore maximized at an end point. At an end point, the maximizing $\chi$ is $1$ or $-1$, which we have already handled; if the end point is $t=0$, then $r^*=1$, $g=z$, and the value is $1\le u$. Hence $f(\chi)\le u+vt^z$ in all cases.
\end{proof}

\begin{lemma}[Nearly optimal centers]\label{thm:stability}
Let $z>1$, let $(u,g,v)$ be a $z$-certificate, and let $\alpha=\min\{(z-1)/z,1/2\}$. There is a constant $C_z$ that depends only on $z$ with the following property. Let $Q$ be a multiset of $w$ points of $\R^d$, and let $a\in\R^d$ satisfy $\Psi_Q(a)\le(1+\eta)\min_c\Psi_Q(c)$ with $\eta\in[0,1]$. Then for all $y_1,\dots,y_w\in\R^d$, some bijection $\phi\colon Q\to\{1,\dots,w\}$ satisfies
\[
\sum_{p\in Q}\|p-y_{\phi(p)}\|^z\le\Bigl(u+\frac{z-1}{z}\omega\Bigr)\Psi_Q(a)+\Bigl(v+\frac\omega z\Bigr)\sum_{p\in Q}\|a-y_{\phi(p)}\|^z,\qquad\text{where }\omega=C_z|g|\eta^\alpha .
\]
In particular, a block whose representative is an optimal center of the block is $(u,v)$-movable.
\end{lemma}
\begin{proof}
Write $A_Q=\Psi_Q(a)$ and $G=\nabla\Psi_Q(a)$, and let $T_Q=\sum_{p\in Q}\|a-y_{\phi(p)}\|^z$, the cost of moving the mass of $Q$ to the slots, which is the same for every bijection $\phi$. If $A_Q=0$, all points of $Q$ coincide with $a$, every bijection costs $T_Q\le vT_Q$, and we are done; so let $A_Q>0$. Since $A_Q-\min_c\Psi_Q(c)\le\eta\min_c\Psi_Q(c)\le\eta A_Q$, we have
\begin{equation}\label{eq:descent}
\Psi_Q(a+h)\ge A_Q-\eta A_Q\qquad\text{for all }h\in\R^d .
\end{equation}

\emph{The cross term.}
We bound the expected cost of a uniformly random bijection $\phi$. Apply \Cref{lem:cert-vector} to $p$, $a$ and $y_{\phi(p)}$ for every $p\in Q$, sum over $p$, and take expectations. Since $\phi(p)$ is uniformly distributed on $\{1,\dots,w\}$ for every $p$, and $\sum_p\|p-a\|^{z-2}(p-a)=-G/z$,
\[
\mathbb E\Bigl[\sum_{p\in Q}\|p-y_{\phi(p)}\|^z\Bigr]\le uA_Q+vT_Q-\frac{g}{zw}\Bigl\langle G,\ \sum_{p\in Q}(a-y_{\phi(p)})\Bigr\rangle\le uA_Q+vT_Q+\frac{|g|}{z}\|G\|\Bigl(\frac{T_Q}w\Bigr)^{1/z}.
\]
Here the sum over $p$ is the same for every $\phi$, and the last step uses $\frac1w\sum_p\|a-y_{\phi(p)}\|\le(\frac1w\sum_p\|a-y_{\phi(p)}\|^z)^{1/z}$. It remains to show that
\begin{equation}\label{eq:gradient}
\|G\|\le z\,C_z\,w^{1/z}\eta^\alpha A_Q^{(z-1)/z}.
\end{equation}
Indeed, by~\eqref{eq:gradient} and the weighted inequality of arithmetic and geometric means, $x^{(z-1)/z}y^{1/z}\le\frac{z-1}zx+\frac1zy$ for $x,y\ge0$, the expected cost is then at most
\[
uA_Q+vT_Q+\omega A_Q^{(z-1)/z}T_Q^{1/z}\le uA_Q+vT_Q+\omega\bigl((z-1)A_Q+T_Q\bigr)/z,
\]
and some bijection costs at most the expectation. We may assume $G\ne0$.

\emph{The gradient for $1<z\le2$.}
We derive~\eqref{eq:gradient} from a bound on the first-order Taylor remainder of $x\mapsto\|x\|^z$. For $1<z<2$, this function has no bounded Hessian near $0$. For $1<z\le2$, however, the remainder is at most a constant times $\|h\|^z$. This bound, \eqref{eq:remainder} below, replaces the Hessian bound that we use for $z\ge2$. Let $\theta_z=\max_{0\le r\le1/2}\bigl((1-r)^z-r^z+zr^{z-1}\bigr)\le3$. We claim that for all $e,h\in\R^d$,
\begin{equation}\label{eq:remainder}
\|e+h\|^z\le\|e\|^z+z\|e\|^{z-2}\langle e,h\rangle+\theta_z\|h\|^z .
\end{equation}
For $h=0$, both sides are equal. Otherwise, by homogeneity, let $\|h\|=1$, $r=\|e\|$ and $\langle e,h\rangle=r\chi$ with $\chi\in[-1,1]$. The difference of the two sides of~\eqref{eq:remainder} without the last term is $(r^2+2r\chi+1)^{z/2}-r^z-zr^{z-1}\chi$, which is concave in $\chi$ with derivative $zr\bigl((r^2+2r\chi+1)^{(z-2)/2}-r^{z-2}\bigr)$. If $r\le1/2$, the derivative is nonpositive, so the maximum over $\chi$ is at $\chi=-1$, where the expression is $(1-r)^z-r^z+zr^{z-1}\le\theta_z$. If $r>1/2$, the maximum is at $\chi=-1/(2r)$, where the expression is $\frac z2r^{z-2}\le z2^{1-z}$, the value of the expression in the definition of $\theta_z$ at $r=1/2$. This proves~\eqref{eq:remainder}. Applying it to $e=p-a$ and $-h$ and summing over $p$ gives
\[
\Psi_Q(a+h)\le A_Q+\langle G,h\rangle+\theta_zw\|h\|^z .
\]
For $h$ a nonpositive multiple of $G$, we have $\langle G,h\rangle=-\|G\|\,\|h\|$, so~\eqref{eq:descent} gives $\eta A_Q\ge\|G\|\,\|h\|-\theta_zw\|h\|^z$. The maximum of the right-hand side over $\|h\|\ge0$ is $\frac{z-1}{z}\|G\|^{z/(z-1)}(z\theta_zw)^{-1/(z-1)}$, which gives~\eqref{eq:gradient} with $C_z=\theta_z^{1/z}(z-1)^{-(z-1)/z}$.

\emph{The gradient for $z\ge2$.}
We derive~\eqref{eq:gradient} from a bound on the Hessian of $\Psi_Q$ near $a$. Let $r=(A_Q/w)^{1/z}$. For $\|h\|\le r$, Minkowski's inequality gives $\Psi_Q(a+h)^{1/z}\le A_Q^{1/z}+w^{1/z}\|h\|\le2A_Q^{1/z}$. The Hessian of $x\mapsto\|x\|^z$ has largest eigenvalue $z(z-1)\|x\|^{z-2}$. By the triangle inequality for the operator norm and H\"older's inequality with exponents $z/(z-2)$ and $z/2$ (for $z=2$, the bound is immediate), the Hessian of $\Psi_Q$ at a point $a+h$ with $\|h\|\le r$ therefore has norm at most
\[
z(z-1)\sum_{p\in Q}\|p-a-h\|^{z-2}\le z(z-1)w^{2/z}\Psi_Q(a+h)^{(z-2)/z}\le z(z-1)2^{z-2}w^{2/z}A_Q^{(z-2)/z}=:\Xi .
\]
By the triangle inequality and H\"older's inequality with exponents $z/(z-1)$ and $z$, $\|G\|\le z\sum_p\|p-a\|^{z-1}\le zw^{1/z}A_Q^{(z-1)/z}=zA_Q/r$, so the point $h=-G/\Xi$ satisfies $\|h\|\le r/((z-1)2^{z-2})\le r$. Taylor's formula and~\eqref{eq:descent} give
\[
A_Q-\eta A_Q\le\Psi_Q(a+h)\le A_Q-\frac{\|G\|^2}\Xi+\frac{\|G\|^2}{2\Xi},
\]
that is, $\|G\|^2\le2\Xi\eta A_Q$. This is~\eqref{eq:gradient} with $C_z=(2^{z-1}(z-1)/z)^{1/2}$.
\end{proof}

By \Cref{thm:stability} with $\eta=0$, the best movability constants for blocks with optimal representatives come from a certificate that minimizes $u+2^zv$, that is, from the constant $\beta^*_z$ defined at the beginning of this subsection. The following proposition characterizes $\beta^*_z$ and shows that no pair $(u,v)$ with $u+2^zv<\beta^*_z$ works for all such blocks.

\begin{proposition}[The constant $\beta^*_z$]\label{prop:constant}
Let $z\ge1$. (a)~$\beta^*_z=\sup\mathbb E|1+\xi|^z$, where the supremum is over all real random variables $\xi$ with finite support, $\mathbb E[\xi]=0$ and $\mathbb E|\xi|^z=2^z$. (b)~$\beta^*_z<3^z$ for every $z>1$; moreover, $\beta^*_1=3$, $\beta^*_2=5$, $\beta^*_{3/2}\approx4.1397$, $\beta^*_3\approx16.8308$ and $\beta^*_4\approx53.4056$. (c)~For all $u\ge1$ and $v\ge0$ with $u+2^zv<\beta^*_z$, there is a finite set $Q\subset\R$ such that the block $Q$ with an optimal center of $Q$ as its representative is not $(u,v)$-movable.
\end{proposition}

The proposition is proved in \Cref{app:constant}.

\subsubsection{Candidate Centers}\label{sec:candidates}

The framework requires a finite set of locations with bounded doubling dimension, but in Euclidean space the optimal center of a block can lie anywhere in $\R^d$. We therefore construct a finite set of candidate centers that contains, for every block, a center that is nearly as good as its optimal center. Since the dimension $d$ may be large, we cannot place grids around the clients, as the approximate centroid sets of Matou{\v{s}}ek~\cite{Mat00} for $k$-means do; the size of such sets grows exponentially with $d$. Instead, we run projected stochastic subgradient descent~\cite{NJLS09,Bub15} on the cost $\Psi_Q$ of a set $Q$ of clients: every step uses the subgradient of the cost of a single random client, and the candidate is the average of the iterates. Enumerating all short sequences of clients instead of sampling them makes the construction deterministic, and the resulting set has size polynomial in $n$ and independent of $d$. The clients of $Q$ can be far from its optimal center, so the subgradients of their individual costs can be large. We therefore use importance sampling. For a point $p_0$ and a scale $s$, every client $x$ is weighted by $(s/(\|x-p_0\|+s))^z$, which bounds all subgradients on the ball of radius $s$ around $p_0$ by $zs^{z-1}$. To compensate, a client $x$ is sampled with probability proportional to $(\|x-p_0\|+s)^z$, so that the expected weighted cost is proportional to $\Psi_Q$.

\begin{lemma}[Candidate centers]\label{lem:candidates}
Let $P$ be a multiset of $n\ge2$ points of $\R^d$, $z\ge1$ and $\eta\in(0,1]$, and let $N=\lceil(z4^z/\eta)^2\rceil$. One can compute a set $\Gamma\subset\R^d$ that contains the points of $P$, with $|\Gamma|\le n^{N+3}\lceil\log_2(4n)\rceil$, using $O(Nd)$ arithmetic operations, square roots and real powers per point of $\Gamma$, such that
\[
\min_{c\in\Gamma}\Psi_Q(c)\le(1+\eta)\min_{c\in\R^d}\Psi_Q(c)\qquad\text{for every nonempty multiset }Q\subseteq P.
\]
\end{lemma}
\begin{proof}
\emph{Construction.} $\Gamma$ contains $P$ and the following points. Consider every $p_0\in P$, every distance between two distinct points of $P$, and every $s$ obtained from this distance by halving it at most $\lceil\log_2(2n)\rceil$ times. For $x\in P$, let $\psi_x(c)=s^z\|x-c\|^z/(\|x-p_0\|+s)^z$, a convex function whose subgradients at points $c\in B(p_0,s)$ have norm at most $zs^z\|x-c\|^{z-1}/(\|x-p_0\|+s)^z\le zs^{z-1}$. For every sequence $(x_1,\dots,x_N)\in P^N$, let $c_1=p_0$, and for $\nu=1,\dots,N$, let $c_{\nu+1}$ be the projection of $c_\nu-\tau G_\nu$ onto $B(p_0,s)$. Here $G_\nu$ is a subgradient of $\psi_{x_\nu}$ at $c_\nu$ (zero if $c_\nu=x_\nu$), and $\tau=s^{2-z}/(z\sqrt N)$ is the step size. Add the average $\bar c=\frac1N\sum_{\nu=1}^Nc_\nu$ to $\Gamma$.

\emph{Analysis.} Fix a nonempty multiset $Q\subseteq P$. If all points of $Q$ coincide, this point lies in $\Gamma$. Otherwise, we first choose $p_0$ and $s$. Let $w=|Q|\ge2$, let $c^*$ be an optimal center of $Q$, and let $r=(\Psi_Q(c^*)/w)^{1/z}$, where $\Psi_Q(c^*)>0$. Let $p_0\in Q$ be a point closest to $c^*$, so $\|p_0-c^*\|\le r$. The diameter of $Q$ is a distance between two distinct points of $P$. It is at least $r$, since placing the center at a point of $Q$ shows that $\Psi_Q(c^*)$ is at most $w-1$ times its $z$-th power. It is at most $2w^{1/z}r\le2nr$, since the two points at this distance alone cost at least $2^{1-z}$ times its $z$-th power. The scales obtained from the diameter by halving it at most $\lceil\log_2(2n)\rceil$ times therefore include some $s$ with $r\le s<2r$, and then $c^*\in B(p_0,s)$.

\emph{The sampling distribution.}
Consider the probability distribution on $Q$ that picks $x$ with probability proportional to its weight $(\|x-p_0\|+s)^z$. By Minkowski's inequality and $\|c^*-p_0\|+s\le3r$, the total weight satisfies
\[
\Bigl(\sum_{x\in Q}(\|x-p_0\|+s)^z\Bigr)^{1/z}\le\Psi_Q(c^*)^{1/z}+w^{1/z}\cdot3r=4\Psi_Q(c^*)^{1/z} .
\]
For a random $x$, the expected value of $\psi_x(c)$ is $s^z\Psi_Q(c)$ divided by the total weight; we denote it by $\bar\psi(c)$.

\emph{Projected subgradient steps.}
We show that independent random points $x_1,\dots,x_N$ with this distribution yield a good point $\bar c$ in expectation. Since $c_\nu$ does not depend on $x_\nu$, the vector $G_\nu$ has conditional expectation in the subdifferential of $\bar\psi$ at $c_\nu$. Projections onto $B(p_0,s)\ni c^*$ do not increase distances to $c^*$, so
\[
\|c_{\nu+1}-c^*\|^2\le\|c_\nu-c^*\|^2-2\tau\langle G_\nu,c_\nu-c^*\rangle+\tau^2(zs^{z-1})^2=\|c_\nu-c^*\|^2-2\tau\langle G_\nu,c_\nu-c^*\rangle+\frac{s^2}N ,
\]
and by convexity $\mathbb E\langle G_\nu,c_\nu-c^*\rangle\ge\mathbb E[\bar\psi(c_\nu)]-\bar\psi(c^*)$. Summing over $\nu$ and using $\|c_1-c^*\|\le s$ and the convexity of $\bar\psi$ gives
\[
\mathbb E[\bar\psi(\bar c)]-\bar\psi(c^*)\le\frac1N\sum_{\nu=1}^N\bigl(\mathbb E[\bar\psi(c_\nu)]-\bar\psi(c^*)\bigr)\le\frac{s^2+s^2}{2\tau N}=\frac{zs^z}{\sqrt N}.
\]
We multiply by the total weight divided by $s^z$, which is at most $4^z\Psi_Q(c^*)/s^z$, and obtain $\mathbb E[\Psi_Q(\bar c)]-\Psi_Q(c^*)\le z4^z\Psi_Q(c^*)/\sqrt N\le\eta\Psi_Q(c^*)$. Some sequence in $Q^N\subseteq P^N$ therefore yields a point of $\Gamma$ with $\Psi_Q(\bar c)\le(1+\eta)\Psi_Q(c^*)$. The bound on $|\Gamma|$ counts the choices of $p_0$, the distance, the number of halvings and the sequence.
\end{proof}

\subsubsection{Proof of \texorpdfstring{\Cref{thm:soft}}{Theorem 1.1} in \texorpdfstring{$\R^d$}{Rd}}\label{sec:euclid-proof}

We now combine the candidate centers with the framework to prove \Cref{thm:soft} in $\R^d$. We may assume $\delta\le1$ and $\OPT>0$. Indeed, $\OPT=0$ if and only if $\sum_x\lceil n_x/U\rceil\le k$, where $n_x$ is the number of clients at point $x$, and then $\lceil n_x/U\rceil$ centers at every such point $x$ cost $0$. If $z=1$, let $(u,g,v)=(1,0,1)$, which corresponds to the triangle inequality: every block is $(1,1)$-movable for any representative, and $u+2v=3=\beta^*_1$ (\Cref{prop:constant}(b)). If $z>1$, let $(u,g,v)$ be a $z$-certificate with $u+2^zv\le\beta^*_z+\delta/4$; it depends only on $z$ and $\delta$ and can be computed by \Cref{lem:compute-certificate} with accuracy $\delta/4$. Choose $\eta$ with
\[
0<\eta\le\min\Bigl\{\frac1{10},\frac{\delta}{32\cdot3^z}\Bigr\}\qquad\text{and}\qquad\frac{z-1+2^z}{z}\,\omega\le\frac\delta4,\quad\text{where }\omega=C_z|g|\eta^\alpha,
\]
with $C_z$ and $\alpha$ from \Cref{thm:stability} (for $z=1$, let $\eta=\min\{1/10,\delta/(32\cdot3^z)\}$ and $\omega=0$). Let $\eta_c=\min\{1/4,\eta/(18z)\}$, so that $(1+3\eta_c)^z\le e^{\eta/6}\le1+\eta/3$.

\emph{Locations.}
Compute the set $\Gamma$ of \Cref{lem:candidates} with accuracy $\eta/3$. Apply \Cref{lem:compress} with the set of points of the clients as the low-dimensional side, $Y=\Gamma$ and accuracy $\eta_c$, and let $\Gamma'=\mu(\Gamma)$. All clients lie in $\Gamma'$, since $\Gamma$ contains them and $\mu$ maps them to themselves. The set $\Gamma'$ therefore has doubling dimension at most the bound $D'$ of \Cref{lem:compress}. By~\eqref{eq:compress}, for every nonempty $Q\subseteq P$,
\begin{equation}\label{eq:locations}
\min_{c\in\Gamma'}\Psi_Q(c)\le(1+\eta/3)\min_{c\in\Gamma}\Psi_Q(c)\le(1+\eta/3)^2\min_{c\in\R^d}\Psi_Q(c)\le(1+\eta)\min_{c\in\R^d}\Psi_Q(c).
\end{equation}

\emph{Reference solution and blocks.}
Take an optimal solution, and replace the center of each of its clusters by a point of $\Gamma'$ of minimum cost for that cluster. By~\eqref{eq:locations}, this gives a soft solution $S^*$ with at most $k$ centers at locations in $\Gamma'$ and $\cost(S^*)\le(1+\eta)\OPT$. By \Cref{thm:frs}, compute an uncapacitated solution with at most $k$ centers in $\Gamma'$ whose cost is at most $(1+\eta)$ times the optimal uncapacitated cost over $\Gamma'$, which is at most $\cost(S^*)$. Split it into blocks as in \Cref{sec:approach-blocks}, and replace the representative of every block $P_i$ by a point $a_i\in\Gamma'$ that minimizes $\Psi_{P_i}$ over $\Gamma'$. This does not increase $A$, so $A\le(1+\eta)\cost(S^*)$, and by~\eqref{eq:locations}, $\Psi_{P_i}(a_i)\le(1+\eta)\min_c\Psi_{P_i}(c)$. By \Cref{thm:stability} (for $z=1$, by the triangle inequality), every block is $(u',v')$-movable for $u'=u+(z-1)\omega/z$ and $v'=v+\omega/z$, and
\[
\beta':=u'+2^zv'=u+2^zv+\frac{z-1+2^z}{z}\,\omega\le\beta^*_z+\frac\delta2\le2\cdot3^z .
\]

\emph{The framework.}
\Cref{thm:framework} with accuracy $\eta$, doubling dimension $D'$, the pair $(u',v')$ and the reference solution $S^*$ computes a soft solution $S'$ with at most $K$ centers at locations in $\Gamma'$ and
\begin{align*}
\cost(S')&\le\bigl(\beta'-(\beta'-1)\eps+4\beta'\eta\bigr)\cost(S^*)\\
&\le\Bigl(\beta^*_z-(\beta^*_z-1)\eps+\frac\delta2+\frac\delta4\Bigr)(1+\eta)\OPT\ \le\ \Bigl(\beta^*_z-(\beta^*_z-1)\eps+\frac{7\delta}8\Bigr)\OPT.
\end{align*}
Here we use $\beta'-(\beta'-1)\eps\le\beta^*_z-(\beta^*_z-1)\eps+(\beta'-\beta^*_z)$, $4\beta'\eta\le8\cdot3^z\eta\le\delta/4$, and $\eta(\beta^*_z+1)\le2\cdot3^z\eta\le\delta/16$.

\emph{Distinct centers.}
Finally, we make the centers distinct at a small cost. Some center of every solution with at most $k$ centers serves clients at two distinct points $p\ne p'$, since otherwise moving every center to the point of its clients would give cost $0$. These clients cost at least $2^{1-z}\|p-p'\|^z$, so $\OPT\ge2^{1-z}r_{\min}^z$, where $r_{\min}$ is the minimum distance between two distinct points of clients. Let $r_0>0$ be such that $(\cost(S')^{1/z}+n^{1/z}r_0)^z=\cost(S')+\frac\delta{16}2^{1-z}r_{\min}^z$. Process the at most $K$ centers one by one. Move the current center from its location $c$ to $c+(r,0,\dots,0)$, where $r$ is one of the $K$ values $r_0/(K+1),2r_0/(K+1),\dots,Kr_0/(K+1)$ for which this point differs from all centers moved before. Each of these fewer than $K$ centers excludes at most one value. Keep the assignment. The centers become distinct, and every client's distance to its center grows by less than $r_0$. By Minkowski's inequality, the cost is at most
\[
\bigl(\cost(S')^{1/z}+n^{1/z}r_0\bigr)^z\le\cost(S')+\frac\delta{16}\OPT<\bigl(\beta^*_z-(\beta^*_z-1)\eps+\delta\bigr)\OPT.
\]

\emph{Running time.}
The set $\Gamma$ has $n^{O(z^216^z/\eta^2)}$ points, each computed with $O(d)$ operations per step. Distances are computed in time $O(d)$. All other steps run in time polynomial in $n$, $|\Gamma|$ and $\log\Delta(P\cup\Gamma')$ for fixed $D'$, $z$, $\eps$ and $\delta$, by \Cref{lem:compress,thm:frs,thm:framework}. The dimension $d$ enters only through the arithmetic on coordinates. Since $D'$ is exponential in $D$, the degree of the polynomial is doubly exponential in $D$ for fixed $z$, $\eps$ and $\delta$. \qed

\section{Non-Uniform Capacities}\label{sec:nonuniform}

This section proves \Cref{thm:nonuniform,cor:bicriteria}. As explained in  \Cref{sec:approach-extensions}, the block construction of \Cref{sec:approach-blocks} fails for non-uniform capacities, so we take the blocks from a near-optimal capacitated solution with more centers. \Cref{sec:augment} computes such a solution by greedy additions, \Cref{sec:diagonal} adapts the transport plan to hard capacities, \Cref{sec:near-k} applies the framework to the resulting blocks, and \Cref{sec:proof-nonuniform} combines these steps.

The algorithms of this section and of \Cref{sec:greedy} rely on the following supermodularity of the assignment cost $\cost(S)$ defined in \Cref{sec:prelim}.

\begin{lemma}[Supermodularity]\label{lem:supermodular}
For finite families $S'\subseteq S$ of centers at locations in $F$ with $\cost(S')<\infty$,
\[
\sum_{x\in S\setminus S'}\bigl(\cost(S\setminus\{x\})-\cost(S)\bigr)\ \le\ \cost(S')-\cost(S).
\]
\end{lemma}
\begin{proof}
\emph{Assignments as matchings.}
Give every center $U$ \emph{slots}. An assignment in which every center serves at most $U$ clients is a matching between the clients and the slots that covers all clients. A client $p$ matched to a slot of a center at location $y$ costs $\dist(p,y)^z$.

\emph{The union of two matchings.}
Let $S'\subseteq S''\subseteq S$ and $x\in S''\setminus S'$. We first show that
\begin{equation}\label{eq:exchange}
\cost(S'')+\cost(S\setminus\{x\})\ \le\ \cost(S)+\cost(S''\setminus\{x\}),
\end{equation}
where all four costs are finite, since the families contain $S'$. Let $\sigma_1$ and $\sigma_2$ be optimal assignments for $S$ and for $S''\setminus\{x\}$. Their union may contain an edge twice. In it, every client has degree $2$, every slot has degree at most $2$, and every vertex of degree $2$ meets one edge of each matching. Every component with an edge is therefore a cycle or a path between two slots, and its edges alternate between $\sigma_1$ and $\sigma_2$. A path between two slots has an even number of edges, so its two end edges lie in different matchings. The slots of $x$ and of the centers of $S\setminus S''$ meet only edges of $\sigma_1$, since $\sigma_2$ uses only slots of $S''\setminus\{x\}$. Such a slot is therefore isolated or an end of a path whose other end edge lies in $\sigma_2$, and every component contains at most one of them.

\emph{Splitting the union.}
In a component that contains a slot of a center of $S\setminus S''$, we give the edges of $\sigma_1$ to $S\setminus\{x\}$ and those of $\sigma_2$ to $S''$. In every other component, we give the edges of $\sigma_1$ to $S''$ and those of $\sigma_2$ to $S\setminus\{x\}$. Every client and every slot then meets at most one edge given to each family, and every edge given to a family ends at a slot of that family. Every client meets an edge given to each family. This yields assignments for $S''$ and for $S\setminus\{x\}$ of total cost $\cost(S)+\cost(S''\setminus\{x\})$, which proves~\eqref{eq:exchange}.

\emph{Deleting one center at a time.}
Now delete the centers of $S\setminus S'$ from $S$ one at a time. The increases of the cost add up to $\cost(S')-\cost(S)$. By~\eqref{eq:exchange}, each of them is at least the increase caused by deleting the same center from $S$.
\end{proof}

Every location $y\in F$ has a capacity $U_y$, a positive integer, and a center at $y$ serves at most $U_y$ clients. Soft and hard capacities are as before. We allow several locations at the same point of $V$, which \Cref{sec:proof-nonuniform} uses for one-sided low dimension. Doubling dimensions refer to the sets of points. As in \Cref{sec:hard}, the cost $\cost(S)$ of a finite family $S$ of centers at locations in $F$ is the minimum cost of an assignment in which a center at $y$ serves at most $U_y$ clients. If there is no such assignment, $\cost(S)=\infty$. With $U_y$ slots for a center at $y$, the proof of \Cref{lem:supermodular} applies verbatim. In particular, let $S\subseteq S'$ be families with $\cost(S)<\infty$, and let $y$ be a center. By~\eqref{eq:exchange}, adding $y$ to $S'$ decreases the cost by at most as much as adding $y$ to $S$.

\subsection{Greedy Additions and a Sparse Start}\label{sec:augment}

\begin{lemma}[Greedy additions]\label{lem:augment}
Let $\OPT$ be the optimal soft or the optimal hard cost, let $S_0$ be a family of centers at locations in $F$ with $\cost(S_0)<\infty$, without repeated locations for hard capacities, and let $K'\ge|S_0|$. Starting from $S=S_0$, as long as $|S|<K'$ and, for hard capacities, $S\ne F$, add to $S$ a center at a location $y$, with $y\notin S$ for hard capacities, that minimizes $\cost(S\cup\{y\})$. This takes $K'-|S_0|$ rounds of $|F|$ minimum-cost flow computations, and at the end, $\cost(S)-\OPT\le(1-1/k)^{K'-|S_0|}(\cost(S_0)-\OPT)$.
\end{lemma}
\begin{proof}
Adding a center never increases the cost. If the process stops with $S=F$, then $\cost(S)\le\OPT$ and $\cost(S)\le\cost(S_0)$, which implies the bound. It therefore suffices to show that a step from a family $S$ with $\cost(S)>\OPT$ decreases $\cost(S)-\OPT$ by a factor of at least $1-1/k$. Let $S^*$ be an optimal solution. Add to $S$, one at a time, the centers of $S^*$ for soft capacities, or centers at the locations of $S^*$ outside $S$ for hard capacities. These are at most $k$ centers, and the resulting family costs at most $\cost(S^*)=\OPT$, so the decreases of the cost add up to at least $\cost(S)-\OPT$. By~\eqref{eq:exchange}, adding one of these centers, at a location $y$, decreases the cost by at most $\cost(S)-\cost(S\cup\{y\})$. One of these locations $y$ therefore satisfies
\[
\cost(S)-\cost(S\cup\{y\})\ \ge\ \frac{\cost(S)-\OPT}k ,
\]
and the greedy step decreases the cost at least as much.
\end{proof}

The next lemma provides a start with at most $2k$ centers.

\begin{lemma}[Sparse start]\label{lem:sparse}
Given at most $k$ points, let $A_0$ be the cost of assigning every client to a closest one. One can compute in polynomial time a family of at most $2k$ centers at locations in $F$, without repeated locations for hard capacities, whose cost is at most $(2A_0^{1/z}+\OPT^{1/z})^z$.
\end{lemma}
\begin{proof}
\emph{The linear program.}
Consider hard capacities first. Assign every client to a closest one of the given points, and let $n_x$ be the number of clients assigned to the given point $x$. The following linear program is over fractional transport plans $\pi$ from the given points to the locations, where $\ell_y:=\sum_x\pi_{xy}$ is the load of $y$:
\begin{align*}
&\min\ \sum_{x,y}\pi_{xy}\dist(x,y)^z\\
&\text{subject to}\quad\sum_{y\in F}\pi_{xy}=n_x\ \ \text{for all }x,\qquad \ell_y\le U_y\ \ (y\in F),\qquad\sum_{y\in F}\frac{\ell_y}{U_y}\le k,\qquad \pi\ge0.
\end{align*}
It has the feasible solution in which $\pi_{xy}$ is the number of clients assigned to $x$ that an optimal hard solution serves at $y$. In this solution, every used location contributes at most $1$ to the last sum. By~\eqref{eq:minkowski} applied to the triples $(x,y,p)$, its cost is at most $(A_0^{1/z}+\OPT^{1/z})^z$.

\emph{A sparse vertex.}
Let $\pi$ be an optimal vertex. It satisfies with equality linearly independent constraints, as many as there are variables. At most as many of them as there are zero variables have the form $\pi_{xy}\ge0$. The number of positive variables is therefore at most the number of the other tight constraints: at most $k$ equations, the constraints $\ell_y\le U_y$ of the \emph{full} locations, with $\ell_y=U_y$, and the last constraint. Every location with $\ell_y>0$ has a positive variable. If all these locations are full, the last constraint allows at most $k$ of them. Otherwise, it allows at most $k-1$ full locations, so at most $k+(k-1)+1=2k$ locations have $\ell_y>0$.

\emph{Rounding.}
Open the locations with $\ell_y>0$. Sending the fraction $\pi_{xy}/n_x$ of every client assigned to $x$ to $y$ respects the capacities. By Minkowski's inequality with these fractions as weights, applied to the triples $(p,y,x)$, it costs at most $(A_0^{1/z}+(A_0^{1/z}+\OPT^{1/z}))^z$. A minimum-cost flow finds an integral assignment that is no more expensive.

\emph{Soft capacities.}
For soft capacities, apply this to the instance in which every location is replaced by $k$ locations at the same point with the same capacity. Its hard solutions with at most $k$ centers are the soft solutions of the original instance with at most $k$ centers, since such a solution uses every location at most $k$ times.
\end{proof}

\begin{corollary}[Few more centers]\label{cor:bicriteria}
Let $z\ge1$, $\rho\ge1$ and $\delta\in(0,1]$. Given an instance with non-uniform soft or hard capacities in a finite metric space and an uncapacitated solution with at most $k$ centers whose cost is at most $\rho$ times the minimum, one can compute in polynomial time at most $2k+\lceil k\ln(3^z\rho/\delta)\rceil$ centers at locations in $F$, at distinct locations for hard capacities, and an assignment in which every center serves at most the capacity of its location, with cost at most $(1+\delta)\cdot\OPT$.
\end{corollary}

\begin{proof}[Proof of \Cref{cor:bicriteria}]
The optimal uncapacitated cost is at most $\OPT$, so the given centers have $A_0\le\rho\OPT$. By \Cref{lem:sparse} with the given centers, we obtain at most $2k$ centers of cost at most $(1+2\rho^{1/z})^z\OPT\le3^z\rho\OPT$. \Cref{lem:augment} with $K'=2k+\lceil k\ln(3^z\rho/\delta)\rceil$ reduces the cost to at most
\[
\OPT+e^{-(K'-2k)/k}3^z\rho\OPT\le(1+\delta)\OPT .\qedhere
\]
\end{proof}

For fixed $z$, local search gives a constant $\rho$~\cite{AGKMMP04,GT08}, so $O(k\log(2/\delta))$ centers suffice.

\subsection{Hard Capacities for \texorpdfstring{$k$}{k}-Median}\label{sec:diagonal}

The framework of \Cref{sec:approach} gives a rescued block a new center of its own, which soft capacities allow. With hard capacities, a rescued block at location $y$ shares the center at $y$ with the clients that the reference solution sends there, so the transport plan must keep room for both. The following lemma ensures this by maximizing the diagonal entries. In a transport plan between two sets $S,S^*\subseteq F$ of locations, the rows are indexed by $S$ and the columns by $S^*$. A location $y\in S\cap S^*$ has a row, a column and a \emph{diagonal} entry $\pi_{yy}$. The entry of the pair $yy'$ costs $\dist(y,y')^z$.

\begin{lemma}[Diagonal]\label{lem:diagonal}
Let $z=1$, and let $\pi^0$ be an integral transport plan between sets $S,S^*\subseteq F$ with row sums $w_y$ and column sums $\ell_y$. There is an integral transport plan $\pi$ with the same row and column sums and no larger cost whose support is a forest, such that $\pi_{yy}=\min\{w_y,\ell_y\}$ for every $y\in S\cap S^*$.
\end{lemma}
\begin{proof}
\emph{Filling the diagonal.}
Start with $\pi=\pi^0$. If $\pi_{yy}<\min\{w_y,\ell_y\}$ for some $y\in S\cap S^*$, then row $y$ has a positive entry $\pi_{yy_2}$ with $y_2\ne y$, and column $y$ has a positive entry $\pi_{y_1y}$ with $y_1\ne y$. Decrease $\pi_{yy_2}$ and $\pi_{y_1y}$ by the smaller of these two entries, and increase $\pi_{yy}$ and $\pi_{y_1y_2}$ by the same amount. This keeps all row and column sums. Since $\dist(y_1,y_2)\le\dist(y_1,y)+\dist(y,y_2)$, it does not increase the cost. No diagonal entry decreases, and their sum increases, so the procedure stops after at most $\sum_yw_y$ steps.

\emph{A forest.}
Finally, apply \Cref{fact:cycle}. A positive diagonal entry $\pi_{yy}=w_y$ is the only positive entry of its row, and $\pi_{yy}=\ell_y$ is the only one of its column. Such an entry lies on no cycle, so it does not change.
\end{proof}

For $z>1$, the diagonal condition can make the plan arbitrarily more expensive. Let $S=\{0,\dots,n-1\}$ and $S^*=\{1,\dots,n\}$ be points of a line, with one client at every point $x\in S$, served at $x$ by $S$ and at $x+1$ by $S^*$. This plan costs $n$, but the only plan with $\pi_{yy}=1$ for $y=1,\dots,n-1$ sends the client at $0$ to $n$, at cost $n^z$.

\subsection{Solutions with Few Extra Centers}\label{sec:near-k}

This subsection applies the framework of \Cref{sec:approach} to the blocks of a given family $S_1$ of centers, using the diagonal lemma for hard capacities; \Cref{sec:proof-nonuniform} obtains $S_1$ by greedy additions and completes the proof of \Cref{thm:nonuniform}.

In this subsection, $P\cup F$ has doubling dimension at most $D$. We are given a family $S_1$ of centers at locations in $F$, without repeated locations for hard capacities, together with an assignment of cost $\cost(S_1)$; it plays the role of the uncapacitated solution $S_1$ of \Cref{sec:approach-blocks}. The clients served by a center of $S_1$ that serves clients form a block, whose representative is the location of the center; let $m\le|S_1|$ be the number of blocks. A block with representative $y$ has at most $U_y$ clients, so a new center at $y$ can serve it. With hard capacities, every location is the representative of at most one block. The total internal cost is $A=\cost(S_1)$, and by the proof of \Cref{lem:triangle-movable}, which uses only the triangle inequality, every block is $(u,v)$-movable with $(u,v)=(3^{z-1},(3/2)^{z-1})$.

\paragraph{Configurations.}
We use the network $\Net$ of \Cref{def:network} on the set $X$ of points of $P\cup F$ and change only the treatment of the points. For a point $x$, let $F_x$ be the set of locations at $x$, and $I_x$ the set of blocks with representative at $x$. A \emph{configuration} at $x$ with $k_x$ centers consists of a set $R_x\subseteq I_x$ of rescued blocks and of $k_x$ centers at locations of $F_x$. With soft capacities, one center is at the representative of every rescued block, and the other $k_x-|R_x|$ centers are at a location of $F_x$ of maximum capacity $U^{\max}_x$. Such a configuration \emph{admits} an inflow $\ell$ if $\ell\le(k_x-|R_x|)U^{\max}_x$. With hard capacities, the centers are at the locations of a set $F'_x\subseteq F_x$ that contains the representative of every rescued block. Such a configuration admits $\ell$ if $\ell\le\sum_{y\in F'_x}U_y-\sum_{i\in R_x}w_i$. If $F_x=\emptyset$, only $\ell=0$ is admitted. The outflow of the configuration is $s_x=\sum_{i\in I_x\setminus R_x}w_i$.

A \emph{network solution} consists of a configuration at every point and an integral flow $f\ge0$ on the arcs of $\Net$. In it, every portal vertex has equal inflow and outflow, every vertex $\src x$ has outflow $s_x$, and the inflow of every vertex $\snk x$ is admitted by the configuration at $x$. It has $\sum_xk_x$ centers and cost
\[
\Phi=\sum_x\sum_{i\in I_x\setminus R_x}(u-1)A_i+v\cost(f).
\]
For uniform soft capacities and $k_x=|R_x|+\lceil\ell/U\rceil$, where $\ell$ is the inflow of $\snk x$, these are the network solutions of \Cref{def:network-solution}.

\begin{lemma}[Network solutions with configurations]\label{lem:configurations}
For every $\gamma\in(0,1]$, a network solution of minimum cost with at most $k+\lfloor\gamma k\rfloor$ centers, if one exists, can be computed in time $(2n+1)^{O(M)}\cdot\mathrm{poly}(|X|,|F|,H)$. Given a network solution with at most $k+\lfloor\gamma k\rfloor$ centers, one can compute in polynomial time a solution with at most $k+\lfloor\gamma k\rfloor$ centers, without repeated locations for hard capacities, whose cost is at most $A+\Phi$.
\end{lemma}
\begin{proof}
\emph{Local tables.}
For a point $x$ and integers $0\le s,\ell\le n$ and $0\le k_x\le k+\lfloor\gamma k\rfloor$, let $\loc_x(s,\ell,k_x)$ be the minimum of $\sum_{i\in I_x\setminus R_x}(u-1)A_i$ over the configurations at $x$ with $k_x$ centers and outflow $s$ that admit $\ell$. For soft capacities, a knapsack over the blocks of $I_x$ computes, as in \Cref{sec:dp}, the minimum cost of a set $R_x$ with outflow $s$ and $q$ blocks for every $q$. The value $\loc_x(s,\ell,k_x)$ is the minimum of these values over all $q\le k_x$ with $(k_x-q)U^{\max}_x\ge\ell$. For hard capacities, a recursion over the locations $y\in F_x$ stores the minimum cost for every outflow, number of open locations and admitted inflow. In it, each location is closed, open, or open with its block rescued, and admitted inflows above $n$ are replaced by $n$.

\emph{The dynamic program.}
In the dynamic program of \Cref{sec:dp}, the vertices $\src x$ and $\snk x$ lie only in the leaf attached to $\{x\}$, so we can make their two forget nodes consecutive and merge them. The merged node keeps the states in which $\src x$ has balance $-s$ and $\snk x$ has balance $\ell\ge0$, chooses $k_x$, and adds $k_x$ to $k'$ and $\loc_x(s,\ell,k_x)$ to the cost. Everything else, including the analysis of the running time, is unchanged.

\emph{From network solutions to solutions.}
For the second part, open the centers of the configurations, and decompose $f$ into unit paths. At every point $x$, distribute the paths that end at $\snk x$ among the centers at $x$ so that a center at $y$ receives at most $U_y$ of them. A center that serves a rescued block $i$ receives at most $U_y-w_i$ of them. This is possible since the configuration admits the inflow of $\snk x$. Serve every rescued block by the center at its representative, and assign the clients of the normal blocks to the ends of their paths as in the proof of \Cref{lem:to-clustering}. As there, the cost is at most $A+\Phi$.
\end{proof}

\begin{theorem}[Few extra centers]\label{thm:near-k}
Fix $D\ge1$, $z\ge1$, $\eta\in(0,1/10]$ and $\gamma\in(0,1]$. There is a deterministic polynomial-time algorithm that, given an instance with non-uniform capacities such that $P\cup F$ has doubling dimension at most $D$ and a family $S_1$ as above with $|S_1|\le(3+\ln(2\cdot3^z/\eta))k$ and $\cost(S_1)\le(1+\eta)\OPT$, outputs a solution with at most $k+\lfloor\gamma k\rfloor$ centers and cost at most $3^ze^\eta(1+\eta)\OPT$ for soft capacities, and, if $z=1$, also without repeated locations for hard capacities.
\end{theorem}
\begin{proof}
\emph{The algorithm.}
Let $\Lambda=2^{13}D(4+\ln(2\cdot3^z/\eta))/\gamma$, larger than in~\eqref{eq:params} because the forest now has up to $m+k$ vertices, and $\lambda=\eta/(8z\Lambda)$. Since $m\le|S_1|\le(3+\ln(2\cdot3^z/\eta))k$, we have $\Lambda\ge2^{13}D(m+k)/(\gamma k)$. For every $\Dec\in\Hfam$, the algorithm computes a network solution of minimum cost with at most $k+\lfloor\gamma k\rfloor$ centers by \Cref{lem:configurations}, and it turns the cheapest one into a solution. The running time is polynomial as in the running-time analysis at the end of \Cref{sec:proof-combine}. It remains to show that some $\Dec\in\Hfam$ admits a network solution with at most $k+\lfloor\gamma k\rfloor$ centers and $A+\Phi\le3^ze^\eta(1+\eta)\OPT$.

\emph{A forest plan.}
Let $S^*$ be an optimal solution, with centers $j\in J$ at locations $c_j$ and loads $\ell_j\le U_{c_j}$. For soft capacities, let $\pi$ be the transport plan of \Cref{lem:transport}; its proof does not use $m\le2k-1$, so $T\le(A^{1/z}+\OPT^{1/z})^z$. For hard capacities and $z=1$, identify every block with its representative and every center of $S^*$ with its location, so that $a_i$ and $c_j$ are the locations $i$ and $j$. Let $\pi^0_{yy'}$ be the number of clients served at $y$ by $S_1$ and at $y'$ by $S^*$; by the triangle inequality, $\pi^0$ costs at most $A+\OPT$. Let $\pi$ be the plan of \Cref{lem:diagonal}, so $T\le A+\OPT$. In both cases, the support of $\pi$ is a forest on at most $m+k$ vertices, so it has fewer than $m+k$ edges.

\emph{Bad edges and routing.}
Call an edge $ij$ \emph{bad} if $a_i$ and $c_j$ are distinct points and the pair $(a_i,c_j)$ is $\Lambda$-bad. By \Cref{cor:few-bad}, some $\Dec\in\Hfam$ has at most $2^{11}D(m+k)/\Lambda<\gamma k$ bad edges, hence at most $\lfloor\gamma k\rfloor$. Let $R$ be the set of blocks incident to them. For every edge $ij$ with $i\notin R$, send $\pi_{ij}$ units from $\src{(a_i)}$ to $\snk{(c_j)}$ along a path of \Cref{lem:routes}(b). By the choice of $\lambda$, this costs at most $e^\eta\dist(a_i,c_j)^z$ per unit. Let $\ell'_j=\sum_{i\notin R}\pi_{ij}$.

\emph{Configurations.}
At every point $x$, the configuration rescues the blocks of $R\cap I_x$. For soft capacities, it has $k_x-|R_x|$ equal to the number of centers $j$ with $c_j\in F_x$ and $\ell'_j>0$. It admits the inflow $\sum_{j:c_j\in F_x}\ell'_j$, since $\ell'_j\le U_{c_j}\le U^{\max}_x$. For hard capacities, $F'_x$ consists of the locations of $F_x$ that lie in $S^*$ or represent a block of $R$. If $y\in S^*$ represents a rescued block, then $\ell'_y\le\ell_y-\pi_{yy}$, and \Cref{lem:diagonal} gives
\[
\ell'_y+w_y\le\ell_y+w_y-\min\{w_y,\ell_y\}=\max\{w_y,\ell_y\}\le U_y .
\]
If $y\in S^*$ represents no rescued block, then $\ell'_y\le\ell_y\le U_y$. If $y\notin S^*$ represents a rescued block, then $w_y\le U_y$. The configuration therefore admits the inflow $\sum_{y\in S^*\cap F_x}\ell'_y$. In both cases, there are at most $k+|R|\le k+\lfloor\gamma k\rfloor$ centers, since $S^*$ has at most $k$ centers.

\emph{The cost.}
The flow satisfies $\cost(f)\le e^\eta\sum_{i\notin R}T_i$, so
\[
A+\Phi\le A+\sum_{i\notin R}\bigl((u-1)A_i+e^\eta vT_i\bigr)\le uA+e^\eta vT .
\]
For soft capacities, $A\le(1+\eta)\OPT$ and $T\le((1+\eta)^{1/z}+1)^z\OPT\le(1+\eta)2^z\OPT$ give $uA+e^\eta vT\le e^\eta(1+\eta)(u+2^zv)\OPT=3^ze^\eta(1+\eta)\OPT$. For hard capacities and $z=1$, we have $u=v=1$ and $A+e^\eta(A+\OPT)\le3e^\eta(1+\eta)\OPT$.
\end{proof}

\subsection{Proof of \texorpdfstring{\Cref{thm:nonuniform}}{Theorem 1.2}}\label{sec:proof-nonuniform}

\emph{Parameters.}
We may assume $\delta\le1$. Let $\beta=3^z$, where $z=1$ for hard capacities. Suppose first that $P\cup F$ has doubling dimension at most $D$. Let $\eta=\min\{1/10,\delta/(4\beta)\}$ and $\gamma=\min\{\eps,\eta\}/4$. If $k<4/\eta$, the algorithm tries every family of at most $k$ centers at locations in $F$, without repeated locations for hard capacities, and returns one of minimum cost. Otherwise, it proceeds as follows.
\begin{enumerate}[label=\arabic*.,leftmargin=1.6em]
\item Compute at most $k$ centers whose uncapacitated cost is $A_0\le(1+\eta)\OPT$ by \Cref{thm:frs}. By \Cref{lem:sparse}, compute a family $S_0$ of at most $2k$ centers with $\cost(S_0)\le(2(1+\eta)^{1/z}+1)^z\OPT\le(1+\eta)\beta\OPT$.
\item Apply \Cref{lem:augment} to $S_0$ with $K'=|S_0|+\lceil k\ln(2\beta/\eta)\rceil$. The result $S_1$ has at most $(3+\ln(2\beta/\eta))k$ centers. Since $(1-1/k)^{K'-|S_0|}\le\eta/(2\beta)$, it satisfies $\cost(S_1)\le\OPT+\frac\eta{2\beta}(1+\eta)\beta\OPT\le(1+\eta)\OPT$.
\item Apply \Cref{thm:near-k}. The result $S_2$ has at most $k+\lfloor\gamma k\rfloor\le K$ centers, since $\gamma\le\eps$, and $\cost(S_2)\le\beta e^\eta(1+\eta)\OPT\le\beta(1+2.2\eta)\OPT$.
\item Apply \Cref{lem:augment} to $S_2$ with $K'=K$, and return the result.
\end{enumerate}

\emph{The cost.}
The result has at most $K$ centers. We have $K-|S_2|\ge(\eps-\gamma)k-1$, and $\gamma+1/k\le\eta/2$ since $\gamma\le\eta/4$ and $k\ge4/\eta$. This gives $(1-1/k)^{K-|S_2|}\le e^{-\eps}e^{\eta/2}\le e^{-\eps}(1+0.6\eta)$. By \Cref{lem:augment} and step~3, the cost of the result is at most
\begin{equation}\label{eq:nonuniform-cost}
\OPT+e^{-\eps}(1+0.6\eta)(\beta-1+2.2\beta\eta)\OPT\le\bigl(1+(\beta-1)e^{-\eps}+3\beta\eta\bigr)\OPT\le\bigl(1+(\beta-1)e^{-\eps}+\delta\bigr)\OPT .
\end{equation}
All steps take polynomial time. As in the running-time analysis at the end of \Cref{sec:proof-combine}, the degree is $(D/(\eps\delta))^{O(D/\delta)}$ for fixed $z$ and $\delta\le1/2$, since $|S_1|/k=O(\log(1/\delta))$ for fixed $z$.

\emph{One-sided low dimension.}
If only $P$ or only $F$ has doubling dimension at most $D$, we compress the other side as in \Cref{sec:proof-soft-onesided}, with accuracy $\eta_c=\min\{1/4,\eta/(18z)\}$ for the $\eta$ above. If $F$ is the other side, every location $y$ is placed at the point $\mu(y)$ and keeps its capacity, so that several locations may share a point. If $P$ is the other side, every client $p$ is moved to $\mu(p)$. The compressed instance has doubling dimension at most $D'$, the same solutions, and an optimal cost of at most $(1+3\eta_c)^z\OPT$. We run the algorithm above on it, with $D'$ in place of $D$. By~\eqref{eq:nonuniform-cost}, its solution costs at most $1+(\beta-1)e^{-\eps}+3\beta\eta\le1.3\beta$ times the optimal cost of the compressed instance. In the original instance, the cost grows by a factor of at most $(1+3\eta_c)^z$, and $(1+3\eta_c)^{2z}\le e^{6z\eta_c}\le e^{\eta/3}\le1+\eta/2$. The cost in the original instance is therefore at most
\[
\bigl(1+(\beta-1)e^{-\eps}+3\beta\eta\bigr)\Bigl(1+\frac\eta2\Bigr)\OPT\le\bigl(1+(\beta-1)e^{-\eps}+3.7\beta\eta\bigr)\OPT\le\bigl(1+(\beta-1)e^{-\eps}+\delta\bigr)\OPT,
\]
where the last step uses $\eta\le\delta/(4\beta)$. \qed

\section{Hard Capacities and General Metrics}\label{sec:hard}

This section proves \Cref{thm:hard,thm:general}. The soft solutions of \Cref{thm:soft} may place several centers at the same location, which hard capacities forbid. Two tools bridge the gap: a projection lemma (\Cref{lem:soft-to-hard}) selects a hard subfamily, and the greedy deletion of \Cref{sec:greedy} reduces the center count in general metrics. We use $\cost(S)$ as defined in \Cref{sec:prelim}. A hard solution is a subset of $F$. Since $F$ contains the centers of an optimal hard solution, $\cost(F)\le\OPT$ for the optimal hard cost $\OPT$.

\subsection{Projection}\label{sec:projection}

For $k$-median and $S=F$, the following lemma converts soft solutions into hard solutions~\cite[Theorem~1.2]{Li15}. We allow any family $S$. Applying Minkowski's inequality at the two places where costs are added extends the lemma to all $z\ge1$.

\begin{lemma}[Projection]\label{lem:soft-to-hard}
Let $S$ be a finite family of centers at locations in $F$ with $\cost(S)<\infty$, and let $S^*$ be a solution whose centers lie at arbitrary points of the metric space, each serving at most $U$ clients. One can compute in polynomial time a subfamily $S'\subseteq S$ with $|S'|\le|S^*|$ and $\cost(S')\le\bigl(\cost(S)^{1/z}+2\cost(S^*)^{1/z}\bigr)^z$.
\end{lemma}
\begin{proof}
\emph{A forest plan.}
We first compare the two assignments by a transport plan. As in \Cref{sec:approach-reference}, let $J$ be the set of centers of $S^*$ that serve clients, where $j\in J$ is at point $c_j$ and serves $\ell_j\le U$ clients. Let $\sigma$ be the assignment of $S^*$, and let $\sigma'$ be an assignment of the clients to $S$ of cost $\cost(S)$ in which every center serves at most $U$ clients. Let $\pi^0$ be the matrix indexed by $S\times J$ with $\pi^0_{xj}=|\{p:\sigma'(p)=x,\ \sigma(p)=j\}|$; its column sums are $\ell_j$, and its row sums are at most $U$. Entry $xj$ costs $\dist(x,c_j)^z$, where $\dist(x,c_j)$ is the distance from the location of $x$ to $c_j$. Applying~\eqref{eq:minkowski} to the triples $(\sigma'(p),c_{\sigma(p)},p)$ gives
\begin{equation}\label{eq:projection-plan}
\cost(\pi^0)\le\bigl(\cost(S)^{1/z}+\cost(S^*)^{1/z}\bigr)^z .
\end{equation}
By \Cref{fact:cycle}, applied with the row sums of $\pi^0$, we obtain a plan $\pi$ with the same row and column sums, no larger cost and a forest as support.

\emph{Shifting load along paths.}
Next we change $\pi$ until every component of its support contains at most one used center of $S$ that is not full. Call a center $x\in S$ \emph{used} if its row sum is positive, and \emph{full} if its row sum is $U$. While some component of the support contains two used centers $x\ne x'$ of $S$ that are not full, consider the path between them in the support. It alternates between centers of $S$ and of $J$ and has even length. Adding $1$ to every other entry along the path and subtracting $1$ from the others keeps all column sums and the row sums of the interior centers. It changes the row sums of $x$ and $x'$ by $+1$ and $-1$, or by $-1$ and $+1$. Both directions are feasible, since the row sums of $x$ and $x'$ lie strictly between $0$ and $U$ and all entries on the path are positive. The cost changes by opposite amounts in the two directions, so one direction does not increase the cost. We repeat this change in that direction as long as it stays feasible. This stops when an entry becomes zero or $x$ or $x'$ becomes full. No step adds an entry to the support. A full center stays full, because only the two end centers of a path change their row sums. Every step therefore removes an entry of the support or creates a new full center, and the process stops after at most $|S|\cdot|J|+|S|$ steps.

\emph{Counting the used centers.}
At the end, every component of the support contains at most one used center of $S$ that is not full. If the column sums in a component add up to $\ell$, the component therefore contains exactly $\lceil\ell/U\rceil$ used centers of $S$. Since every column sum is at most $U$, it also contains at least $\lceil\ell/U\rceil$ centers of $J$. Summing over the components shows that the set $S'$ of used centers satisfies $|S'|\le|J|\le|S^*|$.

\emph{The cost.}
Assign the clients of every $j\in J$ to the $\ell_j$ slots given by column $j$ of the final plan; every center of $S'$ receives its row sum, which is at most $U$. The final plan costs at most $\cost(\pi^0)$. Applying~\eqref{eq:minkowski} to the triples $(p,x,c_{\sigma(p)})$, where $p$ is assigned to $x$, and then~\eqref{eq:projection-plan} bounds the cost by
\[
\Bigl(\cost(S^*)^{1/z}+\bigl(\cost(S)^{1/z}+\cost(S^*)^{1/z}\bigr)\Bigr)^z=\bigl(\cost(S)^{1/z}+2\cost(S^*)^{1/z}\bigr)^z.\qedhere
\]
\end{proof}

For $k$-means in Euclidean space, the random matching of \Cref{sec:certificates} sharpens the bound.

\begin{lemma}[Euclidean projection]\label{lem:euclid-projection}
Let $z=2$ and $V=\R^d$, and suppose that every center of $S^*$ is at the centroid of the clients that it serves. Then the subfamily $S'$ of \Cref{lem:soft-to-hard} satisfies $\cost(S')\le\cost(S)+2\cost(S^*)+2\sqrt{\cost(S)\cost(S^*)}$.
\end{lemma}
\begin{proof}
In the last step of the proof of \Cref{lem:soft-to-hard}, assign the $\ell_j$ clients of every $j\in J$ to the $\ell_j$ slots of column $j$ by a uniformly random bijection. Since $c_j$ is the centroid of these clients, $\sum_{p:\sigma(p)=j}(p-c_j)=0$. Together with
\[
\|p-x\|^2=\|p-c_j\|^2+2\langle p-c_j,c_j-x\rangle+\|c_j-x\|^2
\]
for a client $p$ assigned to $x$, this shows that the expected cost of these clients is $\sum_{p:\sigma(p)=j}\|p-c_j\|^2+\sum_{x}\pi_{xj}\|x-c_j\|^2$. Some bijection costs at most the expectation. Summing over $j$ and using $\cost(\pi)\le\cost(\pi^0)$ and~\eqref{eq:projection-plan} gives
\[
\cost(S')\le\cost(S^*)+\cost(\pi)\le\cost(S^*)+\bigl(\cost(S)^{1/2}+\cost(S^*)^{1/2}\bigr)^2.\qedhere
\]
\end{proof}

\begin{proof}[Proof of \Cref{thm:hard}]
\emph{$k$-median.}
Uniform capacities are a special case of non-uniform capacities. For $k$-median, the algorithm of \Cref{thm:nonuniform} for hard capacities therefore has ratio $1+2e^{-\eps}$.

\emph{A soft solution.}
For every $z\ge1$, we also obtain the second ratio of \Cref{thm:hard} as follows. Let $\OPT$ be the optimal hard cost, and let $\beta=3^z$, or $\beta=\beta^*_z$ for instances with $P\cup F\subset\R^d$ and $P$ of doubling dimension at most $D$. In low dimension, every hard solution is a soft solution. The algorithm of \Cref{thm:soft} with accuracy $\delta/(2\cdot3^{z-1})$ therefore computes a soft solution with at most $K$ centers at locations in $F$ and cost at most $(\beta-(\beta-1)\eps+\delta/(2\cdot3^{z-1}))\OPT$. For instances in $\R^d$ with a finite set $F\subset\R^d$, every hard solution is a solution of the instance in which centers may be placed anywhere. The algorithm of \Cref{thm:soft} for centers anywhere in $\R^d$, with the same accuracy, therefore computes at most $K$ centers in $\R^d$ and an assignment with the same cost bound. In both cases, every center serves at most $U$ clients.

\emph{Projection.}
Apply \Cref{lem:soft-to-hard} with $S=F$ and this solution as $S^*$, and recall that $\cost(F)\le\OPT$. The derivative of $w\mapsto(1+2w^{1/z})^z$ is $2(2+w^{-1/z})^{z-1}\le2\cdot3^{z-1}$ for $w\ge1$, and $\beta-(\beta-1)\eps\ge1$. The cost is therefore at most
\[
\Bigl(1+2\Bigl(\beta-(\beta-1)\eps+\frac{\delta}{2\cdot3^{z-1}}\Bigr)^{1/z}\Bigr)^z\OPT\le\Bigl(\bigl(1+2(\beta-(\beta-1)\eps)^{1/z}\bigr)^z+\delta\Bigr)\OPT .
\]

\emph{$k$-means in $\R^d$.}
We first move every center of the solution for centers anywhere in $\R^d$ to the centroid of its clients, which does not increase the cost, and apply \Cref{lem:euclid-projection} instead. For $z=2$, the accuracy is $\delta/6$, and the cost is at most
\[
\Bigl(1+2\Bigl(5-4\eps+\frac\delta6\Bigr)+2\Bigl(5-4\eps+\frac\delta6\Bigr)^{1/2}\Bigr)\OPT\le\bigl(11-8\eps+2\sqrt{5-4\eps}+\delta\bigr)\OPT.\qedhere
\]
\end{proof}

\subsection{Greedy Deletion}\label{sec:greedy}

The greedy deletion follows the reverse greedy algorithm of Chrobak, Kenyon and Young~\cite{CKY06}. Its key ingredient is the supermodularity of the assignment cost (\Cref{lem:supermodular}), the substitutes property of the assignment problem~\cite{Sha62}.

\begin{lemma}[One deletion]\label{lem:deletion}
Let $S$ be a family of $k+q$ centers at locations in $F$ with $q\ge1$ and $\cost(S)<\infty$, and let $S^*$ be a solution with at most $k$ centers at arbitrary points. Some center $x\in S$ satisfies
\[
q\bigl(\cost(S\setminus\{x\})-\cost(S)\bigr)\ \le\ \bigl(\cost(S)^{1/z}+2\cost(S^*)^{1/z}\bigr)^z-\cost(S).
\]
\end{lemma}
\begin{proof}
By \Cref{lem:soft-to-hard}, some $S'\subseteq S$ with $|S'|\le k$ has $\cost(S')\le(\cost(S)^{1/z}+2\cost(S^*)^{1/z})^z$. Adding centers of $S$ to $S'$ does not increase $\cost(S')$, so we may assume that $|S'|=k$. The family $S\setminus S'$ then has $q$ centers. By \Cref{lem:supermodular}, the increases $\cost(S\setminus\{x\})-\cost(S)$ caused by deleting them add up to at most $\cost(S')-\cost(S)$, and the smallest of these $q$ increases is at most their average.
\end{proof}

For $z=1$, the right side is $2\cost(S^*)$. For $z>1$, it grows with $\cost(S)$, and we compare the deletions with a differential equation.

\begin{lemma}[Greedy deletion]\label{lem:greedy}
Let $\eps,\eta\in(0,1]$ with $k\ge4/(\eps\eta)$, let $\rho\ge1$, and let $S_0$ be a family of at most $2k-1$ centers at locations in $F$ with $\cost(S_0)\le\rho\OPT$, where $\OPT$ is the optimal soft or the optimal hard cost. Starting from $S=S_0$, as long as $|S|>K$, delete a center $x\in S$ that minimizes $\cost(S\setminus\{x\})$. This takes polynomial time, keeps $S\subseteq F$ if $S_0\subseteq F$, and ends with $\cost(S)\le\zeta\OPT$ for some $\zeta\ge\rho$ with
\begin{equation}\label{eq:greedy-integral}
\int_\rho^{\zeta}\frac{dw}{(w^{1/z}+2)^z-w}\ \le\ \ln\frac1\eps+\frac\eta2 .
\end{equation}
In particular, $\zeta^{1/z}\le\rho^{1/z}+C\,(\ln(1/\eps)+\eta/2)$, where $C=2$ for $z=1$, $C=3$ for $z=2$ and $\rho\ge4$, and $C=2\cdot3^{z-1}$ in general.
\end{lemma}
\begin{proof}
\emph{Running time and trivial cases.}
There are fewer than $2k$ deletions, each found by fewer than $2k$ minimum-cost flow computations. If $|S_0|\le K$, take $\zeta=\rho$. Otherwise, $K-k=\lfloor\eps k\rfloor$ satisfies $K-k<|S_0|-k\le k-1$ and $K-k\ge\eps k-1\ge1$. If $\OPT=0$, then $\cost(S_0)=0$, and by \Cref{lem:deletion} with an optimal solution as $S^*$, every deletion keeps the cost $0$, so $\zeta=\rho$ works. From now on, let $\OPT>0$.

\emph{The recurrence.}
Let $\zeta(q)=\cost(S)/\OPT$ when $k+q$ centers remain; the deletions happen at $q=|S_0|-k,\dots,K-k+1$. The greedy step deletes a center that minimizes $\cost(S\setminus\{x\})$, so \Cref{lem:deletion} with an optimal solution as $S^*$ gives
\[
\zeta(q-1)\ \le\ \zeta(q)+\frac{(\zeta(q)^{1/z}+2)^z-\zeta(q)}q .
\]

\emph{Comparison with the integral.}
The function $w\mapsto(w^{1/z}+2)^z-w$ is positive and nondecreasing on $[0,\infty)$, since its derivative is $(1+2w^{-1/z})^{z-1}-1\ge0$. By the last display, the integral of $1/((w^{1/z}+2)^z-w)$ from $\zeta(q)$ to $\zeta(q-1)$ is therefore at most $1/q$, also if $\zeta(q-1)<\zeta(q)$. Since $\zeta(|S_0|-k)\le\rho$, summing over $q$ shows that the integral from $\rho$ to $\zeta(K-k)$ is at most $\sum_{q=K-k+1}^{k-1}1/q\le\ln((k-1)/(K-k))$. As $\eps k\ge4/\eta\ge4$, we have
\[
\ln\frac{k-1}{K-k}\ \le\ \ln\frac{k}{\eps k-1}\ =\ \ln\frac1\eps+\ln\Bigl(1+\frac1{\eps k-1}\Bigr)\ \le\ \ln\frac1\eps+\frac2{\eps k}\ \le\ \ln\frac1\eps+\frac\eta2 .
\]
The value $\zeta=\max\{\zeta(K-k),\rho\}$ therefore satisfies~\eqref{eq:greedy-integral}.

\emph{The explicit bound.}
Substituting $w=\theta^z$ turns the integral in~\eqref{eq:greedy-integral} into the integral of $z\theta^{z-1}/((\theta+2)^z-\theta^z)$ from $\rho^{1/z}$ to $\zeta^{1/z}$. Since $(\theta+2)^z-\theta^z\le2z(\theta+2)^{z-1}$, the integrand is at least $\frac12(\theta/(\theta+2))^{z-1}$, which is $\frac12$ for $z=1$ and at least $\frac12\cdot3^{1-z}$ for $\theta\ge1$. For $z=2$, the integrand equals $\frac12\cdot\theta/(\theta+1)$, which is at least $\frac13$ for $\theta\ge2$. As $\theta\ge\rho^{1/z}\ge1$, and $\theta\ge2$ if $z=2$ and $\rho\ge4$, the integrand is at least $1/C$. By~\eqref{eq:greedy-integral}, we obtain
\[
\frac{\zeta^{1/z}-\rho^{1/z}}C\ \le\ \int_{\rho^{1/z}}^{\zeta^{1/z}}\frac{z\theta^{z-1}\,d\theta}{(\theta+2)^z-\theta^z}\ \le\ \ln\frac1\eps+\frac\eta2 .\qedhere
\]
\end{proof}

The greedy deletion of \Cref{lem:greedy} yields the following result for general metrics.

\begin{corollary}[General metrics]\label{thm:general}
For all fixed $z\ge1$, $\rho\ge1$ and $\eps\in(0,1]$, there is an algorithm for uniform capacities in finite metric spaces that, given an uncapacitated solution with at most $k$ centers whose cost is at most $\rho$ times the minimum, has ratio $(\rho^{1/z}+C_{\mathrm s}\ln(1/\eps))^z$ for soft and $(1+2\rho^{1/z}+C_{\mathrm h}\ln(1/\eps))^z$ for hard capacities, where $C_{\mathrm s}$ and $C_{\mathrm h}$ are the constant $C$ of \Cref{lem:greedy} for the starting ratios $\rho$ and $(1+2\rho^{1/z})^z$, respectively; both are at most $2\cdot3^{z-1}$. For fixed $z$ and $\rho$, both ratios are $O(1+\log^z(1/\eps))$. For $k$-median, $C_{\mathrm s}=C_{\mathrm h}=2$, so the ratios are $\rho+2\ln(1/\eps)$ and $1+2\rho+2\ln(1/\eps)$.
\end{corollary}

For $z=2$, the hard ratio is $(1+2\sqrt\rho+3\ln(1/\eps))^2$, since $(1+2\sqrt\rho)^2\ge4$, and the soft ratio is $(\sqrt\rho+3\ln(1/\eps))^2$ if $\rho\ge4$.

\begin{proof}
\emph{Parameters.}
Let $\theta_0=\rho^{1/z}$ and $C_0=C_{\mathrm s}$ for soft capacities, and $\theta_0=1+2\rho^{1/z}$ and $C_0=C_{\mathrm h}$ for hard capacities, so that the claimed ratio is $(\theta_0+C_0\ln(1/\eps))^z$ in both cases and $C_0$ is the constant $C$ of \Cref{lem:greedy} for the starting ratio $\theta_0^z$. Let $\eta\in(0,1]$ be so small that $\eta$ times this ratio is at most $\delta$, which is used in \Cref{app:bits}, and that $(\theta_0+C_0(\ln(1/\eps)+\eta/2))^z\le(\theta_0+C_0\ln(1/\eps))^z+\delta$.

\emph{Small $k$.}
If $k<4/(\eps\eta)$, the algorithm tries every family of at most $k$ centers at locations in $F$, with distinct locations for hard capacities, and returns one of minimum cost. There are at most $(|F|+1)^{4/(\eps\eta)}$ of them.

\emph{The start.}
Otherwise, the given uncapacitated solution has cost $A\le\rho\OPT$, since the optimal uncapacitated cost is at most $\OPT$ for soft and for hard capacities. Splitting its clusters into blocks as in \Cref{sec:approach-blocks} gives at most $2k-1$ blocks, and serving every block by a center at the center of its cluster costs $A$. For soft capacities, let $S_0$ be the family of these centers, so $\cost(S_0)\le A\le\theta_0^z\OPT$. For hard capacities, apply \Cref{lem:soft-to-hard} with $S=F$ and this solution as $S^*$. Since $\cost(F)\le\OPT$, this gives a set $S_0\subseteq F$ with $|S_0|\le2k-1$ and
\[
\cost(S_0)\le\bigl(\OPT^{1/z}+2A^{1/z}\bigr)^z\le\theta_0^z\OPT .
\]
\emph{Greedy deletion.}
By \Cref{lem:greedy} with $\theta_0^z$ in place of $\rho$, the greedy deletions end with cost at most
\[
\bigl(\theta_0+C_0(\ln(1/\eps)+\eta/2)\bigr)^z\OPT\le\Bigl(\bigl(\theta_0+C_0\ln(1/\eps)\bigr)^z+\delta\Bigr)\OPT .
\]
Since $C_0\le2\cdot3^{z-1}$ and $\theta_0\le3\rho^{1/z}$, the ratio $(\theta_0+C_0\ln(1/\eps))^z$ is $O(1+\log^z(1/\eps))$ for fixed $z$ and $\rho$.
\end{proof}

\Cref{app:sharper} replaces the explicit bounds above by the exact bound $\zeta_z(\rho,\eps)$ of greedy deletion, the value at time $\ln(1/\eps)$ of the solution of $w'=(w^{1/z}+2)^z-w$ with $w(0)=\rho$. For $k$-means, this improves the ratios for moderate $\eps$; with soft capacities, $\rho=9$ and $\eps=1/2$, it gives $\zeta_2(9,1/2)\approx22.6$ instead of $(3+3\ln2)^2\approx25.8$. In low dimension, greedy deletion can start from a hard solution of ratio arbitrarily close to $3^z$, obtained by projection. For $z>1$, this improves the hard ratio of \Cref{thm:hard} to $\min\{(1+2(\beta-(\beta-1)\eps)^{1/z})^z,\ \zeta_z(3^z,\eps)\}$; for $k$-means and $\eps=1/2$, it gives $\zeta_2(9,1/2)\approx22.6$ instead of $(1+2\sqrt5)^2\approx29.9$. \Cref{tab:results} lists the resulting ratios.

\section{Related Work}\label{sec:related}

We survey related work on capacitated clustering in general metrics, approximation schemes in low dimension, parameterized algorithms, and coresets.

\paragraph{Capacitated clustering in general metrics.}
The best known polynomial-time ratio for capacitated $k$-median with $k$ centers and exact capacities is $O(\log k)$~\cite{ABMMW19}. For uniform capacities, the reverse greedy algorithm, which deletes centers one at a time, has ratio $O(\log n)$~\cite{AGGN10}. Its analysis builds on that of Chrobak, Kenyon and Young~\cite{CKY06} for uncapacitated $k$-median, and \Cref{thm:general} runs it from the blocks instead of from all locations. Constant ratios are known if the capacities may be exceeded by a constant factor~\cite{CGTS02,CR05,BFRS15,ShLi14} or by a factor arbitrarily close to $1$~\cite{BRU16,DL16,DaiLi26}. In capacitated facility location, where every center has an opening cost and the number of centers is unbounded, local search and LP-based algorithms give constant ratios~\cite{KPR00,PTW01,BGG12,ASS17}.

Constant ratios are also known if more than $k$ centers may be opened. Local search gives ratio $1+\delta$ with $(5+5/\delta)k$ centers~\cite{KPR00}, and ratio $7+\delta$ is possible with $2k-1$ centers~\cite{ABGL15}. With $(1+\eps)k$ centers, Li gave ratio $\exp(O(1/\eps^2))$ for uniform capacities~\cite{Li15} and $O(\eps^{-2}\log(1/\eps))$ for non-uniform soft capacities~\cite{Li16}. Dai and Li~\cite{DaiLi26} recently gave ratio $O(\eps^{-5})$ for non-uniform hard capacities in time polynomial in $n$ and $1/\eps$. For uniform capacities, soft and hard capacities are equivalent up to a constant factor~\cite{Li15}, so the ratio of~\cite{Li16} carries over to hard capacities. For capacitated $k$-means, Han, Xu, Du and Zhang~\cite{HXDZ22} gave a constant ratio in $\R^d$ with a constant factor more centers. Adding centers greedily, as in \Cref{thm:nonuniform}, is known as greedy augmentation for uncapacitated facility location~\cite{GK99,CG05}. With capacities, the supermodularity of the assignment cost~\cite{Sha62} makes greedy augmentation give ratio $1+\delta$ with $O(k\log(2/\delta))$ centers in every metric, also for non-uniform capacities, as we show in \Cref{cor:bicriteria}.

For comparison, without capacities, $k$-median admits ratio $2+\delta$~\cite{CGLSS25}, and local search gives ratio $3+\delta$~\cite{AGKMMP04}. For $k$-means without capacities, the ratio $9+\delta$~\cite{ANSW20} was improved to $3+2\sqrt2+\delta$~\cite{CCGGLW25} and $4.9+\delta$~\cite{ACCGGLSW27}. For every fixed $z$, local search gives a constant ratio~\cite{GT08}.

Opening more than $k$ centers is a common relaxation also without capacities. In general metrics, $k$-median admits ratio $2(1+1/\eps)$ with $(1+\eps)k$ centers~\cite{LV92b}, and local search gives ratio $1+\delta$ with $(3+5/\delta)k$ centers~\cite{KPR00}. For $k$-means, the ratio of Makarychev, Makarychev, Sviridenko and Ward~\cite{MMSW16} decreases from $9+\delta$ with $k$ centers to below $2.59$ with $2k$ centers, and sampling-based algorithms achieve constant ratios with $O(k)$ or $O(k\log k)$ centers~\cite{ADK09,AJM09,Wei16,MRS20}. Without the triangle inequality, Lin and Vitter~\cite{LV92a} achieved ratio $1+\delta$ with $O((1+1/\delta)k\log n)$ centers and showed that a logarithmic factor is then essentially necessary. Li and Svensson~\cite{LS16} showed that for $k$-median, an algorithm that opens $k+O(1)$ centers can be turned into one that opens $k$ centers with nearly the same ratio. Extra centers are also used for fair clustering~\cite{ABV21,MV21}.

\paragraph{Approximation schemes in low dimension.}
Without capacities, approximation schemes in low dimension use dynamic programming over hierarchical decompositions~\cite{ARR98,KR07,Tal04,CFS21} or local search~\cite{CKM19,FRS19}. They are also known in planar and minor-free graphs~\cite{CKM19,CPP19a}, in graphs of bounded highway dimension~\cite{FFKP18,FS21} and, in time $f(k,\delta)\cdot n^{O(1)}$, in metrics of bounded scatter dimension~\cite{ABBCGKMSS23}. For $k$-means in fixed dimension and in doubling metrics, near-linear time is possible~\cite{CA18,CFS21}. With $(1+\eps)k$ centers, local search is an approximation scheme for $k$-median in the plane~\cite{CAM15} and for $k$-means in fixed dimension~\cite{BV16}, and there are approximation schemes for $k$-median and $k$-means with outliers in doubling metrics~\cite{FKRS19} and for $k$-median in planar graphs in time $f(\eps)\cdot n^{O(1)}$~\cite{CPP19b}.

With capacities, a dynamic program must count the clients that enter or leave every cell, and local search can be arbitrarily bad~\cite{CohenAddad20}. Arora, Raghavan and Rao~\cite{ARR98} gave a quasi-polynomial-time scheme for capacitated $k$-median in the plane that exceeds the capacities by a factor arbitrarily close to $1$. Cohen-Addad~\cite{CohenAddad20} gave schemes that respect uniform capacities, in quasi-polynomial time for $k$-median and $k$-means in doubling metrics and in polynomial time for $k$-median in the plane. Like the scheme of Cohen-Addad, Feldmann and Saulpic~\cite{CFS21} without capacities, the dynamic program of~\cite{CohenAddad20} treats the points that the decomposition cuts badly separately, but it uses $(\log n/\delta)^{O(D)}$ portals per cell in doubling dimension $D$, which leads to quasi-polynomial time. For planar and minor-free metrics, the quasi-polynomial-time schemes for capacitated $k$-median and capacitated facility location~\cite{CLPP23,CCCLPP25} embed the metric into a graph of polylogarithmic treewidth with expected distortion $1+\delta$. Capacitated clustering with outliers, where a given number of clients may be left unserved, is studied in~\cite{DGI25,JK23}.

\paragraph{Parameterized algorithms.}
Capacitated $k$-median is W[2]-hard parameterized by $k$~\cite{ABMMW19}. In time $f(k,\delta)\cdot n^{O(1)}$, it admits ratio $7+\delta$~\cite{ABMMW19} and $3+\delta$~\cite{CL19}, and capacitated $k$-means admits ratio $9+\delta$~\cite{CL19}; see also~\cite{XMXZZ20}. In the same running time, Dai, Li and Peng~\cite{DLP26} obtained ratio $3+\delta$ for capacitated clustering with general norm objectives. This ratio is tight for the class of all norms, since it includes the $k$-supplier objective, for which no such algorithm achieves a ratio better than $3$ unless $\mathrm{FPT}=\mathrm{W}[2]$~\cite{GJ23}. For $k$-median and $k$-means alone, under the gap exponential time hypothesis, no algorithm with such a running time achieves a ratio better than $1+2/e$ and $1+8/e$, respectively, even without capacities~\cite{CGKLL19}. Also in this running time, capacitated clustering admits approximation schemes in Euclidean space of any dimension~\cite{CL19,BFS24} and in metrics of bounded scatter dimension, which include doubling metrics and planar graphs~\cite{BGI25}. Several frameworks handle general assignment constraints~\cite{DX15,BJK18,GJK20,FZHXW20,BGJK20}.

\paragraph{Coresets.}
A coreset is a small weighted set of points on which the cost of every solution is preserved up to a factor of $1\pm\delta$. Har-Peled and Mazumdar~\cite{HM04} gave coresets for $k$-median and $k$-means in Euclidean space of fixed dimension, Feldman and Langberg~\cite{FL11} gave a unified framework based on sensitivity sampling, and Cohen-Addad, Saulpic and Schwiegelshohn~\cite{CSS21} gave coresets whose size is independent of both $n$ and the dimension. In doubling metrics, there are coresets whose size is independent of $n$~\cite{HJLW18}. With capacities, a coreset must preserve the cost of the best assignment that respects the capacities, not only the cost of assigning every client to a closest center. Coresets that preserve the cost of such constrained assignments were developed for fair clustering, where the clusters must contain points of each color in given proportions~\cite{SSS19,HJV19,BFS24}. For capacitated clustering in general metrics, there are coresets of size polynomial in $k$, $1/\delta$ and $\log n$~\cite{CL19}. For capacitated $k$-median in Euclidean space, there are coresets whose size is independent of $n$~\cite{BCJKSTW22,HLLW25}; see also~\cite{BFS24}.

\section{Conclusion and Open Questions}\label{sec:open}

Two ideas drive our results. The rescue lemma (\Cref{lem:rescue}) is a statement about transport plans whose support is a forest and uses no geometry; only the dynamic program that finds a good set of rescued blocks needs low dimension. The second idea pays with a reserve of extra centers for the pairs that the hierarchical decomposition cuts badly. Cohen-Addad, Feldmann and Saulpic~\cite{CFS21} and Cohen-Addad~\cite{CohenAddad20} also treat such pairs separately, but in the capacitated setting this still required $(\log n/\delta)^{O(D)}$ portals per cell, whereas the reserve allows a constant number. A natural next step is capacitated $k$-median in planar graphs, where approximation schemes are known only in quasi-polynomial time~\cite{CLPP23,CCCLPP25}, and in graphs of bounded highway dimension, where polynomial-time schemes are known only without capacities~\cite{FS21}. Can $(1+\eps)k$ centers yield a polynomial-time algorithm in these metrics with a ratio independent of $\eps$? Our dynamic program does not apply directly, since a cell of small diameter in such a graph may need arbitrarily many portals; a star with $n$ leaves, for instance, has diameter $2$, but a set of points that contains a point at distance less than $1$ from every leaf must contain all $n$ leaves. The known schemes instead embed these graphs into graphs of small treewidth~\cite{FFKP18,CLPP23,CCCLPP25}, and it is open whether the reserve of extra centers can be combined with such embeddings.

In low dimension, our ratios tend to a constant as $\eps\to0$, namely $3^z$ in general and $\beta^*_z$ in Euclidean space for uniform soft capacities, and they decrease linearly in $\eps$. We highlight four further open questions.
\begin{enumerate}[label=\arabic*.,leftmargin=1.6em]
\item \emph{Approximation schemes.} With $k$ centers and uniform capacities, Cohen-Addad~\cite{CohenAddad20} gave a polynomial-time approximation scheme for $k$-median in the plane and quasi-polynomial-time schemes in doubling metrics. Is there a polynomial-time approximation scheme for uniform capacitated $k$-median or $k$-means in $\R^d$ for a fixed $d\ge3$, at least with $(1+\eps)k$ centers? Without capacities, local search with $(1+\eps)k$ centers gives such schemes~\cite{CAM15,BV16}, but with capacities, local search can be arbitrarily bad~\cite{CohenAddad20}.
\item \emph{Better constants.} No lower bound above $1$ is known in low dimension, so the gap lies between our constants and $1$. In Euclidean space, $\beta^*_z$ is optimal for movability with optimal representatives (\Cref{prop:constant}(c)). A smaller constant therefore needs a different comparison with the optimal solution, for instance a better bound on the transport cost $T$ than \Cref{lem:transport}. In doubling metrics, does $k$-median admit ratio $3-c$ for some constant $c>0$ with $(1+\eps)k$ centers for every $\eps>0$?
\item \emph{Non-uniform capacities.} For non-uniform capacities, our ratios decrease with $\eps$ only as $1+(3^z-1)e^{-\eps}$, and for hard capacities, \Cref{thm:nonuniform} applies only to $z=1$. Does the linear trade-off of \Cref{thm:soft} extend to non-uniform capacities, and is there a ratio bounded independently of $\eps$ for $k$-means with non-uniform hard capacities? In general metrics, we know no analogue of \Cref{thm:general} for non-uniform capacities: the block construction of \Cref{sec:approach-blocks} needs uniform capacities, and so does the step of our projection lemma that counts the centers in a component.
\item \emph{General metrics.} In general metrics, our ratios grow as $(2+o(1))^z\ln^z(1/\eps)$ as $\eps\to0$ (\Cref{lem:Lz}); for $k$-median this is $3+2\ln(1/\eps)$ with soft capacities, compared with $O(\eps^{-2}\log(1/\eps))$ for non-uniform capacities~\cite{Li16}. Is there a ratio that stays bounded as $\eps\to0$, even for uniform $k$-median? Such a ratio cannot be proved against the natural LP relaxation, whose integrality gap is unbounded with fewer than $2k$ centers~\cite{Li15,BFRS15}.
\end{enumerate}

\section*{AI Methodology}
\addcontentsline{toc}{section}{AI Methodology}
AI methodology: The problem, the approach and the main ideas of this paper are due to the authors: splitting a near-optimal uncapacitated solution into capacity-respecting blocks, rescuing blocks with extra centers (the rescue lemma), paying with a reserve of extra centers for the pairs that the decomposition cuts badly, and the framework that combines these steps. We used ChatGPT (OpenAI) to help work out technical proofs in the derandomization of the decomposition, in the extensions of the framework to other settings and in the appendices, and to sharpen the constants in our ratios. The authors verified the correctness and originality of all content, including references, and take full responsibility for the content of the paper.

\appendix
\crefalias{section}{appendix}
\section{The Constant \texorpdfstring{$\beta^*_z$}{Bz}}\label{app:constant}

This appendix proves \Cref{prop:constant} and shows how to compute a nearly optimal certificate. All random variables are real and have finite support. Taking expectations in~\eqref{eq:certificate} gives \emph{weak duality}: for every $z$-certificate $(u,g,v)$ and every random variable $\xi$ with $\mathbb E[\xi]=0$,
\begin{equation}\label{eq:weak-duality}
\mathbb E|1+\xi|^z\le u+v\,\mathbb E|\xi|^z .
\end{equation}

\paragraph{Proof of \Cref{prop:constant}(a).}
For $\tau\ge0$, let $\psi(\tau)$ be the supremum of $\mathbb E|1+\xi|^z$ over all $\xi$ with $\mathbb E[\xi]=0$ and $\mathbb E|\xi|^z=\tau$; so $\psi(2^z)$ is the supremum in~(a). By~\eqref{eq:weak-duality}, every certificate has $u+2^zv\ge\psi(2^z)$.

\emph{Concavity.}
We first show that $\psi$ is finite and concave. By Minkowski's inequality, $\mathbb E|1+\xi|^z\le(1+(\mathbb E|\xi|^z)^{1/z})^z$, so $\psi$ is finite, and $\psi(0)=1$. If $\xi_1$ and $\xi_2$ are feasible for $\tau_1$ and $\tau_2$, a mixture that follows $\xi_1$ with some probability and $\xi_2$ otherwise is feasible for the same convex combination of $\tau_1$ and $\tau_2$. Its value is the same convex combination of $\mathbb E|1+\xi_1|^z$ and $\mathbb E|1+\xi_2|^z$.

\emph{The certificate.}
We now construct a certificate with $u+2^zv=\psi(2^z)$. A finite concave function on $[0,\infty)$ has a supergradient at the interior point $2^z$: there is $v$ with $\psi(\tau)\le\psi(2^z)+v(\tau-2^z)$ for all $\tau\ge0$. Let $u=\psi(2^z)-2^zv$, and let $h(t)=|1+t|^z-u-v|t|^z$. For every $\xi$ with $\mathbb E[\xi]=0$, we have $\mathbb E\,h(\xi)\le\psi(\mathbb E|\xi|^z)-u-v\,\mathbb E|\xi|^z\le0$. With $\xi=0$, this gives $h(0)\le0$. For $t_1>0>t_2$, the variable that takes the values $t_1$ and $t_2$ with probabilities $-t_2/(t_1-t_2)$ and $t_1/(t_1-t_2)$ has mean zero, which gives $-t_2h(t_1)+t_1h(t_2)\le0$, that is, $h(t_1)/t_1\le h(t_2)/t_2$. The number $g=\sup_{t>0}h(t)/t$ is therefore finite, and $h(t)\le gt$ for all $t$: for $t>0$ by the definition of $g$, and for $t<0$ because $g\le h(t)/t$. The triple $(u,g,v)$ is thus a certificate with $u+2^zv=\psi(2^z)$. In particular, the minimum in the definition of $\beta^*_z$ is attained. \qed

\paragraph{Proof of \Cref{prop:constant}(c).}
Let $u+2^zv<\beta^*_z$.

\emph{Weighted points.}
We first construct weighted points on the line with optimal center $0$ whose cost at $-1$ exceeds $u$ times their cost at $0$ plus $v$. By~(a), some $\xi$ with $\mathbb E[\xi]=0$ and $\mathbb E|\xi|^z=2^z$ has $\mathbb E|1+\xi|^z>u+2^zv$. We may assume $\Pr[\xi=0]=0$. Indeed, replace an atom at $0$ by two atoms of half its mass each, placed symmetrically around $0$, and scale $\xi$ to restore $\mathbb E|\xi|^z=2^z$. This changes $\mathbb E|1+\xi|^z$ by an amount that tends to $0$ as the two atoms approach $0$. For every value $x$ of $\xi$, place the point $1/x\in\R$ with weight $\theta_x=\Pr[\xi=x]\,|x|^z/2^z$. The weights sum to $1$, and
\[
\sum_x\theta_x|1/x|^z=2^{-z},\qquad\sum_x\theta_x|1/x+1|^z=2^{-z}\,\mathbb E|1+\xi|^z>2^{-z}u+v .
\]
The point $0$ is an optimal center of the weighted points. For $z>1$, the derivative of $c\mapsto\sum_x\theta_x|1/x-c|^z$ at $c=0$ is $-z\sum_x\theta_x|1/x|^{z-2}/x=-z2^{-z}\,\mathbb E[\xi]=0$. For $z=1$, the points $1/x$ with $x>0$ have total weight $\mathbb E[\max\{\xi,0\}]/2=\frac14\mathbb E|\xi|=\frac12$, so $0$ is a weighted median.

\emph{A block.}
We turn the weighted points into a block that is not $(u,v)$-movable. Approximate the weights by multiples of $1/w$, and replace every point by the corresponding number of distinct points nearby. Translate the resulting set $Q$ of $w$ points so that $0$ is an optimal center of $Q$. For $z=1$, we keep $w/2$ points on each side of $0$, and for $z>1$, the optimal center of the perturbed set tends to $0$ by strict convexity. By continuity, for large $w$ and small perturbations,
\[
\sum_{p\in Q}|p+1|^z>u\sum_{p\in Q}|p|^z+vw .
\]
Take $0$ as the representative of the block $Q$ and $w$ slots at $-1$. Every bijection then costs $\sum_{p\in Q}|p+1|^z$, which is more than $u$ times the internal cost $\sum_{p\in Q}|p|^z$ plus $v$ times the cost $w$ of moving the mass of the block to the slots. The block $Q$ is therefore not $(u,v)$-movable. \qed

\paragraph{Proof of \Cref{prop:constant}(b).}
\emph{$\beta^*_z<3^z$ for $z>1$.} By convexity, $(3^{z-1},0,(3/2)^{z-1})$ is a certificate (\Cref{lem:triangle-movable}), so $h_0(t)=|1+t|^z-3^{z-1}-(3/2)^{z-1}|t|^z\le0$. Equality requires $|1+t|=1+|t|$, so $t\ge0$, and equality in the convexity step, so $3=\frac32|t|$; hence $h_0(t)<0$ for $t\ne2$. As $t\to-\infty$, $-h_0(t)$ grows like $((3/2)^{z-1}-1)|t|^z$, and $-h_0(1)>0$. The function $-h_0(t)/(1-t)$ is therefore positive and continuous on $(-\infty,1)$ and tends to $\infty$ at both ends; let $g>0$ be its minimum. Then $(3^{z-1}-g,\ g,\ (3/2)^{z-1})$ is a certificate, since $h_0(t)\le g(t-1)$ for all $t$, and its value is $3^z-g$.

\emph{$z=1$.} The certificate $(1,0,1)$ gives $\beta^*_1\le3$. Conversely, the certificate inequality at $t=0$ gives $u\ge1$, and dividing it by $|t|$ and letting $t\to\infty$ and $t\to-\infty$ gives $v+g\ge1$ and $v-g\ge1$; so $u+2v\ge3$.

\emph{$z=2$.} The certificate $(1,2,1)$ gives $\beta^*_2\le5$, and $u,v\ge1$ (\Cref{lem:cert-vector}) gives $\beta^*_2\ge5$.

\emph{$z=4$.} For $q>0$, let $a_q=3/q+2/q^2$. Expanding both sides shows that for all $t\in\R$,
\[
(1+qa_q^2)+(4+4a_q)t+(1+q)t^4-(1+t)^4=q\Bigl(t^2-\frac{2t}q-a_q\Bigr)^2\ge0,
\]
so $(1+qa_q^2,\,4+4a_q,\,1+q)$ is a certificate with value $17+16q+9/q+12/q^2+4/q^3$. Let $q^*$ be the positive root of $16q^4-9q^2-24q-12$. The roots of $t^2-2t/q^*-a_{q^*}$ are $t_1>0>t_2$ with $t_1+t_2=2/q^*$ and $t_1t_2=-a_{q^*}$. The variable $\xi$ that takes the values $t_1$ and $t_2$ with probabilities $-t_2/(t_1-t_2)$ and $t_1/(t_1-t_2)$ has mean zero and
\[
\mathbb E\,\xi^4=-t_1t_2(t_1^2+t_1t_2+t_2^2)=a_{q^*}\bigl(a_{q^*}+4/(q^*)^2\bigr)=\frac9{(q^*)^2}+\frac{24}{(q^*)^3}+\frac{12}{(q^*)^4}=16 .
\]
The identity above is an equality at $t_1$ and $t_2$, so $\mathbb E(1+\xi)^4$ equals the value of the certificate for $q=q^*$. By weak duality, $\beta^*_4$ equals this value, which is about $53.4056$.

\emph{$z=3/2$ and $z=3$.} The triples
\[
\Bigl(\frac{1107}{1000},\frac{2809}{4000},\frac{4289}{4000}\Bigr)\quad\text{for }z=\frac32,\qquad\Bigl(\frac{21263}{5000},\frac{5767}{1000},\frac{15723}{10000}\Bigr)\quad\text{for }z=3
\]
are certificates, with values $\frac{1107}{1000}+\frac{4289}{2000}\sqrt2<4.1398$ and $16.831$. For $z=3$, the certificate inequality on the intervals $t\ge0$, $-1\le t\le0$ and $t\le-1$ states that three cubic polynomials with rational coefficients are nonnegative on intervals. For $z=3/2$, we write the certificate inequality for $t\ge0$ and for $t\le0$ in the variable $|t|^{1/2}$. Its right-hand side $u\pm g|t|+v|t|^{3/2}$ is positive, and squaring the inequality turns it into the nonnegativity of three polynomials of degree $6$ in $|t|^{1/2}$ with rational coefficients on $[0,\infty)$, $[0,1]$ and $[1,\infty)$. We verified all six statements with exact rational arithmetic, by checking with Sturm sequences that each polynomial has no root in its interval and is positive at one end point. For the lower bounds, scale the variable that takes the values $0.4954$ and $-26.15$ (for $z=3/2$), or $2.9486$ and $-1.1093$ (for $z=3$), with probabilities that give mean zero, so that $\mathbb E|\xi|^z=2^z$. Then $\mathbb E|1+\xi|^z$ exceeds $4.1396$ and $16.8308$, respectively. \qed

\begin{figure}[t]
\centering
\begin{tikzpicture}
\begin{axis}[width=0.55\textwidth, height=4.4cm,
  tick label style={font=\scriptsize}, label style={font=\footnotesize},
  every axis plot/.append style={line width=1.1pt},
  xmin=1, xmax=4, ymode=log, ymin=2, ymax=110,
  xtick={1,2,3,4}, ytick={3,5,9,27,81}, yticklabels={$3$,$5$,$9$,$27$,$81$},
  xlabel={$z$},
  legend style={font=\scriptsize, draw=none, fill=none, at={(0.03,0.97)}, anchor=north west, cells={anchor=west}, row sep=-1pt}]
\addplot[domain=1:4, samples=31, sepcol] {3^x}; \addlegendentry{$3^z$}
\addplot[blockcol] table[row sep=\\] {
1.00 3.000000\\ 1.10 3.261450\\ 1.20 3.502556\\ 1.30 3.727941\\ 1.40 3.939784\\ 1.50 4.139656\\ 1.60 4.328855\\ 1.70 4.508516\\ 1.80 4.679660\\ 1.90 4.843208\\ 1.95 4.922400\\ 2.00 5.000000\\ 2.05 5.328364\\ 2.10 5.675685\\ 2.20 6.431785\\ 2.30 7.278129\\ 2.40 8.225676\\ 2.50 9.286652\\ 2.60 10.474711\\ 2.70 11.805101\\ 2.80 13.294854\\ 2.90 14.963006\\ 3.00 16.830827\\ 3.10 18.922089\\ 3.20 21.263359\\ 3.30 23.884334\\ 3.40 26.818203\\ 3.50 30.102061\\ 3.60 33.777370\\ 3.70 37.890470\\ 3.80 42.493150\\ 3.90 47.643294\\ 4.00 53.405588\\
}; \addlegendentry{$\beta^*_z$}
\addplot[only marks, mark=*, mark size=1.4pt, blockcol] coordinates {(1,3) (2,5) (3,16.830827) (4,53.405588)};
\end{axis}
\end{tikzpicture}
\caption{The constant $\beta^*_z$, computed numerically, and the constant $3^z$ of the triangle bound, on a logarithmic scale. The dots mark $\beta^*_1=3$, $\beta^*_2=5$, $\beta^*_3\approx16.83$ and $\beta^*_4\approx53.41$.}
\label{fig:Bz}
\end{figure}
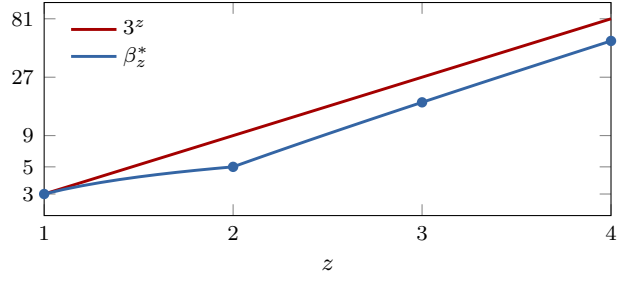

\Cref{fig:Bz} compares $\beta^*_z$ with the constant $3^z$ of the triangle bound.

\paragraph{Computing a certificate.}
We now turn to the algorithmic question of finding such a certificate efficiently. \Cref{thm:soft} in $\R^d$ needs a certificate whose value is close to $\beta^*_z$.

\begin{lemma}\label{lem:compute-certificate}
For fixed $z>1$ and $\eta\in(0,1]$, a $z$-certificate $(u,g,v)$ with $u+2^zv\le\beta^*_z+\eta$ can be computed in time that depends only on $z$ and $\eta$.
\end{lemma}
\begin{proof}
\emph{Bounded coefficients.}
We first bound the coefficients of an optimal certificate $(u^*,g^*,v^*)$. Since $u^*+2^zv^*\le3^z$ and $u^*,v^*\ge1$, we have $u^*,v^*\le3^z$, and the certificate inequality at $t=-1$ gives $g^*\le u^*+v^*\le2\cdot3^z$. We may assume $g^*\ge0$: if $g^*<0$, then $(u^*,0,v^*)$ is also a certificate, since $|1-t|^z\le(1+t)^z\le u^*+g^*t+v^*t^z$ for $t\ge0$.

\emph{Large $|t|$.}
We show that the certificate inequality holds automatically for large $|t|$. Let $\nu=\eta/2^{z+2}$ and
\[
t_0=\max\Bigl\{1,\ \frac{z2^{z+1}}{\nu},\ \Bigl(\frac{4\cdot3^z}{\nu}\Bigr)^{1/(z-1)}\Bigr\}.
\]
For $|t|\ge t_0$, we have $|1+t|^z\le|t|^z+z2^{z-1}|t|^{z-1}\le|t|^z+\frac\nu2|t|^z$ and $2\cdot3^z|t|\le\frac\nu2|t|^z$. Every triple with $u\ge1$, $|g|\le2\cdot3^z$ and $v\ge1+\nu$ therefore satisfies the certificate inequality for $|t|\ge t_0$.

\emph{The linear program.}
We compute a nearly optimal certificate by a linear program over a grid. On $[-t_0,t_0]$, the function $t\mapsto|1+t|^z-gt-v|t|^z$ is Lipschitz with constant $z(1+t_0)^{z-1}+2\cdot3^z+(3^z+1)zt_0^{z-1}$ for all $|g|\le2\cdot3^z$ and $v\le3^z+1$. Take the grid in $[-t_0,t_0]$ whose spacing is $\eta/4$ divided by this Lipschitz constant. Solve the linear program that minimizes $u+2^zv$ subject to $1\le u\le3^z$, $0\le g\le2\cdot3^z$, $1+\nu\le v\le3^z+\nu$ and the certificate inequality at all grid points. The triple $(u^*,g^*,v^*+\nu)$ is feasible, so the optimum is at most $\beta^*_z+2^z\nu\le\beta^*_z+\eta/4$. If $(u,g,v)$ is an optimal solution, every $t\in[-t_0,t_0]$ is within one grid spacing of a grid point, so $(u+\eta/4,g,v)$ is a certificate with value at most $\beta^*_z+\eta/2$. The size of the linear program depends only on $z$ and $\eta$.

\emph{Rational numbers.}
To use rational numbers only, we replace $|1+t|^z$ by an upper bound and $|t|^z$ by a lower bound, both rational and with error at most $\eta/(16\cdot3^z)$. Every solution of this linear program still satisfies the certificate inequality at the grid points, because $v\ge0$. The triple $(u^*+\eta/8,g^*,v^*+\nu)$ is feasible, because the errors change the constraints by at most $(1+v^*+\nu)\eta/(16\cdot3^z)\le\eta/8$. The same argument therefore gives a certificate with value at most $\beta^*_z+5\eta/8$.
\end{proof}

\section{Sharper Ratios}\label{app:sharper}

This appendix sharpens the ratios of \Cref{thm:general} and, for $z>1$, of \Cref{thm:hard}. It also discusses other uncapacitated algorithms as starting points.

\paragraph{The exact bound of greedy deletion.}
For $\rho\ge1$ and $\eps\in(0,1]$, let $\zeta_z(\rho,\eps)\ge\rho$ be defined by
\[
\int_\rho^{\zeta_z(\rho,\eps)}\frac{dw}{(w^{1/z}+2)^z-w}=\ln\frac1\eps ,
\]
that is, $\zeta_z(\rho,\eps)$ is the value at time $\ln(1/\eps)$ of the solution of $w'=(w^{1/z}+2)^z-w$ with $w(0)=\rho$. By~\eqref{eq:greedy-integral}, the greedy deletion of \Cref{lem:greedy} ends with cost at most $\zeta_z(\rho,\eps e^{-\eta/2})\OPT$. Since $\zeta_z$ depends continuously on $\eps$, the proof of \Cref{thm:general} with a small enough $\eta$ gives the ratios $\zeta_z(\rho,\eps)$ for soft and $\zeta_z((1+2\rho^{1/z})^z,\eps)$ for hard capacities. For $z=1$, these are the ratios of \Cref{thm:general}, since $\zeta_1(\rho,\eps)=\rho+2\ln(1/\eps)$.

\begin{lemma}[Growth of $\zeta_z$]\label{lem:Lz}
Let $\rho\ge1$, $\eps\in(0,1]$ and $\tau=\ln(1/\eps)$. Then $\zeta_z(\rho,\eps)\ge(\rho^{1/z}+2\tau)^z$, with equality for $z=1$, and $\zeta_z(\rho,\eps)=(2+o(1))^z\tau^z$ as $\eps\to0$ for fixed $z$ and $\rho$. For $z=2$, $\theta=\zeta_2(\rho,\eps)^{1/2}$ solves $\theta-\ln(1+\theta)=\sqrt\rho-\ln(1+\sqrt\rho)+2\tau$, and if $\rho\ge4$, then $\zeta_2(\rho,\eps)\le(\sqrt\rho+2\tau+\ln(1+\tau))^2$.
\end{lemma}
\begin{proof}
\emph{Bounds for all $z$.}
Substituting $w=\theta^z$ as in the proof of \Cref{lem:greedy} gives
\[
\int_{\rho^{1/z}}^{\zeta_z(\rho,\eps)^{1/z}}\frac{z\theta^{z-1}\,d\theta}{(\theta+2)^z-\theta^z}=\tau .
\]
Since $2z\theta^{z-1}\le(\theta+2)^z-\theta^z\le2z(\theta+2)^{z-1}$, the integrand lies between $\frac12(\theta/(\theta+2))^{z-1}$ and $\frac12$, and it equals $\frac12$ for $z=1$. This gives the lower bound and the case $z=1$. By the lower bound, $\zeta_z(\rho,\eps)\to\infty$ as $\eps\to0$. Since $(\theta/(\theta+2))^{z-1}\to1$ as $\theta\to\infty$, it follows that $\zeta_z(\rho,\eps)^{1/z}=(2+o(1))\tau$.

\emph{The case $z=2$.}
For $z=2$, the integrand is $\frac12\cdot \theta/(\theta+1)$, whose integral is $\frac12(\theta-\ln(1+\theta))$. With $\theta_0=\sqrt\rho$ and $\theta=\zeta_2(\rho,\eps)^{1/2}$, this gives the equation for $\theta$, which we write as
\[
\theta-\theta_0-\ln\Bigl(1+\frac{\theta-\theta_0}{1+\theta_0}\Bigr)=2\tau .
\]
The left side increases with $\theta$. If $\theta_0\ge2$, it is at least $2\tau$ at $\theta=\theta_0+2\tau+\ln(1+\tau)$, since $(2\tau+\ln(1+\tau))/(1+\theta_0)\le\tau$. Therefore $\zeta_2(\rho,\eps)^{1/2}=\theta\le\sqrt\rho+2\tau+\ln(1+\tau)$ for $\rho\ge4$.
\end{proof}

For $k$-means, starting from ratio $9$~\cite{ANSW20}, the ratios are $\zeta_2(9,\eps)$ for soft and $\zeta_2(49,\eps)$ for hard capacities. With $\tau=\ln(1/\eps)$, \Cref{lem:Lz} bounds them by $(3+2\tau+\ln(1+\tau))^2$ and $(7+2\tau+\ln(1+\tau))^2$, which are $(4+o(1))\tau^2$ as $\eps\to0$. The bounds $(3+3\tau)^2$ and $(7+3\tau)^2$ of \Cref{thm:general} are $(9+o(1))\tau^2$. For $\eps=1/2$, $\zeta_2(9,\eps)\approx22.6$ and $\zeta_2(49,\eps)\approx73.4$, instead of $25.8$ and $82.4$.

\paragraph{Other uncapacitated algorithms.}
\Cref{thm:general} can start from any uncapacitated algorithm. The randomized algorithm of~\cite{CGLSS25} has ratio $2+\delta$ for $k$-median, which gives the ratios $2+2\ln(1/\eps)$ and $5+2\ln(1/\eps)$ with high probability. For $k$-means, the ratio $4.9+\delta$ of~\cite{ACCGGLSW27} gives the ratios $\zeta_2(4.9,\eps)$ and $\zeta_2((1+2\sqrt{4.9})^2,\eps)$, which are about $16.4$ and $49.5$ for $\eps=1/2$.

\paragraph{Hard capacities in low dimension.}
In low dimension, \Cref{thm:general} applies with $\rho$ arbitrarily close to $1$. If $P\cup F$ has doubling dimension at most $D$, \Cref{thm:frs} computes such a solution. Otherwise, $P$ or $F$ has doubling dimension at most $D$, which includes the instances in $\R^d$ with clients of bounded doubling dimension and a finite set $F$. As in \Cref{sec:proof-soft-onesided}, we compress the other side by \Cref{lem:compress} with accuracy $\eta_c=\min\{1/4,\eta/(18z)\}$ for a small $\eta>0$. We then apply \Cref{thm:frs} with accuracy $\eta_c$ to the compressed instance; the computed centers are locations in $F$. As there, measured with the original points, the result costs at most $(1+3\eta_c)^{2z}(1+\eta_c)$ times the optimal uncapacitated cost, which tends to $1$ as $\eta\to0$. Since $\zeta_z(\rho,\eps)$ depends continuously on $\rho$, this gives the ratio $\zeta_z(3^z,\eps)$ for hard capacities. For $z>1$ and with $\beta$ as in \Cref{thm:hard}, the ratio of \Cref{thm:hard} therefore improves to
\[
\min\bigl\{(1+2(\beta-(\beta-1)\eps)^{1/z})^z,\ \zeta_z(3^z,\eps)\bigr\}.
\]
For $k$-means in doubling metrics, the second term is smaller if $\eps>0.27$. For $\eps=1/2$, it is $\zeta_2(9,1/2)\approx22.6$ instead of $(1+2\sqrt5)^2\approx29.9$.

\begin{table}[t]
\centering
\footnotesize
\renewcommand{\arraystretch}{1.12}
\setlength{\tabcolsep}{5pt}
\begin{tabular}{@{}llll@{}}
\toprule
Setting & $k$-median & $k$-means & general $z$\\
\midrule
\multicolumn{4}{@{}l}{\emph{uniform capacities}}\\
$\R^d$, soft & $\bm{3-2\eps}$ & $\bm{5-4\eps}$ & $\bm{\beta^*_z-(\beta^*_z-1)\eps}$\\
$\R^d$, hard & $\bm{1+2e^{-\eps}}$ & $\bm{11-8\eps+2\sqrt{5-4\eps}}$ & $\bm{\min\{(1+2(\beta^*_z-(\beta^*_z-1)\eps)^{1/z})^z,\zeta_z(3^z,\eps)\}}$\\
doubling, soft & $\bm{3-2\eps}$ & $\bm{9-8\eps}$ & $\bm{3^z-(3^z-1)\eps}$\\
doubling, hard & $\bm{1+2e^{-\eps}}$ & $\bm{\min\{(1+2\sqrt{9-8\eps})^2,\zeta_2(9,\eps)\}}$ & $\bm{\min\{(1+2(3^z-(3^z-1)\eps)^{1/z})^z,\zeta_z(3^z,\eps)\}}$\\
any metric, soft & $3+2\ln\frac1\eps$ & $\zeta_2(9,\eps)$ & $\zeta_z(\rho,\eps)$\\
any metric, hard & $7+2\ln\frac1\eps$ & $\zeta_2(49,\eps)$ & $\zeta_z((1+2\rho^{1/z})^z,\eps)$\\
\midrule
\multicolumn{4}{@{}l}{\emph{non-uniform capacities}}\\
doubling, soft & $\bm{1+2e^{-\eps}}$ & $\bm{1+8e^{-\eps}}$ & $\bm{1+(3^z-1)e^{-\eps}}$\\
doubling, hard & $\bm{1+2e^{-\eps}}$ & --- & ---\\
\bottomrule
\end{tabular}
\caption{All ratios with $K=k+\lfloor\eps k\rfloor$ centers and exact capacities: those of \Cref{thm:soft,thm:hard,thm:nonuniform,thm:general}, sharpened by this appendix. Here $\zeta_z(\rho,\eps)$ is the exact bound of greedy deletion, and $\rho$ is the ratio of the uncapacitated solution from which \Cref{thm:general} starts; for $k$-means, $\rho=9$~\cite{ANSW20}. By \Cref{lem:Lz}, $\zeta_z(\rho,\eps)=O(1+\log^z\frac1\eps)$ for fixed $z$ and $\rho$, and $\zeta_2(9,\eps)\le(3+3\ln\frac1\eps)^2$ and $\zeta_2(49,\eps)\le(7+3\ln\frac1\eps)^2$. Bold entries are bounded by a constant as $\eps\to0$.}
\label{tab:results}
\end{table}

\section{Implementation with Rational Numbers}\label{app:bits}

In this appendix, the input consists of rational numbers: the coordinates of the clients in the Euclidean setting, or the matrix of distances between the points of $P\cup F$ otherwise. We also assume that $z$ is rational, or more generally that $w\mapsto w^z$ can be approximated to any given precision in time polynomial in the precision and the input size. We list the changes needed to implement the algorithms in time polynomial in the input size. Let $\eta$ be the accuracy parameter in the proof of the theorem at hand, and $\eta=\delta$ for \Cref{cor:bicriteria}. Every change weakens an inequality of the exact analysis by a factor of $1+O(\eta)$ or by an additive $O(\eta)\cdot\OPT$, with absolute constants in the $O$-terms. We collect these losses at the end. Rounding errors are additive and are charged to a positive lower bound $\uOPT$ on $\OPT$.

\paragraph{A lower bound.}
Whether $\OPT=0$ can be checked directly, since every client must then be at the location of its center. A point $x$ with $n_x$ clients needs $\lceil n_x/U\rceil$ centers when centers may share a location, and $n_x\le U$ is needed with hard capacities; if $\OPT=0$, we output such a solution. Otherwise, in the Euclidean setting, $\OPT\ge\uOPT:=2^{1-z}r_{\min}^z$ by \Cref{sec:euclid-proof}, where $r_{\min}$ is the minimum distance between two distinct points of clients. In the other settings, $\OPT\ge\uOPT$, the $z$-th power of the minimum positive distance between a client and a location.

\paragraph{Working metric.}
Distances may be irrational, for instance Euclidean distances between rational points. In every setting, let $\eta_c=\min\{1/4,\eta/(18z)\}$, which is the accuracy of the compression lemma where this lemma is used; note $\eta_c\le\eta$. For every pair of distinct points $x,y$ that the algorithm uses, compute a rational upper approximation of $\dist(x,y)$ with relative error at most $\eta_c/z$. Let $\dist'$ be the shortest-path metric of the complete graph on these points with these edge lengths. Every path between $x$ and $y$ has length at least $\dist(x,y)$ by the triangle inequality, and the direct edge has length at most $(1+\eta_c/z)\dist(x,y)$, so $\dist\le\dist'\le(1+\eta_c/z)\dist$. If a set has doubling dimension at most $D$ under $\dist$, it has doubling dimension at most $2D$ under $\dist'$. Indeed, a $\dist'$-ball of radius $r$ lies in the $\dist$-ball of radius $r$, which is covered by $2^{2D}$ $\dist$-balls of radius $r/4$. Each of them lies in the $\dist'$-ball of radius $(1+\eta_c/z)r/4\le r/2$ around the same point. We run the compression lemma, the construction of the decompositions and the network with $\dist'$, with $D$ replaced by $2D$. In particular, the framework uses $\Lambda=2^{14}D/\gamma$ and $M=\lceil(4/\lambda)^{2D}\rceil$, so that \Cref{lem:family} and~\eqref{eq:portals} hold for $\dist'$. \Cref{lem:routes}(a) still gives path costs of at least $\dist'(x,y)^z\ge\dist(x,y)^z$. For pairs that are not $\Lambda$-bad, \Cref{lem:routes}(b) gives at most $((1+\eta/z)\dist'(x,y))^z\le e^{2\eta}\dist(x,y)^z$ instead of $e^\eta\dist(x,y)^z$. In the compression lemma, the distortion with respect to $\dist$ becomes $(1+3\eta_c)(1+\eta_c/z)$ instead of $1+3\eta_c$. Movability and all costs of solutions refer to the original distances.

\paragraph{Costs.}
The dynamic program uses rational upper approximations of the arc costs, with additive error at most $\eta\uOPT/(4n\max\{1,v\})$, and of the values $(u-1)A_i$, with additive error at most $\eta\uOPT/(4n)$. The flow $f$ of a network solution consists of at most $n$ unit paths, each of which uses at most one crossing arc, and all other arcs cost $0$. The approximate value of $v\cost(f)$ therefore exceeds the true one by at most $\eta\uOPT/4$. As there are $m\le n$ blocks, the approximate cost of every network solution exceeds its true cost by at most $\eta\uOPT/2$. The dynamic program minimizes the approximate cost exactly, so the cost of its solution is at most the cost of the routed rescued solution plus $\eta\uOPT/2$. The matchings in \Cref{lem:to-clustering}, the minimum-cost flows, and the changes of transport plans in the proof of \Cref{lem:soft-to-hard} compare costs by rational upper approximations of $\dist(x,y)^z$ with additive error at most $\eta\uOPT/(4n)$. Each of these computations minimizes the approximate cost, or never increases it, and the plans and assignments involved move at most $n$ units. The result therefore costs at most $\eta\uOPT/4$ more than in the exact computation. The local search of \Cref{thm:frs} is run with costs approximated to relative error $\eta/(4k)$, accepting a swap only if it decreases the approximate cost by a factor of at least $1-\eta/(4k)$. When the search stops, every swap leaves the true cost at least $(1-\eta/(4k))/(1+\eta/(4k))\ge1-\eta/(2k)$ times the current cost. The analysis of~\cite{FRS19} combines the local-optimality inequalities of at most $k$ test swaps, so for such an approximate local optimum, its guarantee grows by a factor of at most $1/(1-\eta/2)\le1+\eta$~\cite[Section~2]{FRS19}. Since every accepted swap decreases the approximate cost by a factor of at least $1-\eta/(4k)$, the number of iterations is polynomial in $k/\eta$ and in the logarithm of the ratio between the initial cost and $\uOPT$.

\paragraph{Candidate centers.}
The construction of \Cref{lem:candidates} uses square roots and real powers. We run it with $N$ replaced by $N'=100N$ and $s$ replaced by a rational number $\tilde s$ with $s\le\tilde s\le(1+1/(8z))s$. Every subgradient is replaced by a rational vector within distance $z\tilde s^{z-1}/\sqrt{N'}$ of it, and every iterate by a rational point within distance $\tilde s/(zN')$ of the exact projected step. The step size is $\tau=\tilde s^{2-z}/(z\sqrt{N'})$. In the proof of \Cref{lem:candidates}, we then have $\|c^*-p_0\|+\tilde s\le(3+\frac1{4z})r$, so the total weight is at most $(4+\frac1{4z})^z\Psi_Q(c^*)\le1.07\cdot4^z\Psi_Q(c^*)$. The iterates lie within distance $\tilde s/(zN')$ of $B(p_0,\tilde s)$, so $\|c_\nu-c^*\|\le3\tilde s$. The bound $z\tilde s^{z-1}$ on the norms of the exact subgradients grows by a factor of at most $(1+1/(zN'))^{z-1}\le1+2/N'$, so the approximate subgradients have norm at most $(1+2/N'+1/\sqrt{N'})z\tilde s^{z-1}$. The inequality for $\|c_{\nu+1}-c^*\|^2$ therefore gains the error term $2\tau\cdot3\tilde s\cdot z\tilde s^{z-1}/\sqrt{N'}=6\tilde s^2/N'$ from the approximate subgradient and the error term $4\tilde s^2/(zN')+(\tilde s/(zN'))^2$ from the rounded iterate. Its last term $\tilde s^2/N'$ grows by the factor $(1+2/N'+1/\sqrt{N'})^2$. Summing over the $N'$ steps and dividing by $2\tau N'$ as before gives
\[
\mathbb E[\bar\psi(\bar c)]-\bar\psi(c^*)\le\Bigl(\frac12+3+\frac12\Bigl(1+\frac2{N'}+\frac1{\sqrt{N'}}\Bigr)^2+\frac2z+\frac1{2z^2N'}\Bigr)\frac{z\tilde s^z}{\sqrt{N'}}\le\frac{8z\tilde s^z}{\sqrt{N'}}.
\]
Multiplying by the total weight divided by $\tilde s^z$ gives $\mathbb E[\Psi_Q(\bar c)]-\Psi_Q(c^*)\le1.07\cdot8z4^z\Psi_Q(c^*)/\sqrt{N'}\le z4^z\Psi_Q(c^*)/\sqrt N$, the bound of the exact analysis. All points of $\Gamma$ are rational with polynomial bit size, so $\log\Delta(P\cup\Gamma)$ is polynomial in the input size.

\paragraph{Other steps.}
In the proof of \Cref{thm:soft} in $\R^d$, the number $r_0$ can be replaced by a smaller positive rational number, which keeps the moved centers rational. The greedy deletion of \Cref{sec:greedy} is run in the working metric $\dist'$, which changes its guarantee by a factor of at most $(1+\eta_c/z)^z\le e^{\eta_c}$. It compares the costs of families by rational upper approximations with relative error at most $\eta/(8k)$. The cost after every chosen deletion is therefore at most $1+\eta/(8k)$ times the cost after the cheapest one. Dividing the cost by $1+\eta/(8k)$ raised to the number of deletions so far gives a sequence that satisfies the recurrence in the proof of \Cref{lem:greedy}, since $((w^{1/z}+2)^z-w)/w=(1+2w^{-1/z})^z-1$ is nonincreasing in $w$. The fewer than $2k$ deletions therefore lose a factor of at most $(1+\eta/(8k))^{2k}\le e^{\eta/4}$. The certificate of \Cref{lem:compute-certificate} is rational. All numbers that appear have bit size polynomial in the input size, and $H=O(\log\Delta(P\cup F))$ is polynomial in the input size.

\paragraph{Non-uniform capacities.}
Whether $\OPT=0$ is checked as above, with the largest capacities of the locations at a point. The linear program of \Cref{lem:sparse} uses rational upper approximations of $\dist(x,y)^z$ with relative error at most $\eta/4$, and an optimal vertex is computed in polynomial time~\cite{GLS88}. This weakens the bound of \Cref{lem:sparse} by a factor of at most $1+\eta/4$. The greedy additions of \Cref{lem:augment} compare costs by rational upper approximations with relative error at most $\theta=\eta/(8K')$. The cost after an addition is then at most $1+\theta$ times $\OPT+(1-1/k)(\cost(S)-\OPT)$, where $S$ is the family before it. By induction, the cost at the end is at most $(1+\theta)^{K'}\le e^{\eta/8}$ times the bound of \Cref{lem:augment}.

\paragraph{Accounting.}
Compared with the exact analysis, the network stretch for pairs that are not $\Lambda$-bad grows from $e^\eta$ to $e^{2\eta}$. The bound on $A$ grows by a factor $1+O(\eta)$ because of the working metric and the approximate local search. The distortion of the compression lemma grows by the factor $1+\eta_c/z$, and the rounding errors add at most $\eta\uOPT\le\eta\OPT$. In the proof of \Cref{thm:soft} in $\R^d$, the representatives are chosen by costs computed with $\dist'$ and rounded up with relative error at most $\eta_c/z$. Since $\eta_c\le\eta/(18z)$, we still have $((1+3\eta_c)(1+\eta_c/z))^z\le1+\eta/3$ in~\eqref{eq:locations} and
\[
\Psi_{P_i}(a_i)\le(1+\eta_c/z)^{2z}(1+\eta/3)^2\min_c\Psi_{P_i}(c)\le(1+\eta)\min_c\Psi_{P_i}(c),
\]
so $\omega$ is unchanged. Following these losses through the proofs, the final cost bounds grow by a factor of $1+O(\eta)$ and by an additive $O(\eta)\cdot\OPT$. In every proof, $\eta$ times the ratio of the theorem is at most $\delta$, so the final bounds grow by $O(\delta)\cdot\OPT$. Running the algorithms with $\delta$ divided by a suitable absolute constant in place of $\delta$ therefore gives the guarantees of \Cref{thm:soft,thm:hard,thm:nonuniform,cor:bicriteria,thm:general}.

\phantomsection
\addcontentsline{toc}{section}{References}
\bibliographystyle{alpha}
\bibliography{refs}

\end{document}